\documentclass[aps,prd,superscriptaddress,showpacs,twocolumn,nofootinbib]{revtex4}
\usepackage{graphicx}
\usepackage{dcolumn}
\usepackage{bm}
\usepackage{natbib}
\usepackage{amsmath}
\usepackage{amssymb}

\newcommand{\be}{\begin{equation}}
\newcommand{\ee}{\end{equation}}
\newcommand{\bea}{\begin{eqnarray}}
\newcommand{\eea}{\end{eqnarray}}

\newcommand{\WMAP}{{\slshape WMAP~}}
\newcommand{\PLANCK}{{\slshape Planck~}}
\newcommand{\WMAPc}{{\slshape WMAP}}
\newcommand{\PLANCKc}{{\slshape Planck}}

\newcommand{\s}{{\rm ~s}}

\newcommand{\microK}{\mu{\rm K}}

\newcommand{\pann}{p_\mathrm{ann}}
\newcommand{\pdec}{p_\mathrm{dec}}

\begin{document}
%\draft

\title{Searching for Dark Matter in the CMB:\\
A Compact Parameterization of Energy Injection from New Physics}

\author{Douglas P. Finkbeiner}
\email{dfinkbeiner@cfa.harvard.edu}
\affiliation{Harvard-Smithsonian Center for Astrophysics, 60 Garden St., Cambridge, MA 02138, USA}
\affiliation{Physics Department, Harvard University, Cambridge, MA 02138, USA}

\author{Silvia Galli}
\email{galli@apc.univ-paris7.fr}
\affiliation{Laboratoire Astroparticule et Cosmologie (APC), Universit\'e Paris Diderot, 75205 Paris cedex 13, France}
\affiliation{Physics Department and INFN, Universita' di Roma ``La Sapienza'', Ple Aldo Moro 2, 00185, Rome, Italy}

\author{Tongyan Lin}
\email{tongyan@physics.harvard.edu}
\affiliation{Physics Department, Harvard University, Cambridge, MA 02138, USA}

\author{Tracy R. Slatyer}
\email{tslatyer@ias.edu}
\affiliation{School of Natural Sciences, Institute for Advanced Study, Princeton, NJ 08540, USA}

%\date{\today}

\begin{abstract} 
High-precision measurements of the temperature and polarization
anisotropies of the cosmic microwave background radiation have been
employed to set robust constraints on dark matter annihilation during
recombination. In this work we improve and generalize these
constraints to apply to energy deposition during the recombination era
with arbitrary redshift dependence. Our approach also provides more
rigorous and model-independent bounds on dark matter annihilation and
decay scenarios. We employ principal component analysis to identify a
basis of weighting functions for the energy deposition. The
coefficients of these weighting functions parameterize any energy
deposition model and can be constrained directly by experiment.  For
generic energy deposition histories that are currently allowed by \WMAP7 data,
up to 3 principal component coefficients are measurable by \PLANCK and
up to 5 coefficients are measurable by an ideal cosmic variance
limited experiment.  
For WIMP dark matter, our analysis demonstrates that the effect on the
CMB is described well by a single (normalization) parameter and a ``universal''
redshift dependence for the energy deposition history. We give \WMAP7
constraints on both generic energy deposition histories and the universal WIMP case.
\end{abstract}


\maketitle

\section{Introduction}

Measurements of the cosmic microwave background (CMB) in the past decade by experiments including \WMAPc, ACBAR and BOOMERANG \cite{Komatsu:2010fb, Reichardt:2008ay,Montroy:2005yx}, and more recently SPT, QUaD and ACT \cite{Carlstrom:2009um, Brown:2009uy, ACT:2010cy}, have provided an unprecedented window onto the universe around redshift 1000. With the advent of the \PLANCK Surveyor \cite{Planck:2006uk}, the successor experiment to \WMAPc, percent-level modifications to recombination will be observable. \PLANCK has already completed three sky surveys and begun a fourth, and cosmological data are expected to be released publicly in 2012-13.

Accurate measurements of the CMB have the potential to probe the physics of dark matter (DM) beyond its gravitational interactions. In the large class of models where the DM is a thermal relic, its cosmological abundance is determined by its annihilation rate in the early universe: the correct relic density ($\sim 22\%$ of the energy density of the universe) is obtained for an $s$-wave annihilation cross section of $\langle \sigma v \rangle \approx 3 \times 10^{-26}$ cm$^3$/s during freezeout.

DM annihilation at this rate modifies the ionization history of the universe and has a potentially measurable effect on the CMB. During the epoch of recombination, DM annihilation produces high-energy photons and electrons, which heat and ionize the hydrogen and helium gas as they cool. The result is an increased residual ionization fraction after recombination, giving rise to a low-redshift tail in the last scattering surface. The broader last scattering surface damps correlations between temperature fluctuations, while enhancing low-$\ell$ correlations between polarization fluctuations.

The resulting constraints on the dark matter annihilation rate have been studied by several authors \cite{Padmanabhan:2005es, Galli:2009zc, Slatyer:2009yq,  Chluba:2009uv, Galli:2010it,Hutsi:2011vx,Galli:2011rz}. These bounds have a notable advantage over other indirect constraints on dark matter annihilation, in that they are independent of the DM distribution in the present day, and do not suffer from uncertainties associated with Galactic astrophysics. They depend only on the cosmological DM density, which is well measured; the DM mass; the annihilation rates to the final states; and the standard physics of recombination. Recombination modeling, while not simple, involves only well-understood conventional physics, and the latest models are thought to be accurate at the sub-percent level required for \PLANCK data \cite{Chluba:2010ca, AliHaimoud:2010dx}. 

Current limits from \WMAP already significantly constrain models of light dark matter with masses of around a few GeV and below, if the annihilation rate at recombination is the thermal relic cross section. Heavier DM is also constrained if the annihilation rate is enhanced at low velocities or for other reasons is much larger than the thermal relic cross section at recombination. Models lying in these general categories are also of significant interest for their possible connection with experimental anomalies. 

The DAMA/LIBRA \cite{Bernabei:2010ke} and CoGeNT \cite{Aalseth:2010vx, Aalseth:2011wp} direct detection experiments have reported excess events and annual modulation that may have a consistent explanation as originating from scattering of $\mathcal{O}(5-10)$ GeV WIMPs (e.g. \cite{Hooper:2010uy}). Results from the XENON10, XENON100 and CDMS experiments are in tension with this interpretation \cite{Savage:2010tg, Ahmed:2010wy}, but there is ongoing debate on the sensitivity of these experiments to the very low-energy nuclear recoils in question (see e.g. \cite{Collar:2010ht,Collar:2011kf}).

The PAMELA \cite{Adriani:2008zr}, \emph{Fermi}  \cite{Abdo:2009zk}, PPB-BETS \cite{Torii:2008xu}, ATIC \cite{ATIClatest} and H.E.S.S \cite{Aharonian:2009ah} have measured electron and positron cosmic rays in the neighborhood of the Earth, and found results consistent with a new primary source of $e^\pm$ in the $10-1000$ GeV energy range. If the signal is attributed to dark matter annihilation then the annihilation rate in the Galactic halo today must be 1-3 orders of magnitude above the canonical thermal relic value \cite{Cholis:2008qq, Cholis:2008wq}. This has motivated models of dark matter with enhanced annihilation at low velocities \cite{ArkaniHamed:2008qn, Pospelov:2008jd}. While this enhancement would not be significant during freezeout, it would be effective at recombination when the typical velocity of dark matter is $v \sim 10^{-8} c$ \cite{Galli:2009zc}.

With the release of data from \PLANCK expected in the next two years, models falling into these categories should either be robustly ruled out, or give rise to a measurable signal \cite{Galli:2009zc, Slatyer:2009yq, Hutsi:2011vx}. If no signal is observed, the sensitivity of \PLANCK will allow us to probe regions of parameter space relevant for supersymmetric models, where the DM is a thermal relic with mass of several tens of GeV. It is timely to explore improvements to these constraints.

The approach of previous studies has been to specify the energy deposition history (redshift dependence) and then calculate the effect on the ionization history and anisotropy spectrum using public codes such as \texttt{RECFAST} and \texttt{CAMB}. A single parameter describing the normalization of the signal is then added to the standard likelihood analysis using \texttt{CosmoMC}, and bounded by \WMAP observations. The redshift dependence has been studied in two cases: in the ``on-the-spot'' case, assuming that the amount of energy deposited to the gas precisely tracks the rate of dark matter annihilation (e.g. \cite{Padmanabhan:2005es, Galli:2009zc, Galli:2011rz}), or employing detailed energy deposition histories for specific models (e.g. \cite{Slatyer:2009yq, Hutsi:2011vx, Galli:2011rz}). In the first case, model-independent constraints are obtained, but without a precise way to connect the bounds to any particular model. The second case only precisely constrains specific models. 

While these analyses have been adequate for simple estimates of whether a model is strongly ruled out, easily allowed, or on the borderline, the upcoming high-precision data from \PLANCK demand a more careful model-independent analysis. Such an analysis can also be applied to more general classes of energy deposition histories during and after recombination: for example, the energy deposited by a late-decaying particle species, decay from an excited state of the dark matter, or dark matter annihilation in models where the redshift dependence of the annihilation rate has an unusual form (as in some models of asymmetric dark matter).
 
In this work we exploit the fact that the effects of energy deposition at different redshifts are not uncorrelated. Any arbitrary energy deposition history can be decomposed into a linear combination of orthogonal basis vectors, with orthogonal effects on the observed CMB power spectra ($C_\ell$'s). For a broad range of smooth energy deposition histories, the vast majority of the effect on the $C_\ell$'s can be described by a small number of independent parameters, corresponding to the coefficients of the first few vectors in a well-chosen basis. These parameters in turn can be expressed as (orthogonal) weighted averages of the energy deposition history over redshift.

We employ principal component analysis (PCA) to make this statement quantitative and derive the relevant weight functions, and the corresponding perturbations to the $C_\ell$ spectra. Our approach in principle generalizes to all possible energy deposition histories. To investigate the number of observable parameters, we consider generic perturbations about two physically interesting fiducial cases. We focus primarily on the example of dark matter annihilation, or any other scenario where the power deposited per volume scales approximately as $(1+z)^6$ (i.e. as density squared), as an energy deposition mechanism, but also show results for the case of dark matter decay, or similar scenarios where the power deposited scales as  $(1+z)^3$.

Our computation of the effects of energy deposition on the CMB anisotropies, and the approximations we use for estimating the significance of these  effects in experimental data, are described in \S \ref{sec:fisher}. In \S \ref{sec:pca} we present our principal component analysis for both ``annihilation-like'' and ``decay-like'' general energy deposition histories\footnote{Files containing the results of these analyses are available online at \texttt{http://nebel.rc.fas.harvard.edu/epsilon/}; see also Appendix \ref{app:website}.}. There are significant degeneracies between energy deposition and perturbations to the cosmological parameters, and so we marginalize over the standard cosmological parameters when deriving the principal components\footnote{We test the effect of including additional cosmological parameters (running of the scalar spectral index, the number of massless neutrino species, and the primordial He fraction) and find no large degeneracy with energy injection, justifying our neglect of these additional parameters in our main analysis.}.

We then address the constraints on and detectability of the principal components in current and future experiments. Given a $C_\ell$ spectrum observed by an experiment (e.g. \PLANCKc), we can measure the residual with respect to the best-fit standard $\Lambda$CDM model, and then project this residual onto the $C_\ell$-space directions corresponding to the principal components. Given any model for the energy deposition history, we can then ask if the reconstructed coefficients for the various principal components are consistent with the model. Of course, for the later principal components the effect on $C_\ell$'s is so small that very little information on their coefficients can be recovered. In \S \ref{sec:detectability}, we make this statement quantitative, and estimate the number of principal components whose coefficients could be detectable in \PLANCK and an ideal cosmic variance limited (CVL) experiment, subject to constraints from \WMAP7. The CVL case presents a hard upper limit on the number of independent parameters describing the energy deposition history that can profitably be retained in the analysis. We also discuss the bias to the standard cosmological parameters, in the case where there is non-zero energy deposition that is neglected in the analysis; in our framework it is straightforward to characterize the biases to the cosmological parameters for an arbitrary energy deposition history.

In \S \ref{subsec:dmpca} we present a separate principal component analysis for the more limited case of conventional GeV-TeV WIMPs annihilating to Standard Model final states. We demonstrate that in this case, all the effect of dark matter annihilation can be captured by one parameter only, i.e. the amplitude of the first principal component.

Finally, in \S \ref{sec:cosmomc}, we estimate the constraints on the principal components obtainable with current (\WMAP7) and future (\PLANCKc, CVL) experiments with a full likelihood analysis using the \texttt{CosmoMC} code. We employ here the principal components obtained with the Fisher matrix analysis -- which assumes  that the effect on the CMB scales linearly with the energy deposition. We illustrate the range of validity of this assumption for the different experiments considered.
We check that the constraints previously obtained with our Fisher matrix analysis -- which assumes Gaussian likelihood functions -- are compatible with the ones obtained with the \texttt{CosmoMC} analysis. 
 We check that the constraints on a given energy deposition history can be reconstructed from the constraints on the principal components.  
 
Appendix \ref{app:validation} considers the effects on the analysis of changing various assumptions and conventions, including the effect of additional cosmological parameters and using different codes to calculate the ionization histories. We find that the only such choice that non-negligibly modifies the early (detectable) principal components is the treatment of Lyman-$\alpha$ photons, although the inclusion of additional cosmological parameters can change the constraints at the $\sim 10\%$ level.
Appendix \ref{app:marg} discusses marginalization over the cosmological parameters. Appendix \ref{app:website} describes the results from this analysis that we have made available online.

%%%%%%%%%%%%%%%%%%%%%%%%%%%%%%%%%%%%%%%%%%%%

\section{The Effect of Energy Injection}

\label{sec:fisher}

\begin{figure}
\includegraphics[width=.5\textwidth]{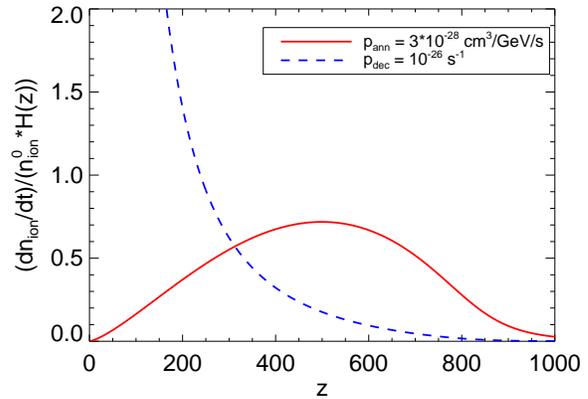}
\caption{\label{fig:DMionization}
Rate of Hydrogen ionization % per volume and per Hubble time 
from energy deposition, relative to the number density of ionized
Hydrogen ($n_{ion}^0$) when there is no energy deposition. The lines
shown are the cases of constant $\pann$ and $\pdec$, corresponding to
on-the-spot energy deposition from dark matter annihilation and dark
matter decay, respectively.
%(As usual, this figure assumes the (Shull/etc) scenarios for the number of ionizations.)
}
\end{figure}

%%%%%%%%%%%%%%%%%%%%%%%%%%%%%%%%%%%%%%%%%%%%

We begin by considering DM annihilation-like or decay-like energy
deposition histories. The energy injection from these sources scales
respectively as density squared and density, so these cases cover the
generic scenarios where energy is injected by two-body or one-body
processes.  It is convenient to express the energy injection as a slowly
varying function of $z$ that depends on the source of the energy injection
(e.g. the WIMP model) and a term containing cosmological parameters.  We
parameterize the energy deposition histories, respectively, as,
\begin{align}
\label{eq:pdef}
 \left( \frac{dE}{dt \, dV} \right)_\mathrm{ann} & = \pann(z) c^2 \Omega_\mathrm{DM}^2  \rho_c^2 (1+z)^6,   \nonumber \\
 \left( \frac{dE}{dt \, dV}\right)_\mathrm{dec} & =  \pdec(z) c^2 \Omega_\mathrm{DM}  \rho_c (1+z)^3,
\end{align}
where $\pann(z)$ (or $\pdec(z)$) contains all of the information about the
source of energy injection and the efficiency with which that
energy ionizes the gas.  We generically refer to $\pann$ and $\pdec$ as the
``energy deposition yield.''
For consistency with \cite{Galli:2011rz}, we express $\pann(z)$ in
units of cm$^3$/s/GeV, while the units of $\pdec(z)$ are s$^{-1}$. If
the energy injection is due to DM annihilation, $\pann = f(z) \langle
\sigma v \rangle / m_\mathrm{DM}$ \cite{Galli:2011rz}, where $f(z)$ is
an $\mathcal{O}(1)$ dimensionless efficiency factor
\cite{Slatyer:2009yq}; if the energy injection is due to DM decay,
$\pdec(z) = f(z)/\tau$, where $\tau$ is the decay lifetime. Other
authors have written $\pann$ in units of m$^3$/s/kg
\cite{Galli:2009zc}, or parameterized the energy deposition in
eV/s/baryon \cite{Padmanabhan:2005es, Slatyer:2009yq, Chluba:2009uv}.
For calibration, the energy deposition from a 100 GeV thermal relic
WIMP with $f(z) = 1$ corresponds to $\pann \approx 3 \times 10^{-28}
\mathrm{cm}^3$/s/GeV $\approx 1.7 \times 10^{-7} \mathrm{m}^3$/s/kg,
or an energy deposition of $2.1 \times 10^{-24}$ eV/s/H, assuming the \WMAP7 best-fit cosmology. Throughout this work, we employ the cosmological parameters from \cite{Larson:2010gs} as a baseline: explicitly, $\omega_b = 2.258 \times 10^{-2}$, $\omega_c = 0.1109$, $A_s$($k$=0.002 Mpc$^{-1}$) $= 2.43 \times 10^{-9}$, $n_s = 0.963$, $\tau = 0.088$, $H_0 = 71.0$ km/s/Mpc.

Energy deposition during recombination primarily affects the CMB through
additional ionizations; the modified ionization history then leads to an
increased width for the surface of last scattering. An alternate approach to
studying generic energy deposition histories might be to study generic
ionization histories \cite{Hu:2003gh}, since the former can be directly mapped to the
latter. We frame the problem in terms of energy deposition histories because
this provides direct constraints on physical energy injection models.

Suppose we are interested primarily in a class of energy deposition
histories for which the energy deposition yield $p(z)$ (that is, $\pann(z)$ or $\pdec(z)$, as
appropriate) is not very rapidly varying. Then we can discretize
$p(z)$ as a sum over a basis of $N$ $\delta$-function-like energy
deposition histories, $p(z) = \sum_{i=1}^N \alpha_i G_i(z)$. The basis
functions $G_i(z)$ are (by default) Gaussians with $\sigma = \Delta z
/ 4$, centered on $z_i$ ($i=1..N$), where $\Delta z$ is the spacing
between the $z_i$. They are normalized such that $\int dz G_i(z) =
\Delta z$. For example, in the limit of large $N$ an energy deposition history
with constant $p(z)=p_0$ corresponds to $\alpha_i=p_0$ for all $i$.

If the energy deposition is small enough, the effect on the CMB
anisotropy power spectrum is linear in the energy depositions at
different redshifts [$\delta C_l(p(z))=\delta C_l(\sum_{i=1}^N \alpha_i G_i(z))=\sum_{i=1}^N\delta C_l(\alpha_i G_i(z))$], and in the amount of energy deposition at any
redshift [$\delta C_l(\sum_{i=1}^N \alpha_i G_i(z))=\sum_{i=1}^N \alpha_i \delta C_l(G_i(z))$]. Then the effect of an arbitrary energy deposition history can be
determined simply from studying the basis functions $G_i(z)$. We will
assume linearity throughout this work, and justify that assumption in
\S \ref{sec:derivs}.

Of course, given any annihilation-like energy deposition history, it can be rewritten in decay-like form with a strongly redshift-dependent $p(z)$, and vice versa. The basis of $G_i(z)$ functions can describe any energy deposition history, at least in the large-$N$ limit. However, the very different ``underlying'' redshift dependence in the two cases, and the uncertainties associated with the annihilation rate at low redshift (due to the onset of structure formation), motivate us to study different redshift ranges in the two cases.

For each $G_i$, we can compute the effect on the ionization history
and the anisotropy spectrum in the limit of small energy deposition. We
determine $\partial C^{TT}_\ell/\partial \alpha_i, \, \partial
C^{EE}_\ell/\partial \alpha_i, \, \partial C^{TE}_\ell/\partial
\alpha_i$ $\forall i, \ell$. In our default analysis we employ the
\texttt{CosmoRec} and \texttt{CAMB} codes, with the prescription for
including the extra energy deposition laid out in \cite{Chen:2003gz,
Padmanabhan:2005es}.  If there are $N$ basis functions and we take
$n_\ell$ spherical harmonics into account, this yields an $n_\ell
\times N$ transfer matrix $T$ whose $(\ell,i)$th element is, 
\begin{equation} \frac{\partial
C_{\ell}}{ \partial \alpha_i} = \left\{\frac{\partial C^{TT}_{\ell}}{\partial
\alpha_i}, \, \frac{\partial C^{EE}_{\ell}}{\partial \alpha_i}, \, \frac{\partial
C^{TE}_{\ell}}{\partial \alpha_i} \right\}. \label{eq:transfermatrix} \end{equation}

In this work we focus primarily on annihilation-like energy deposition histories,
for which we restrict ourselves to the $80 < z < 1300$ range; as a
default, we will take 50 redshift bins covering this range. At higher
redshifts the universe is ionized and so the effect of energy
deposition on the ionization history is negligible, while at lower
redshifts the DM number density becomes so small that the energy
injected from annihilation is insignificant, as shown in Figure
\ref{fig:DMionization}. This in turn justifies our neglect of DM
structure formation: while for $z \lesssim 100$, DM clumps start to
form and the annihilation rate no longer tracks the square of the
average relic density, the energy injection is already sufficiently
suppressed that the signal remains negligible.

For DM decay, the signal is not nearly so suppressed at low redshifts,
and so we consider the redshift range $10 < z < 1300$. With this
expanded redshift range, we switch from linear to log binning, with 90
bins covering this redshift range\footnote{Log binning can of course
also be employed for the annihilation-like case; there is no clear
best choice there, so we will use linear binning as the default but
show results for both options. See Appendix \ref{app:validation} for a
discussion.}; we take the basis functions $G_i(\ln(1+z))$ to be
Gaussians in $\ln(1+z)$, normalized so that their integral with
respect to $d \ln (1+z)$ is given by the spacing between the log bins
$\Delta \ln (1+z)$. With these choices, again an energy deposition history with
constant $p(z)=p_0$ corresponds (in the large-$N$ limit) to $p_0
\sum_{i=1} G_i$.

We again ignore structure formation in the decay-like case, where the
total power injected depends only on the average density. The universe
is rather transparent to the products of DM decay and annihilation at
these redshifts, so even a very spatially non-uniform distribution of
energy injection would not be expected to cause ionization or
temperature hot-spots (at least for particles injected at weak-scale
energies; de-excitation of nearly-degenerate states or annihilation of
very light DM might change this conclusion to some degree). Modeling
of reionization may pose a more significant challenge for analyses
relying on low redshifts ($z \sim 10$); note, however, that the
transparency of the universe at these redshifts means that in
realistic scenarios (even decay-like scenarios) the bulk of the effect
on the CMB comes from earlier times.

%%%%%%%%%%%%%%%%%%%%%%%%%%%%%%%%%%%%%%%%%%%%

\subsection{Brief review of the Fisher matrix}
\label{subsec:fisher}

The degree to which energy deposition is observable in the CMB can be
captured by the Fisher matrix for energy deposition, denoted $F_e$,
which is obtained by contracting the transfer matrix $T$ (Equation
\ref{eq:transfermatrix}) with the appropriate covariance matrix for
the $C_\ell$'s (e.g. \cite{jungman96b,tegmark97,Verde:2009tu}),
\begin{align} 
\Sigma_\ell & = \frac{2}{2 \ell + 1} \times \nonumber \\
&  \left( \begin{array}{ccc} 
\left( C_\ell^{TT} \right)^2 &  \left( C_\ell^{TE} \right)^2 &  C_\ell^{TT}    C_\ell^{TE} \\ 
\left( C_\ell^{TE} \right)^2 & \left( C_\ell^{EE} \right)^2 & C_\ell^{EE}    C_\ell^{TE} \\ 
C_\ell^{TT}    C_\ell^{TE} & C_\ell^{EE}    C_\ell^{TE} & \left[ \left( C_\ell^{TE} \right)^2 + C_\ell^{TT} C_\ell^{EE} \right] \end{array} \right), \nonumber \\
 & \left(F_e\right)_{ij} = \sum_\ell \left( \frac{\partial C_{\ell}}{\partial \alpha_i} \right)^T \cdot \Sigma_\ell^{-1} \cdot \frac{ \partial C_{\ell}}{\partial \alpha_j}. \label{eq:fisher} 
\end{align}
For experiments other than the perfect cosmic variance limited (CVL)
case, noise is included by replacing $C_\ell^{TT,EE} \rightarrow
C_\ell^{TT,EE} + N_\ell^{TT,EE}$, where $N_\ell$ is the effective
noise power spectrum and is given by:
\begin{equation}
  \label{eq:noise}
  N_\ell = (\omega_p)^{-1} e^{\ell(\ell+1)\theta^2}
\end{equation}
Here $\theta$ describes the beam width (FWHM = $ \theta \sqrt{8 \ln
2}$), and the raw sensitivity is $(\omega_p)^{-1} = (\Delta T \times
FWHM)^2$, with all angles in radians.  The standard deviation of the
parameter $\alpha_i$, marginalized over uncertainties in the other
parameters, is given by
$\sigma_{\alpha_i}\geq(F_e^{-1})^{1/2}_{ii}$. The parameter $\alpha_i$
is then detectable at $1 \sigma$ if its signal-to-noise
$\alpha_i/\sigma_{\alpha_i}$ is larger than 1.

So far, we have not taken into account covariance between the standard
cosmological parameters and the energy deposition parameters, but in
fact there are significant degeneracies between them. In particular,
shifting the primordial scalar spectral index $n_s$ can absorb much of
the effect of energy deposition \cite{Padmanabhan:2005es,
Galli:2009zc}. Therefore we must marginalize over the cosmological
parameters, since the naively most measurable energy deposition history
may be strongly degenerate with them and thus difficult to
constrain. We parameterize the usual six-dimensional cosmological
parameter space by the following set of parameters: the physical
baryon density, $\omega_b \equiv \Omega_b h^2$, the physical CDM
density, $\omega_c \equiv \Omega_c h^2$, the primordial scalar
spectral index, $n_s$, the normalization, $A_s$(k = 0.002/Mpc), the
optical depth to reionization, $\tau$, and the Hubble parameter $H_0$.

Using exactly the same machinery as described above for the energy
deposition histories, we determine the derivatives of the $C_\ell$'s with
respect to changes in the cosmological parameters, again assuming that
these changes are in the linear regime. Then these $C_\ell$
derivatives are vectors spanning an $n_c$-dimensional subspace of the
space of all $C_\ell$ derivatives (where for the standard parameter
set $n_c=6$); only directions orthogonal to this subspace can be
constrained. We can regard marginalization over the cosmological
parameters as simply projecting out the components of the energy
deposition derivatives orthogonal to this subspace\footnote{See
Appendix \ref{app:marg} for a detailed explanation of this projection
and how it relates to the standard marginalization prescription.}.

In analogy with Equation \ref{eq:fisher}, we now use the derivatives
with respect to both energy deposition and the cosmological parameters
to construct the full Fisher matrix,
\begin{equation} 
  F_0 = \left( \begin{array}{cc} F_e & F_v \\ F_v^T & F_c \end{array} \right), 
\end{equation}
where $F_e$ is the Fisher matrix for solely the energy deposition
parameters, $F_c$ is the Fisher matrix of the cosmological parameters,
and $F_v$ contains the cross terms. The usual prescription for
marginalization is to invert the Fisher matrix, remove the rows and
columns corresponding to the cosmological parameters, and invert the
resulting submatrix to obtain the marginalized Fisher matrix $F$
(e.g. \cite{Verde:2009tu}). When the number of energy deposition
parameters is much greater than the number of cosmological parameters,
it is convenient to take advantage of the block-matrix inversion,
\begin{widetext} 
\begin{equation} F_0^{-1} = \left( \begin{array}{cc} \left(F_e - F_v F_c^{-1} F_v^T \right)^{-1} &  - \left(F_e - F_v F_c^{-1} F_v^T \right)^{-1} F_v F_c^{-1}  \\ - F_c^{-1} F_v^T \left(F_e - F_v F_c^{-1} F_v^T \right)^{-1} & F_c^{-1} \left( 1 + F_v^T \left(F_e - F_v F_c^{-1} F_v^T \right)^{-1} F_v F_c^{-1} \right) \end{array} \right).
\label{eq:fisher_inversion}
\end{equation} \end{widetext}
We can now read off the marginalized Fisher matrix as $F = F_e - F_v
F_c^{-1} F_v^T$ (note that $F$ has the same units as
$F_e$). 

The Fisher matrix approach to estimate detectability is optimistic in
the sense that it assumes the likelihood function is Gaussian about
its maximum; for non-Gaussian likelihoods, the significance of a given
energy deposition history will generally be smaller, and any constraints on the
amount of energy deposition will be weakened \cite{Verde:2009tu}. We
verify that the Fisher matrix method gives results consistent with
previous studies of \WMAP limits on constant $\pann$ in
\S\ref{subsec:sensitivity_discussion}.

%%%%%%%%%%%%%%%%%%%%%%%%%%%%%%%%%%%%%%%%%%%%

\subsection{Experimental parameters}
\label{subsec:expts}

For comparison to the existing literature and constraint forecasting,
we consider the \WMAP5, \WMAP7 and \PLANCK experiments, as well as a
theoretical experiment that is CVL up to $\ell=2500$. The beam width
and sensitivity parameters for \WMAP and \PLANCK are given in Table
\ref{tab:exptproperties}. We use only the $W$ band for \WMAP and the
143 GHz band for \PLANCKc, under the conservative assumption that the
other bands will be used to remove systematics. The effect of partial
sky coverage is included by dividing $\Sigma_\ell$ by $f_\mathrm{sky}
= 0.65$.

\begin{table}[t]
\begin{tabular}{cccc}
\hline
Experiment & Beam & $10^{6} \Delta T/T$ & $10^{6} \Delta T/T$ \\
& FWHM (arcmin) & (I) & (Q,U) \\
\hline
\WMAP (5 yr, Q band) & 29 & 6.7 & 9.5 \\
\WMAP (5 yr, V band) & 20 & 7.9 & 11.1 \\
\WMAP (5 yr, W band) & 13 & 7.6 & 10.7 \\
\PLANCK (100 GHz) & 10 & 2.5 & 4.0 \\
\PLANCK (143 GHz) & 7.1 & 2.2 & 4.2 \\
\PLANCK (217 GHz) & 5.0 & 4.8 & 9.8 \\
\hline
\end{tabular}
\caption{\label{tab:exptproperties} Detector sensitivities and beams
for different CMB temperature and polarization experiments. Results
for \WMAP temperature sensitivity are taken from
\cite{Jarosik:2003fe}, with the noise reduced by $\sqrt{5/4}$
($\sqrt{7/4}$ for \WMAP7) to account for the longer integration
time. The polarization noise for \WMAP is taken to be $\sqrt{2}
\times$ the temperature noise. \WMAP beam widths are taken from
\cite{Bennett:2003bz}. The sensitivity and beam width for \PLANCK are
taken from the Planck Blue Book, available at
\texttt{http://www.rssd.esa.int/SA/PLANCK/docs}, and assume 14 months
of \PLANCK data.}
\end{table}

%%%%%%%%%%%%%%%%%%%%%%%%%%%%%%%%%%%%%%%%%%%%

\subsection{Comparison to previous results \label{subsec:sensitivity_discussion}}

\begin{figure}[bt]
\includegraphics[width=0.45\textwidth]{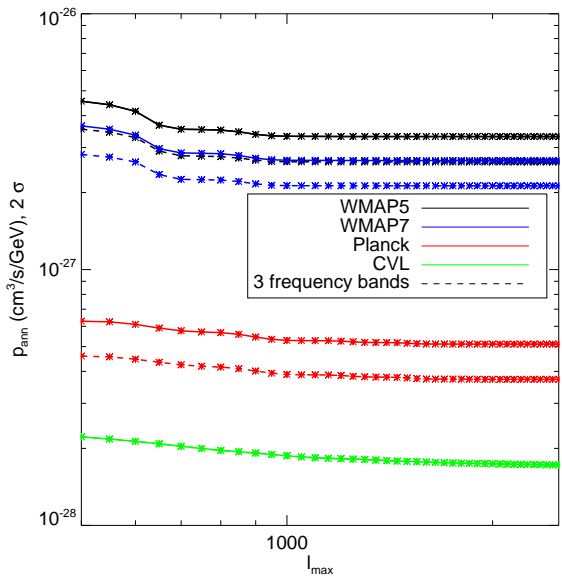}
\caption{The effect of the number of included $\ell$'s, and the number of included frequency bands, on the constraint on a constant-$p_\mathrm{ann}$ energy deposition history; here we show the value of $p_\mathrm{ann}$ corresponding to a $2 \sigma$ signal.}
\label{fig:lmaxsensitivity}
\end{figure}

Constraints on energy deposition from \WMAP7 have been studied
previously in the case where $\pann(z) = \pann$ is constant
\cite{Galli:2009zc,Galli:2011rz}. We have obtained
analogous constraints using the estimates for experimental sensitivity
described in \S \ref{subsec:fisher}-\ref{subsec:expts}, taking the six
standard cosmological parameters and $\pann$ as our parameter set, and
marginalizing over the cosmological parameters.

Note that in this simplified scenario there is no need for the Gaussian basis functions described at the start of the section, and rather than describe the effect of an arbitrary energy deposition history by the ``transfer matrix'' $T$ (Equation \ref{eq:transfermatrix}) we simply compute the effect on the $C_\ell$'s of each of a range of non-zero $\pann$ values using \texttt{CosmoRec} and \texttt{CAMB}. Since no evidence for energy deposition has been found to date, we use the predictions of the best-fit standard cosmological model (as determined by \WMAP7 \cite{Larson:2010gs}) as a proxy for the \WMAP data. Using the covariance matrix and prescription for marginalization laid out in \S \ref{subsec:fisher}, we compute the $\Delta \chi^2$ with the null hypothesis for each $p_\mathrm{ann}$, and by interpolation determine how large a $p_\mathrm{ann}$ would be disfavored at $2 \sigma$.  Our results are shown in Figure
\ref{fig:lmaxsensitivity}, for both \WMAP5 and \WMAP7 noise parameters, with use of
one or three frequency bands (the former is our standard conservative
approach, the latter is more optimistic), as a function of the maximum $\ell$ included in the analysis ($\ell_\mathrm{min}$ is always 2).

In earlier work, \cite{Galli:2009zc} found a \WMAP5 limit of
$p_\mathrm{ann} < 2 \times 10^{-6}$ m$^3$/s/kg $= 3.6 \times 10^{-27}$
cm$^3$/s/GeV, using \WMAP5 data and the \texttt{CosmoMC} code to
perform a full likelihood analysis, in reasonable agreement with our
estimate for the \WMAP5 1-band case.  Using \WMAP7 data,
\cite{Galli:2011rz} provided an updated constraint $p_\mathrm{ann} <
2.4 \times 10^{-27}$ cm$^3$/s/GeV, also in very good agreement with
our estimate. Note that \cite{Galli:2009zc,Galli:2011rz} included
Lyman-$\alpha$ photons in their analysis, which we have not done here
and which would strengthen the constraints slightly: the small
differences between our constraints and those in the literature can
probably be ascribed to the combination of this factor, our use of
\texttt{CosmoRec} rather than \texttt{RECFAST}, the fact that we use
the best-fit standard cosmological model as a proxy for the real \WMAP
data, and the optimism inherent in a Fisher matrix analysis. It is
reassuring that none of these factors seem to give rise to large
discrepancies in the results.

We see that the \WMAP limits are essentially
unaffected by $\ell_\mathrm{max}$ for $\ell_\mathrm{max} \gtrsim
1000$, and the projected \PLANCK bound appears stable for
$\ell_\mathrm{max} \gtrsim 1500$. For the CVL case, of course, higher
$\ell$'s will always yield more information, but the rate of
improvement with $\ell_\mathrm{max}$ is quite slow for the $\ell$'s we
are considering.

%%%%%%%%%%%%%%%%%%%%%%%%%%%%%%%%%%%%%%%%%%%%

\subsection{Numerical stability of derivatives and linearity}
\label{sec:derivs}

When dealing with general energy deposition histories, we hope to work in a regime where the effect of deposition on the CMB is \emph{linear}, so that the effect of a general energy deposition history can be described in terms of a linear combination of basis energy deposition histories. This is the idea behind characterizing the effect of new parameters entirely in terms of the transfer matrix of derivatives, $T$, and the Fisher matrix $F$ derived from it. Equivalently, linearity means it is sensible to speak of a single transfer matrix $T$ largely independent of the ``fiducial'' energy deposition history about which the derivatives $\partial
C_{\ell} / \partial \alpha_i$ are taken (our default assumption is that this ``fiducial'' energy deposition is zero). If the energy deposition history being studied is too great a perturbation away from the fiducial, the first derivatives will no longer accurately describe its effect on the $C_\ell$'s,  and the Fisher matrix estimate of its significance will break down. In this subsection we discuss the numerical stability of the derivatives, and the degree to which they describe the effect of arbitrary energy deposition histories on the $C_\ell$'s.

%%%%%%%%%%%%%%%%%%%%%%%%%%%%%%%%%%%%%%%%%%%%

To obtain derivatives of the $C_\ell$ spectra, we calculate a grid of
$C_\ell$ values for each of the parameters considered in the Fisher
matrix. For each $\ell$ and each parameter, the derivative is
extracted from a polynomial fit to $C_\ell$ as a function of the
parameter values.

We estimate that we have calculated derivatives to a precision of
$\sim1\%$ for cosmological parameters. For energy deposition, the
error is less than $\sim2\%$ for the most relevant redshifts of $z <
800$ and rises to $\sim5\%$ for $z < 1100$. While the error in the derivatives becomes larger for higher
redshifts, the effect on the PCA is small; the difference in signal-to-noise is at the percent level for \WMAP7 and at the few percent level for \PLANCK and the CVL case. The dominant numerical
error here comes from the numerical accuracy limitations of
\texttt{CosmoRec} and \texttt{CAMB}. We estimate the error by
comparing the derivatives obtained from two different accuracy
settings\footnote{The two settings we use are (1) \texttt{runmode=0}
and \texttt{cosmorecacc=0} and (2) \texttt{runmode=1} and
\texttt{cosmorecacc=2}.  In general, we use CAMB accuracy settings
\texttt{accuracy\_boost=4, l\_accuracy\_boost=4, l\_sample\_boost=4}
for computing derivatives.  For the MCMC runs, we set all three
parameters to 1 for speed.}.

The derivatives used in the Fisher matrix are evaluated at the
fiducial cosmology (with no energy deposition). The assumption of
linearity is that these derivatives are still correct away from the
fiducial. For the standard set of six cosmological parameters, the
biases to the cosmological parameters induced by the maximum permitted
energy deposition from \WMAP5 generally lie well within the linear
regime.

For large energy deposition, the effect on the $C_\ell$'s is
nonlinear, i.e. not directly proportional to the deposited power as
parameterized by the $\alpha_i$; equivalently, the derivatives about a
fiducial large energy deposition are not the same as for zero energy
deposition. Our polynomial fits for the derivatives, described above,
also allow us to check the extent to which nonlinearity may become
important: that is, the extent to which $\mathcal{O}(\alpha_i^{2})$
corrections to the effect on the $C_\ell$'s are non-negligible.

The amount of energy deposition such that nonlinearities become
important depends on redshift $z$. This can be estimated by the
fractional rate of ionization per Hubble time,
$(dn_{ion}/dt)/(n^0_{ion} H(z))$, arising from the energy deposition
(where $dn_{ion}/dt$ is related to $dE/dtdV$ according to the
prescription of \cite{1985ApJ...298..268S}). For two fiducial cases
this quantity is shown in Figure~\ref{fig:DMionization}. Conversely,
the energy deposition at redshift $z$ such that
$(dn_{ion}/dt)/(n^0_{ion} H(z)) = 1$ gives a measure of what energy
deposition is required before nonlinearities may become
significant. For each redshift bin, we use the polynomial fits of
$\delta C_\ell(\alpha_i)$ to numerically calculate the derivatives at
this level of energy deposition. We then find $1\%$ corrections
(averaged over $\ell$) to the fiducial derivative $(\partial
C_\ell/\partial \alpha_i)|_{\alpha_i = 0}$.

\begin{figure}
\includegraphics[width=0.45\textwidth]{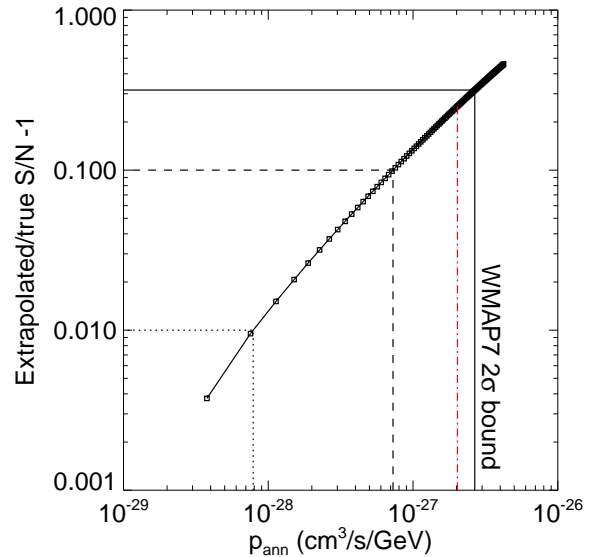}
\caption{The degree of nonlinearity in the computed significance of a sample energy deposition history, for $p_\mathrm{ann}$ constant, using \WMAP7 noise parameters. We show the ratio of (1) the S/N estimated by a linear extrapolation from small energy deposition to (2) the ``true'' S/N (estimated as in \S \ref{subsec:sensitivity_discussion}), as a function of $p_\mathrm{ann}$. The solid, dashed and dotted lines indicate the \WMAP7 2$\sigma$ upper limit on $p_\mathrm{ann}$, the value of $p_\mathrm{ann}$ for which the nonlinearity is $10\%$, and the value for which the nonlinearity is $1\%$, respectively. The red dot-dashed line indicates the 2$\sigma$ upper limit on $p_\mathrm{ann}$ that would be obtained by linearly extrapolating the significance from small energy deposition, which overestimates the significance and hence leads to a too-strong constraint.}
\label{fig:nonlinearity}
\end{figure}

One simple test of the effect of nonlinearity is the degree to which the ``true'' bound on constant $p_\mathrm{ann}$ from a given experiment, evaluated without assuming linearity as described in \S \ref{subsec:sensitivity_discussion}, differs from the bound we would obtain by taking derivatives $\partial C_{\ell} / \partial \pann$ at $\pann = 0$, and assuming linearity, i.e. taking the effect on the $C_\ell$'s to be given by $\pann \left( \partial C_{\ell} / \partial \pann \right)|_{\pann=0}$. Equivalently, we can compare the signal-to-noise estimate for the two methods, as a function of $\pann$.

In Figure \ref{fig:nonlinearity} we show an example of this test using \WMAP7 noise parameters. For each value of $p_\mathrm{ann}$, we compute both the S/N of the resulting signal directly (as in  \S \ref{subsec:sensitivity_discussion}), and the significance that would be obtained by a linear extrapolation from small $p_\mathrm{ann}$. The ratio of the extrapolated S/N to the true S/N is $1 + \epsilon$, and $\epsilon$ provides a measure of nonlinearity. We see that $\epsilon$ approaches 0.3 close to the $2\sigma$ upper bound from \WMAP7, so the estimated $95\%$ bound from \WMAP7 is roughly $30\%$ weaker than would be expected from a linear extrapolation from small energy deposition, but the degree of nonlinearity falls rapidly at lower energy deposition. We will see later that for the energy deposition probed by \PLANCKc, $\epsilon$ is only a few percent. Note that even for \WMAP7, the difference in the bound is  almost entirely due to an overall $\ell$-independent normalization factor; the \emph{shape} of the $\delta C_\ell$'s is almost unchanged.

%%%%%%%%%%%%%%%%%%%%%%%%%%%%%%%%%%%%%%%

\begin{figure*}[tb]
\includegraphics[width=0.32\textwidth]{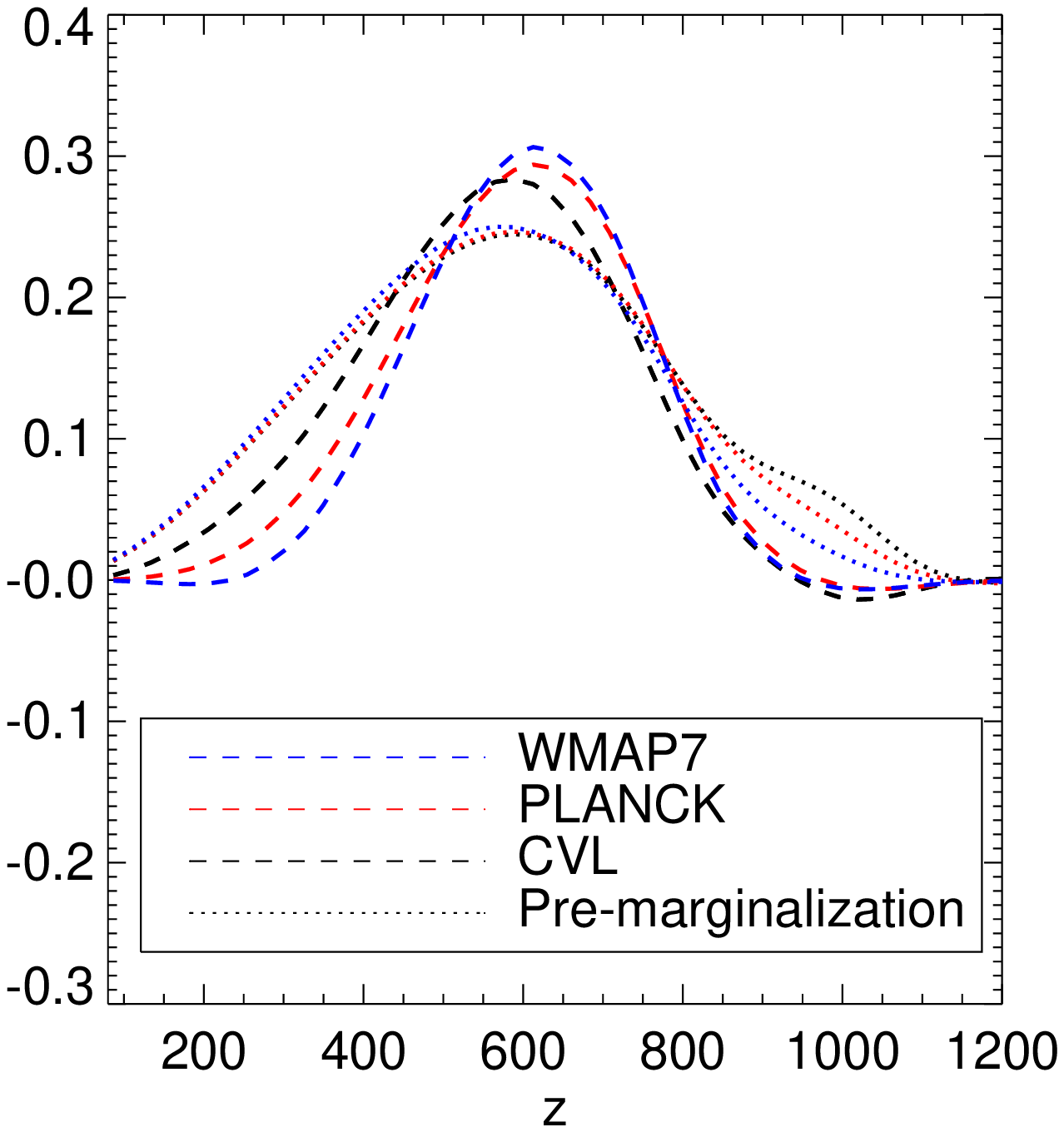}
\includegraphics[width=0.32\textwidth]{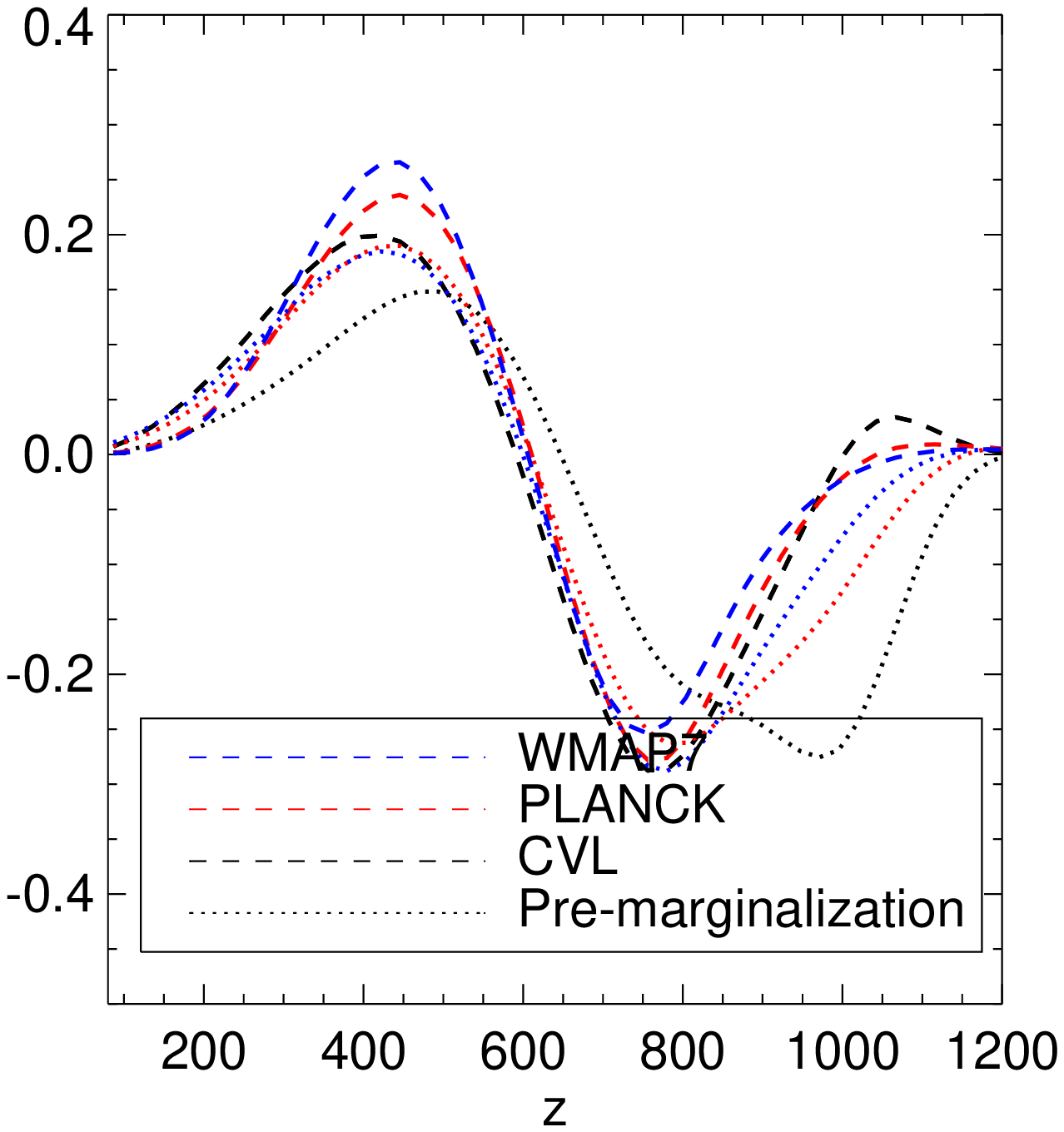}
\includegraphics[width=0.32\textwidth]{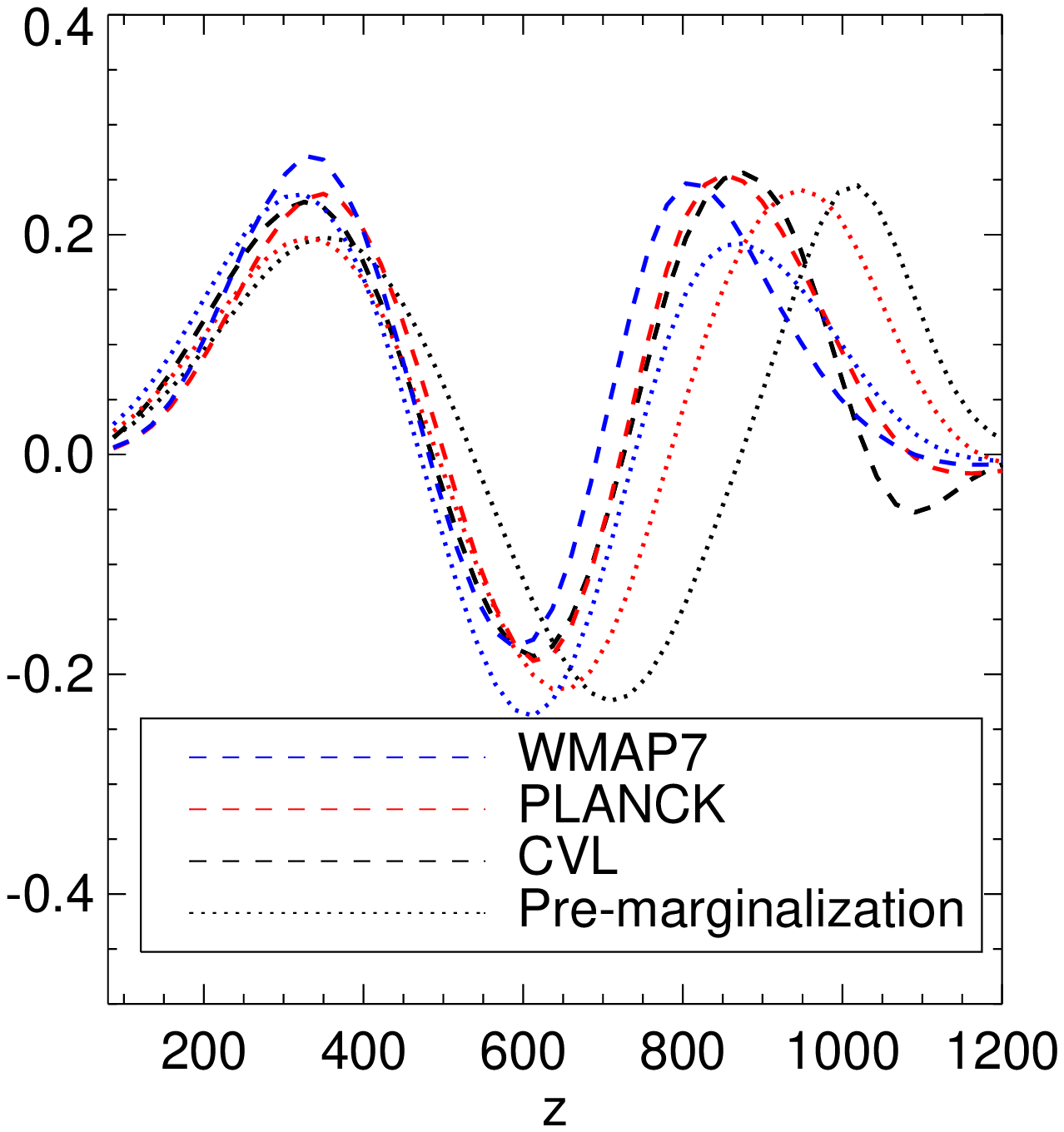} 
\caption{The first three principal components for \WMAP7, \PLANCK and a CVL experiment, both before and after marginalization over the cosmological parameters.}
\label{fig:margcomparison}
\end{figure*}

%%%%%%%%%%%%%%%%%%%%%%%%%%%%%%%%%%%%%%%%%%%%

\section{Principal Component Analysis}
\label{sec:pca}

The effects of energy deposition at different redshifts on the $C_\ell$'s are highly correlated, and so the effects of a large class of energy deposition histories can be characterized by a small number of parameters. Principal component analysis provides a convenient basis into which  energy deposition histories can be decomposed, with the later terms in the decomposition contributing almost nothing to the effect on the $C_\ell$'s. It thus allows generalization of constraints on energy deposition to a wide range of models (subject to the linearity assumption discussed above).

\subsection{The principal components}

\begin{figure*}
\includegraphics[width=0.32\textwidth]{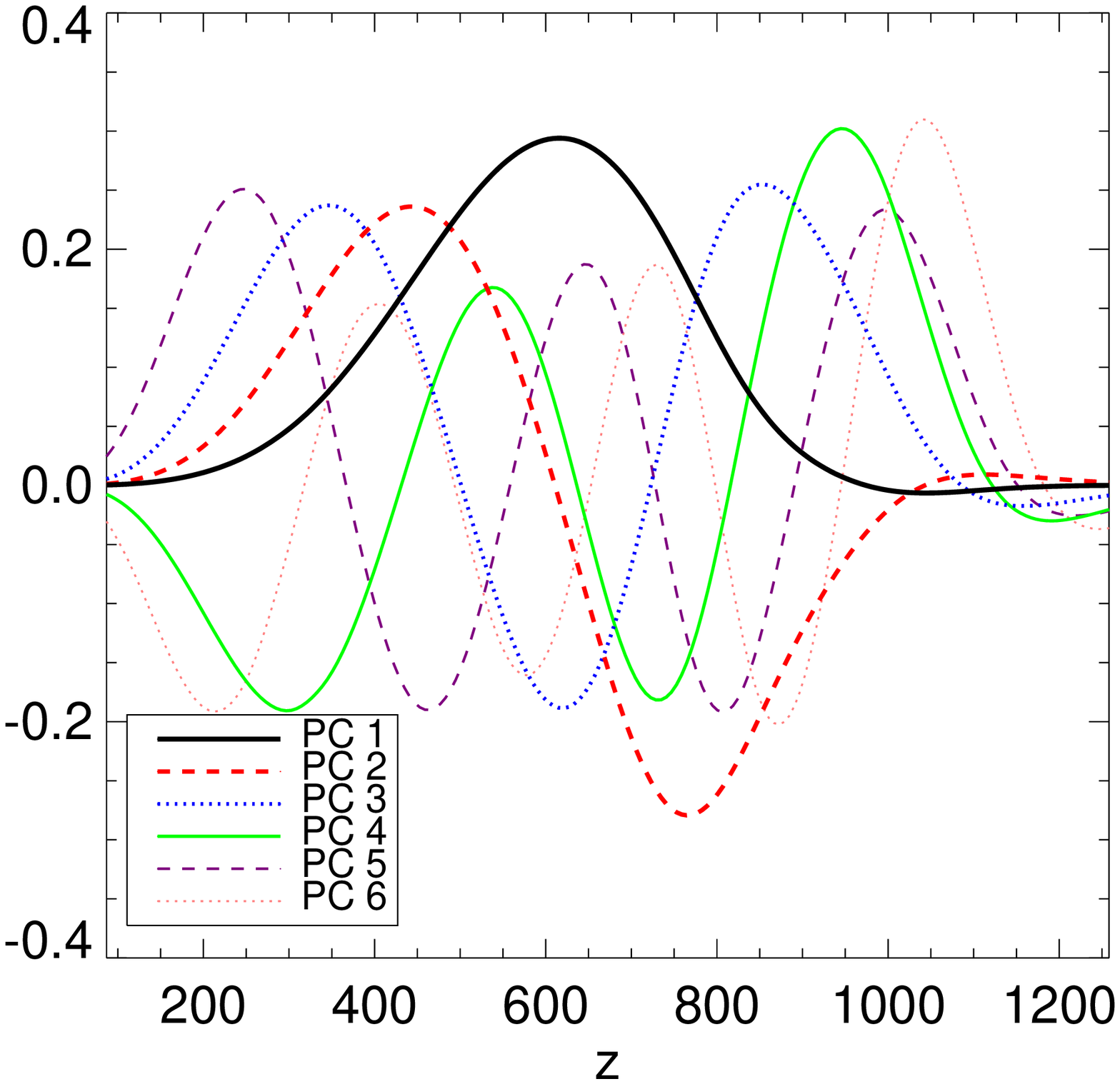}
\includegraphics[width=0.32\textwidth]{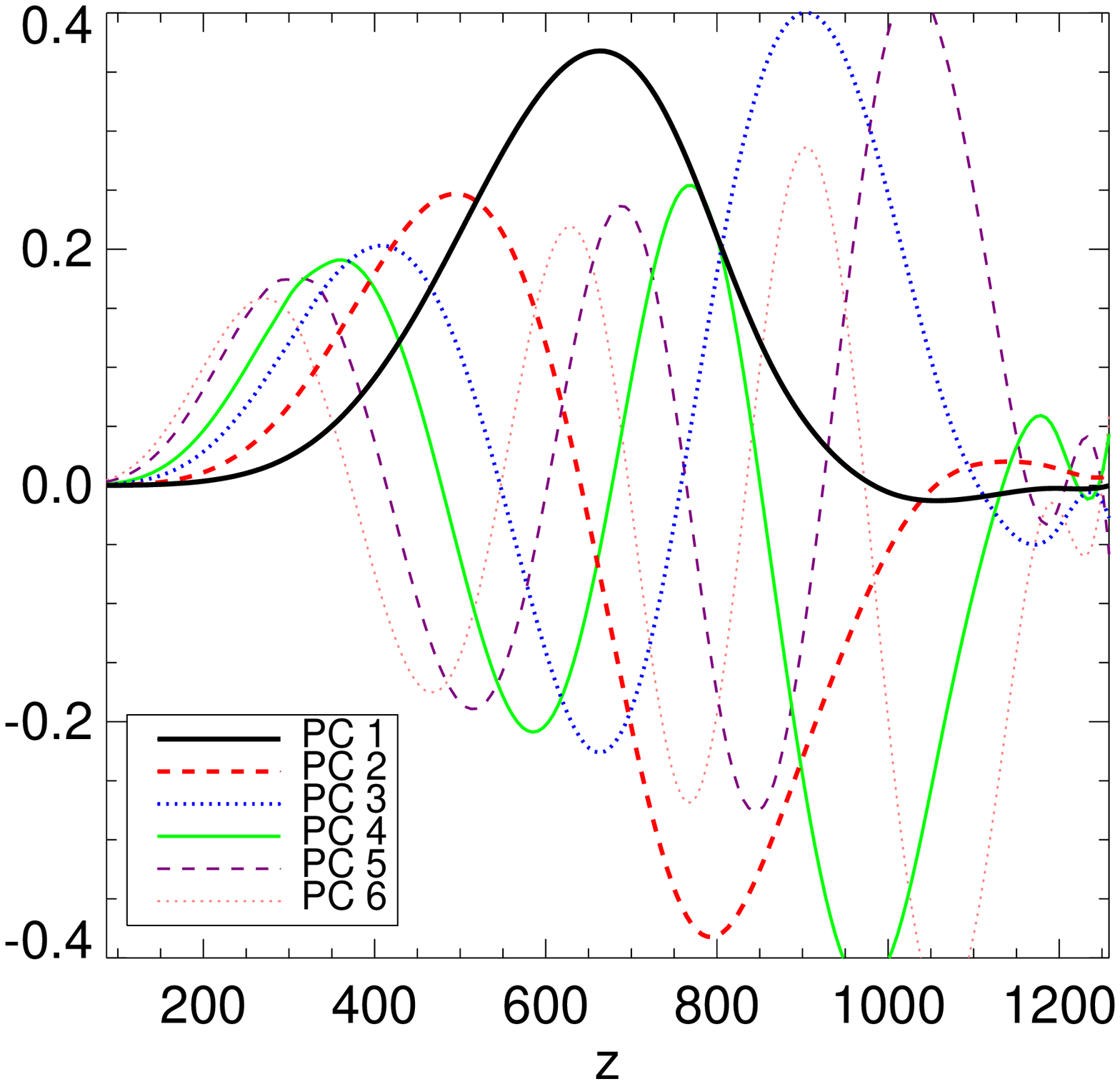}
\includegraphics[width=0.32\textwidth]{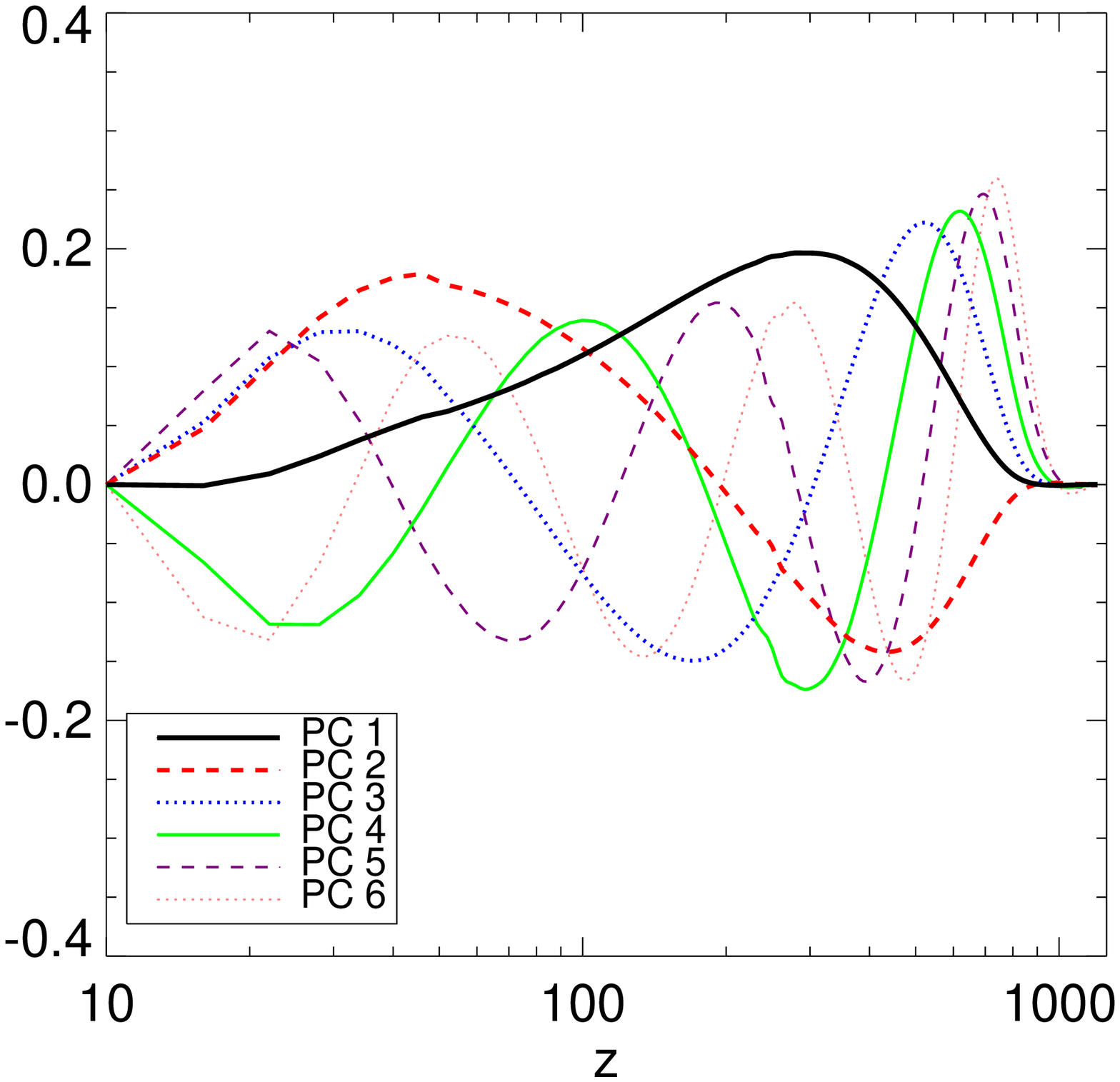}
\caption{The first six principal components for \PLANCK after
marginalization, in the case of (\emph{left}) annihilation-like
redshift dependence with linear binning, (\emph{center})
annihilation-like redshift dependence with log binning, and
(\emph{right}) decay-like redshift dependence with log binning. Note
that for decay-like energy deposition histories, the redshift range is
extended down to $z=10$ in order to fully capture the effect on the
CMB - see \S\ref{sec:fisher}. This larger redshift range makes linear
binning impractical. }
\label{fig:pcs}
\end{figure*}
%%%%%%%%%%%%%%%%%%%%%%%%%%%%%%%%%%%%%%%%%%%%

Having obtained the marginalized Fisher matrix $F$, diagonalizing $F$:
\begin{equation}
  F=W^T \Lambda W,\quad \Lambda=\rm diag(\lambda_1,\lambda_2,....,\lambda_N)
\end{equation} 
yields a convenient basis of eigenvectors or ``principal
components''. $W$ is an orthogonal matrix in which the $i$-th row
contains the eigenvector corresponding to the eigenvalue $\lambda_i$.
If we compute derivatives for $N$ redshift bins, then the $N \times N$
Fisher matrix has $N$ principal components. The eigenvectors are
orthonormal in the space of vectors $\{\alpha_i\}$, $i=1..N$. Let us
label these vectors $e_i$, with corresponding eigenvalues $\lambda_i$,
$i=1..N$. Our convention is to rank the principal components by
decreasing eigenvalue, such that $e_1$ has the largest eigenvalue.

Note that the principal components may be significantly different
from the \emph{unmarginalized} principal components, or the eigenvectors of
$F_e$. Figure \ref{fig:margcomparison} shows the first three principal
components for \WMAP7, \PLANCK and a CVL experiment, both before and
after marginalization, for the annihilation-like case ($dE/dt \propto
p_\mathrm{ann}(z) (1+z)^6$) with 50 linearly-spaced redshift bins. We
see that while the shapes of the PCs are qualitatively similar,
marginalization produces noticeable changes to the PCs, as does
changing from one set of experiment parameters to another. The
differences become more pronounced for higher PCs.

Note that the shapes of the principal components can be affected by a number of other different factors: choice of binning, choice of ionization history calculator, energy deposition model, fiducial cosmological model considered, etc. We discuss these effects in Appendix \ref{app:validation}.

In Figure \ref{fig:pcs} we show the first six marginalized PCs for
\PLANCKc, for annihilation-like ($dE/dt \propto p_\mathrm{ann}(z) (1+z)^6$) and
decay-like ($dE/dt \propto p_\mathrm{dec}(z) (1+z)^3$) energy
deposition histories. We show the annihilation-like case with both log and linear binning.  We note that the first principal component is always
largely or completely non-negative, and (in the annihilating case)
peaked around redshift 600. The first PC can be thought of as a
weighting function, describing the effect of energy deposition on the
CMB (orthogonal to the effect of shifting the cosmological
parameters), as a function of redshift\footnote{Note that the shift in the peak position between log and linear binning is to be expected, as one ``weighting function'' would be integrated over $dz$ and the other over $d \ln (1+z)$; see Appendix \ref{app:validation} for further discussion.}.

\begin{figure}
\includegraphics[width=0.45\textwidth]{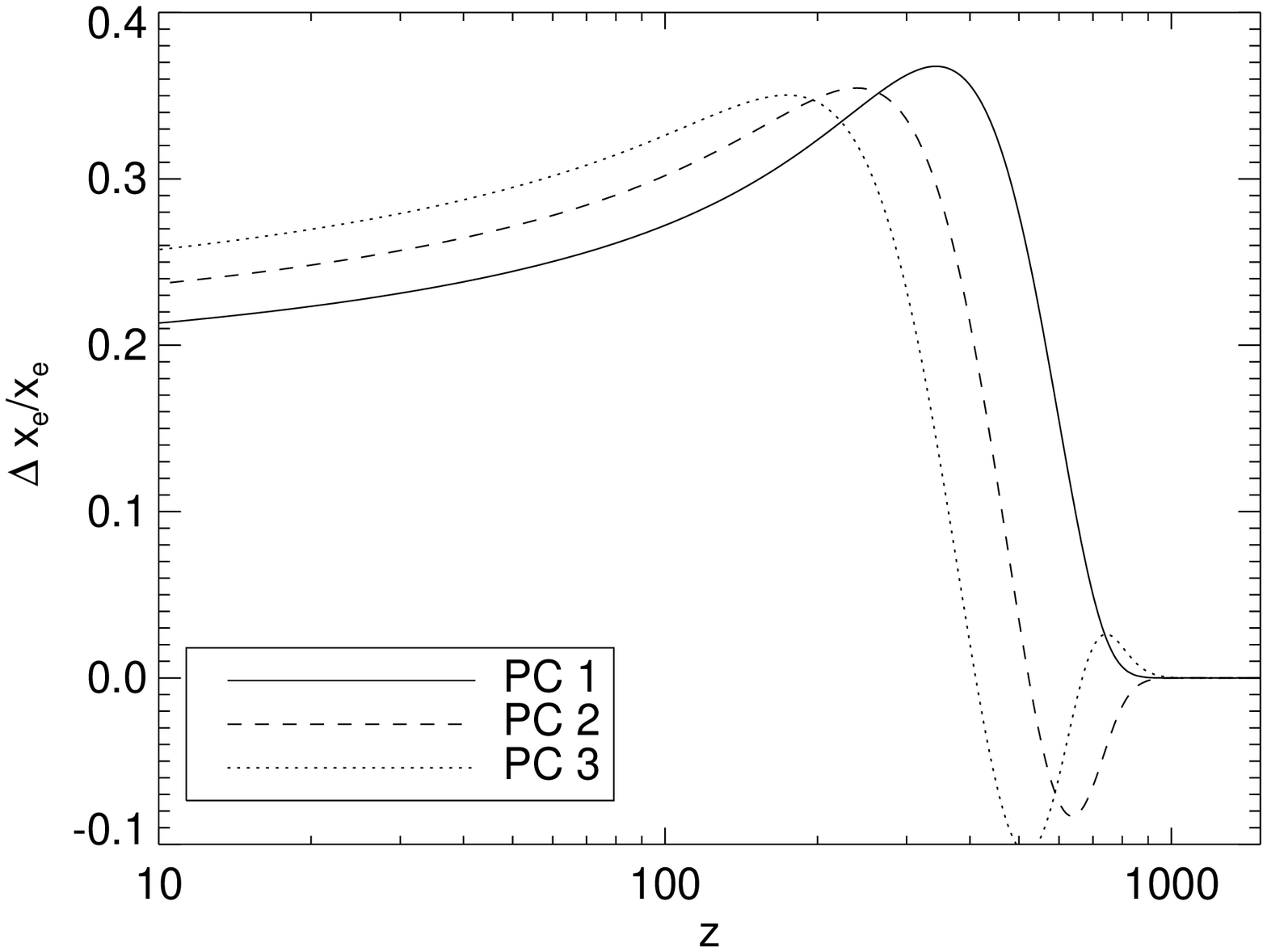}
\caption{Fractional change to the ionization fraction $x_e$ in the presence of energy deposition, for the first three (marginalized) principal components in \PLANCKc. The curve shown is extrapolated from the linear (small energy deposition) regime, with normalization factor $\varepsilon_{1,2,3} = 2 \times 10^{-27}$ cm$^3$/s/GeV.}
\label{fig:frac_delta_xe}
\end{figure}

In Figure \ref{fig:frac_delta_xe} we show the effect on the ionization history for the first three \PLANCK PCs in the annihilation case, with each PC multiplied by an energy deposition coefficient of $\varepsilon = 2 \times 10^{-27}$ cm$^3$/s/GeV to obtain $p_\mathrm{ann}(z)$. Note that this energy deposition is too large to be strictly in the linear regime; this figure illustrates the shape and size of the effect in the linear regime, the true effect for this value of $\varepsilon$ will be somewhat smaller.

As previously, we have considered ``annihilation-like'' and ``decay-like'' energy deposition histories separately. If both analyses were performed over the same redshift range, then while the principal components might appear different, they would span the same space of energy deposition histories. If all principal components were retained, the difference between the two would simply be equivalent to a change of basis, and provided \emph{sufficient} principal components are retained, this will still be approximately true. However, a particular energy deposition history may be described by the early principal components much better in one case than in the other; in particular, energy deposition histories for which the effect on the CMB is dominated by low redshifts will not be well described by the (first few of the) default annihilation-like PCs. Thus we present results for both cases.

\subsection{Mapping into $\delta C_\ell$ space}
\label{sec:pcunits}

%%%%%%%%%%%%%%%%%%%%%%%%%%%%%%%%%%%%%%%%%%%%

\begin{figure*}[t]
\includegraphics[width=\textwidth]{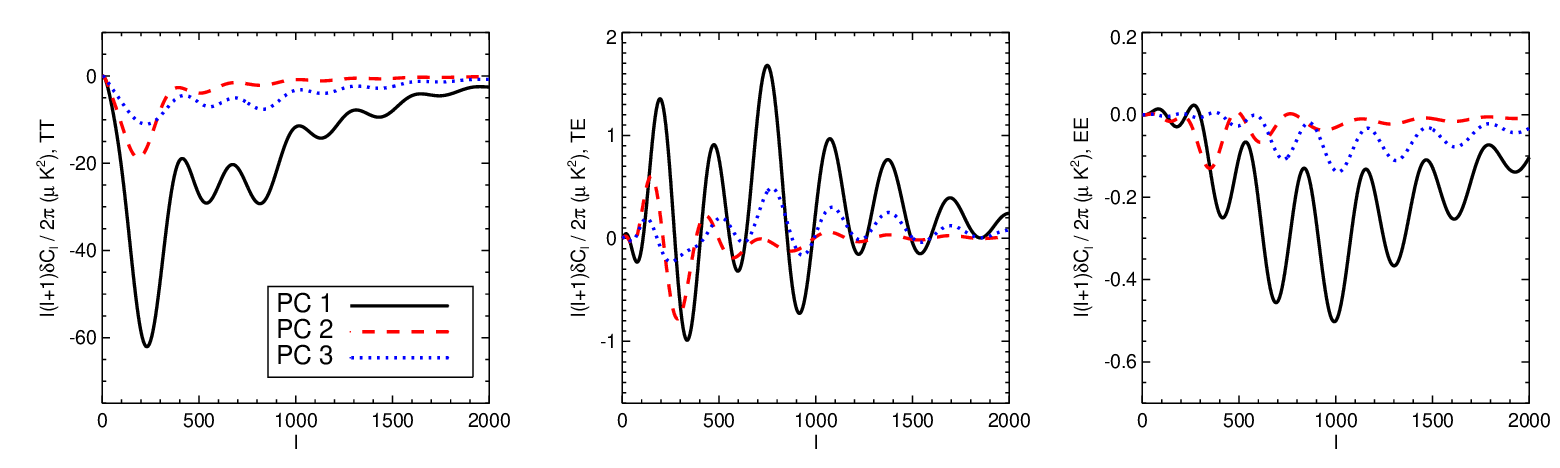}
\caption{The mapping of the first three principal components for \PLANCKc, after marginalization, into $\delta C_\ell$ space. The PCs are multiplied by $\varepsilon_i(z) = 2 \times 10^{-27}$ cm$^3$/s/GeV for all $i$, to fix the normalization of the $\delta C_\ell$'s.}
\label{fig:deltaclpcs}
\end{figure*}

\begin{figure*}
\includegraphics[width=\textwidth]{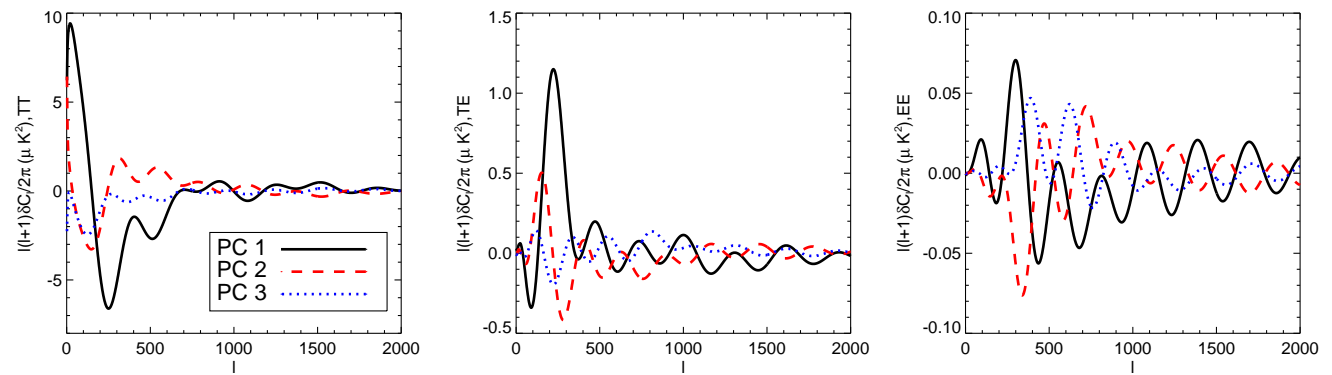}
\caption{The $\perp$ components of the first three principal components for \PLANCKc, after marginalization, mapped into $\delta C_\ell$ space. The normalization is the same as for Figure \ref{fig:deltaclpcs}.}
\label{fig:deltaclpcsperp}
\end{figure*}

\begin{figure*}
\hspace{-.5cm}\includegraphics[width=\textwidth]{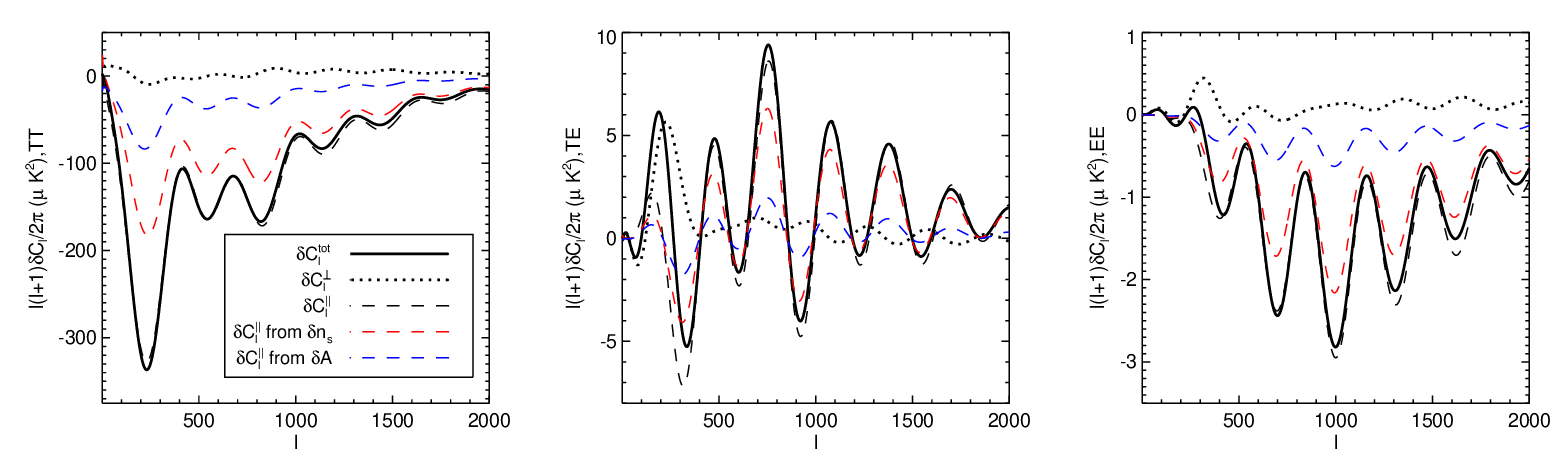}
\caption{\label{fig:deltaCl}Decomposition of $\delta C_\ell$ from energy deposition with constant $p_\mathrm{ann}(z)$  into parallel ($||$) components which can be absorbed by changes in the cosmological parameters, and perpendicular ($\perp$) components that cannot be absorbed by such changes.
The overall effect of the energy deposition is suppression of high-$\ell$ modes, due to the increased optical depth, and enhancement of low-$\ell$ polarization modes, as discussed in \cite{Padmanabhan:2005es}. The suppression at high $\ell$ is clearly seen in the TT and EE spectra; the effect is also present in the TE spectra, with the peaks of $\delta C_\ell^{TE}$ occurring at the troughs of $C_\ell^{TE}$, and vice versa.
The normalization here is $p_\mathrm{ann} = 2 \times 10^{-27}$ cm$^3$/s/GeV,
comparable to the latest limits from \WMAP7+ACT \cite{Galli:2011rz}. This decomposition depends on the sensitivity of the experiment; the case shown is \WMAP7 single band.}
\end{figure*}

%%%%%%%%%%%%%%%%%%%%%%%%%%%%%%%%%%%%%%%%%%%%

Let us consider the mapping into $\delta C_\ell$ space of these
marginalized principal components. Applying the transfer matrix $T$ (Equation \ref{eq:transfermatrix}) to the
eigenvectors yields a set of $N$ vectors in the space of $C_\ell$
perturbations, $\delta C_\ell = T e_i = h_i$. The
$h_i$'s should be understood as $\delta C_\ell$'s per energy
deposition, and have units of $C_\ell/p(z)$. 

We can define a dot product on the space of $\delta C_\ell$'s by
\begin{equation} 
  h_i \cdot h_j = {\sum_\ell h_{i\ell}^T \Sigma_\ell^{-1} h_{j\ell}} = {e_i^T F_e e_j}
\label{eq:dotproduct} 
\end{equation}
We then see that while the PCs are orthogonal, the $h_i$ are in
general not orthogonal to each other, nor to the $\delta C_\ell$'s
from the cosmological parameters. They correspond to actual energy
deposition histories, and in general, there is no such history that is
precisely orthogonal to all the cosmological parameters.

However, we may decompose the $h_i$ into components parallel and
perpendicular to the space spanned by varying the cosmological
parameters, and denote the perpendicular components $h_i^\perp$. The
projection operator that implements this decomposition is given in
Appendix \ref{app:marg}. The $h_i^\perp$ vectors are now
orthogonal amongst themselves, as well as to the cosmological
parameters, and their norms are given by the square root of the
marginalized eigenvalues $\lambda_i$. It is these $h_i^\perp$'s which
determine the detectability of the marginalized principal components,
and which form an orthogonal basis for residuals which cannot be
absorbed by varying the cosmological parameters. The addition of the
parallel components, to recover the $h_i$'s from the $h_i^\perp$'s,
ensures that the $h_i$'s correspond to energy deposition histories, and
so provide an orthogonal basis in redshift space.

In Figure \ref{fig:deltaclpcs}, we show the mapping of the first three
(marginalized) PCs for \PLANCK into the space of $\delta C_\ell$'s; in
Figure \ref{fig:deltaclpcsperp}, we show the components of these
$\delta C_\ell$'s which are orthogonal to the space spanned by varying
the cosmological parameters.  Figure \ref{fig:deltaCl} demonstrates
this projection for a sample DM annihilation model, summing over
principle components, and decomposing the effect on the $C_\ell$'s
into components perpendicular and parallel to the cosmological
parameters.

The eigenvectors of the Fisher matrix $\{e_i\}$ thus provide an
orthogonal basis in both relevant spaces, and their eigenvalues
precisely describe the measurability of a ``unit norm'' energy
deposition history with $z$-dependence given by the eigenvector. For an
arbitrary energy deposition history which we now write as 
\begin{equation}
p(z) = \sum_{i=1}^N\varepsilon_i e_i(z),
\label{eq:decomposedpz}
\end{equation}
 the expected $\Delta \chi^2$ relative to the
null hypothesis of no energy deposition is $\sum_i
\varepsilon_i^2 \lambda_i$. If the $\varepsilon_i$ coefficients are
comparable, the relative sizes of the eigenvalues describe the
fractional variance attributable to each principal
component (eigenvector).

A brief comment on unit conventions: we take the $\{e_i\}$ and
$\{G_i\}$ to be dimensionless, with the units of $p(z)$ (cm$^3$/s/GeV)
carried by the coefficients $\alpha_i, \, \varepsilon_i$. The
derivatives (and transfer matrix) then have units of $C_\ell/p(z)$,
and the Fisher matrix and its eigenvalues have units of $1/p(z)^2$
(since the covariance matrix $\Sigma$ has units of $C_\ell^2$). 
Note also that due to the units of the covariance matrix, the dot
product defined above takes two vectors in $C_\ell$-space to a
dimensionless number (if the vectors have units of $C_\ell$).

%%%%%%%%%%%%%%%%%%%%%%%%%%%%%%%%%%%%%%%%%%%%

\section{Detectability}
\label{sec:detectability}

For a general energy deposition history, the PCs provide a basis in which, by construction, the basis vectors are ranked by the significance of their effect on the $C_\ell$'s. The measurability of a generic (smooth, non-negative) energy deposition history can thus be accurately described by the first few PCs\footnote{It is in principle possible for the coefficients $\varepsilon_i$ to be zero for $i < n$ for some $n$, but  if $n$ is large this implies a very unphysical energy deposition history that oscillates rapidly between positive and negative values. While ``negative energy deposition'' might perhaps have a physical interpretation in terms of increased absorption of free electrons, such an interpretation is not at all obvious, and so we focus on smooth, non-negative energy deposition histories.}. Equivalently, the coefficients of later principal components have extremely large error bars, and will be challenging to measure or constrain.

We now outline the method for reconstructing and constraining the PC coefficients, or any specific energy deposition history, using the PCA formalism. We investigate the number of PCs that can generically be measured at $\ge 1\sigma$ by \PLANCK and a CVL experiment, for arbitrary energy deposition, and show results for broad classes of example models. We also consider the biases to the cosmological parameters that are induced if energy deposition is present but ignored; we present results for each principal component, so the biases due to an arbitrary energy deposition history can be immediately calculated. Our estimates of detectability and the biases will be verified using \texttt{CosmoMC} in the \S \ref{sec:cosmomc}.

%%%%%%%%%%%%%%%%%%%%%%%%%%%%%%%%%%%%%%%%%%%%

\subsection{Estimating limits from the Fisher matrix}

As mentioned previously, the perpendicular components of the $\delta
C_\ell$'s, $h_i^\perp$, are orthogonal with norms
$\sqrt{\lambda_i}$. They are also orthogonal to the space spanned by
varying the cosmological parameters. Given these results and a
measurement of the temperature and polarization anisotropies, it is
straightforward to estimate general constraints on the energy
deposition history from the Fisher matrix formalism. Note that in a careful
study, one would instead use \texttt{CosmoMC} to perform a full
likelihood analysis, using the Fisher matrix results only to determine
the optimal principal components, as we demonstrate in \S
\ref{sec:cosmomc}. We outline this simple method only to help build
intuition and to clarify later comparisons between the Fisher matrix
method and the \texttt{CosmoMC} results.

The first step is to extract any residual between the data and the best-fit model using the standard cosmological parameters; let us denote this residual by $R_\ell^{TT,EE,TE}$. Then we take the dot product (as defined in Equation \ref{eq:dotproduct}) of this residual with the $h_i^\perp$ vectors, normalizing by the corresponding eigenvalues (this normalization is required because the $h_i^\perp$'s are orthogonal, but not orthonormal; see Appendix \ref{app:marg}):
\begin{equation} \bar{\varepsilon}_i = \frac{R \cdot h_i^\perp}{\lambda_i}. \end{equation}
The resulting $\bar{\varepsilon}_i$ are the model-independent reconstructed coefficients for the marginalized principal components. In the absence of energy deposition, we expect them to be zero (within uncertainties).

The individual $1\sigma$ uncertainties on each of these coefficients are $1/\sqrt{\lambda_i}$, in the sense that if a single coefficient is perturbed away from its best-fit value by $1/\sqrt{\lambda_i}$, the corresponding energy deposition history will be disfavored at $1 \sigma$. Thus it is possible to set a \emph{very} general model-independent constraint on each of the coefficients, $\varepsilon_i = \bar{\varepsilon}_i \pm \frac{1}{\sqrt{\lambda_i}}$ (at $1 \sigma$).

Given an arbitrary energy deposition history, we can decompose it into the principal components, each with its own coefficient, and compare these coefficients $\varepsilon_i$ to the bounds. For any particular model, a stronger constraint can be set by noting that,
\begin{equation} \label{deltachi} \Delta \chi^2 = \sum_i \lambda_i \left( \varepsilon_i -  \bar{\varepsilon}_i\right)^2. \end{equation} 
This $\Delta \chi^2$ is relative to the best-fit model including both energy deposition and the standard cosmological parameters; the $\Delta \chi^2$ relative to the best-fit standard cosmological model\footnote{Of course, if the best-fit energy deposition history is everywhere zero, i.e $\bar{\varepsilon}_i \approx 0$ for all $i$, these two quantities are identical.} is simply $\sum_i \lambda_i \varepsilon_i(\varepsilon_i - 2 \bar{\varepsilon}_i )$.

This method has the usual deficiencies of the Fisher matrix approach: it assumes a Gaussian likelihood and also linearity of the derivatives, and so can only be used for an estimate. In \S \ref{sec:cosmomc} we will go beyond the Fisher matrix approach and present constraints derived from a likelihood analysis using \texttt{CosmoMC}: in the same way as this estimate, those limits can be expressed as bounds on (a simple combination of) the PC coefficients, and will therefore be immediately applicable to a wide range of models for energy deposition.

%%%%%%%%%%%%%%%%%%%%%%%%%%%%%%%%%%%%%%%%%%%%

\subsection{Sensitivity of future experiments}
\label{sec:sensitivity}

%\vspace{-1.25in}
\begin{figure*}[htpb]
\centering
(a) \begin{minipage}[ct]{.97\linewidth}
  \includegraphics[width=.45\textwidth]{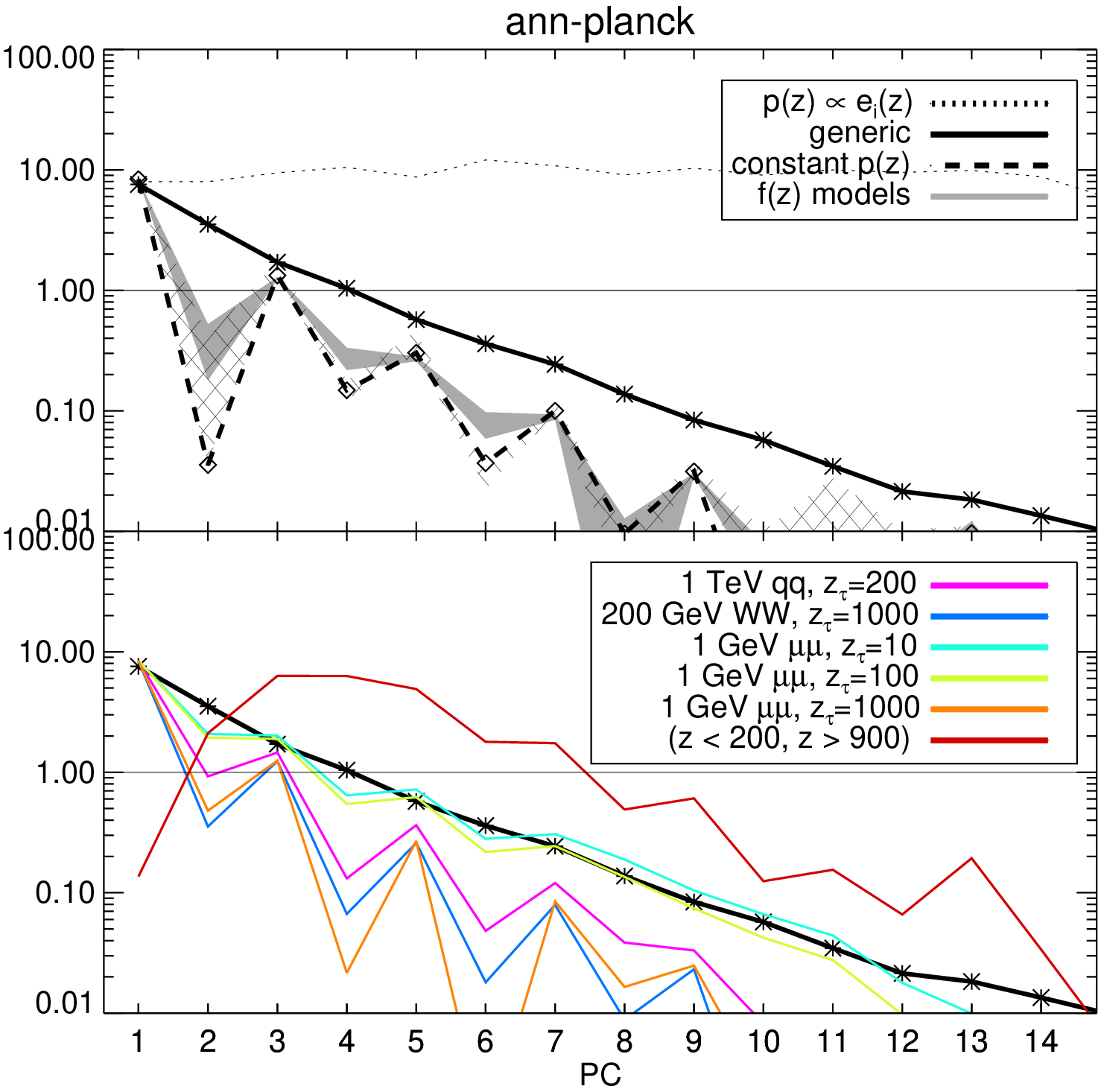}
  \hspace{0.2in}
  \includegraphics[width=.45\textwidth]{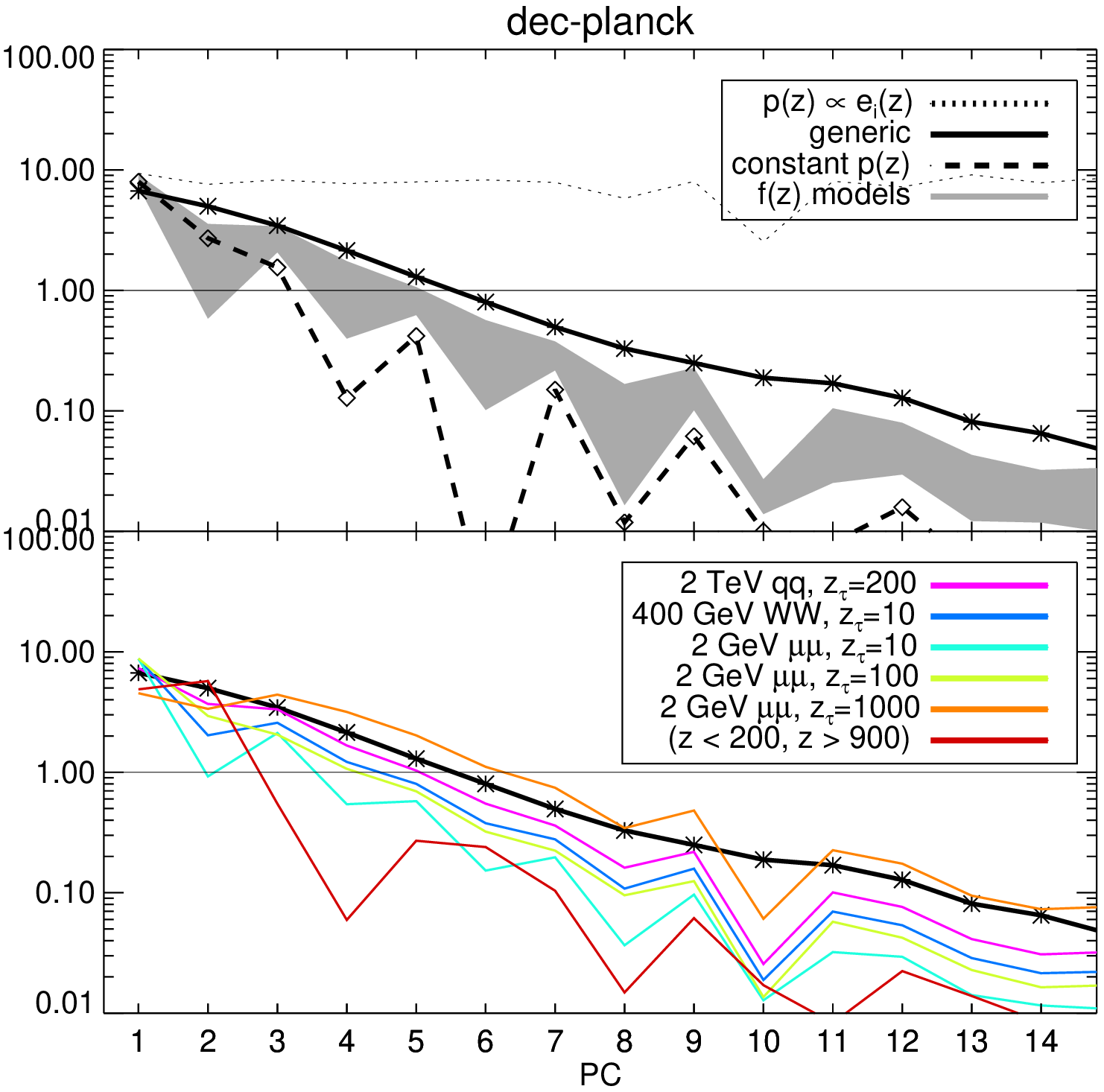}
    \end{minipage}
(b) \begin{minipage}[ct]{.97\linewidth}
  \includegraphics[width=.45\textwidth]{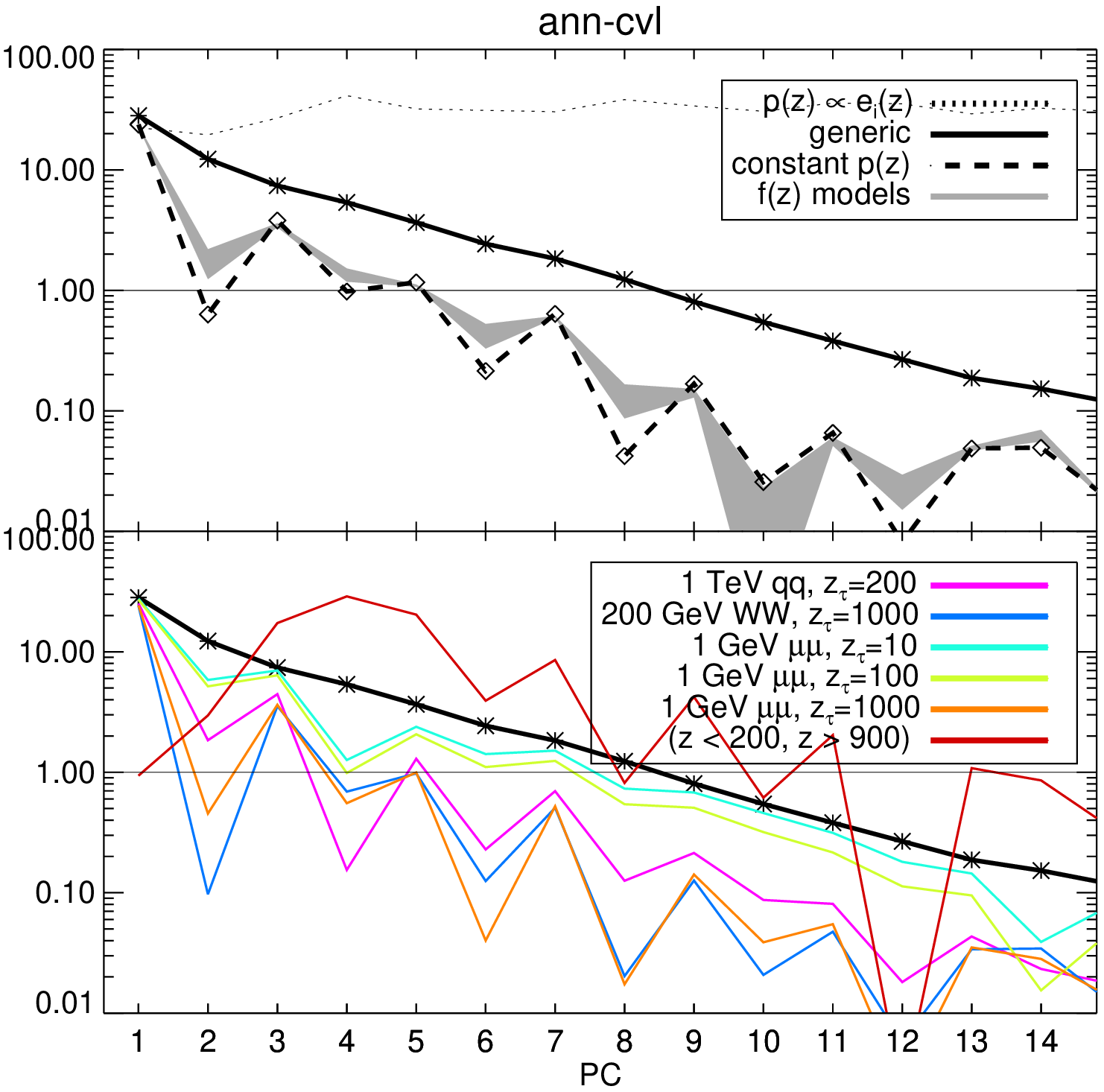}
  \hspace{0.2in}
  \includegraphics[width=.45\textwidth]{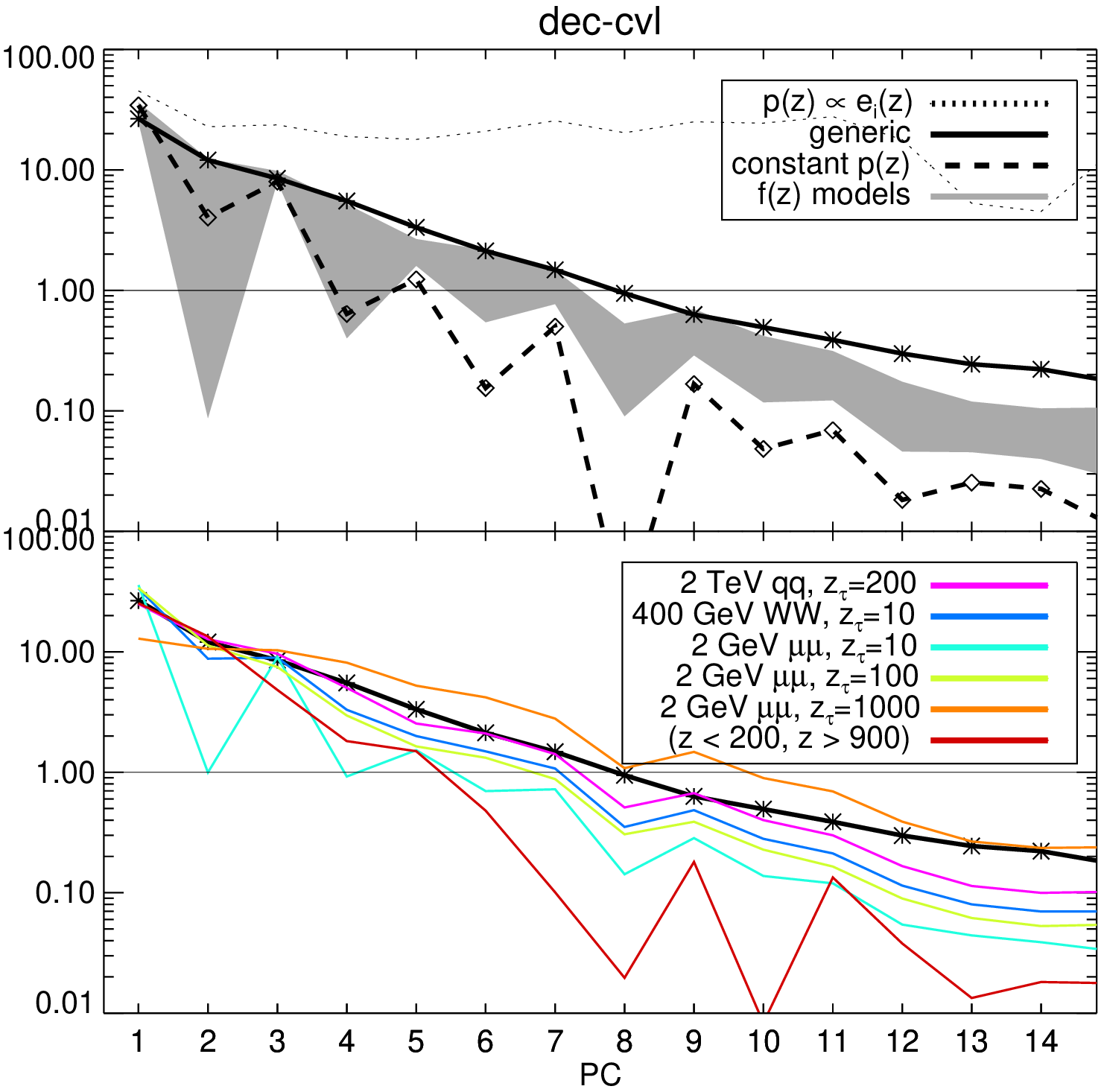}
    \end{minipage}
(c) \begin{minipage}[ct]{.97\linewidth}
 \includegraphics[width=.46\textwidth]{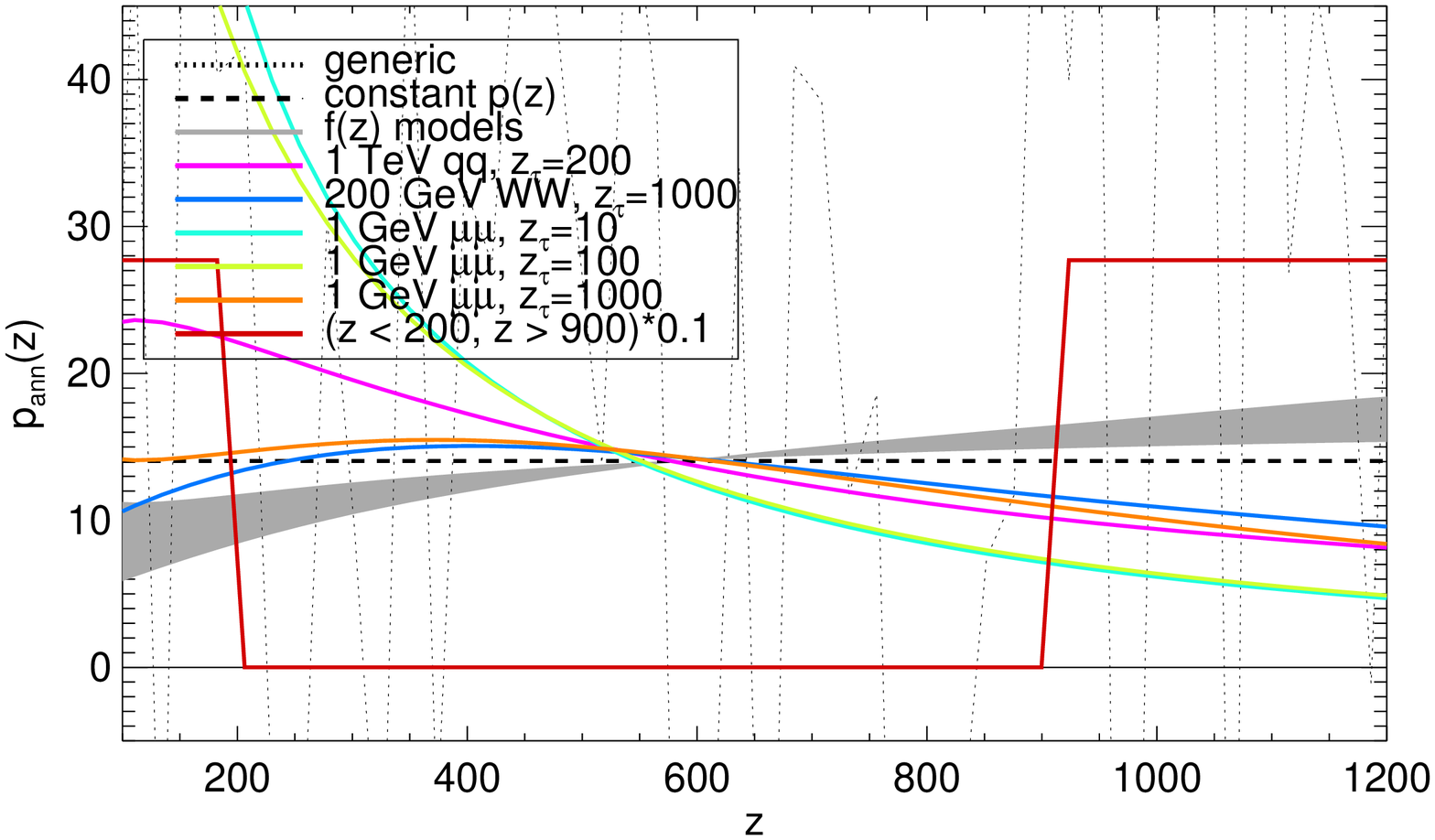}
  \hspace{0.15in}
 \includegraphics[width=.46\textwidth]{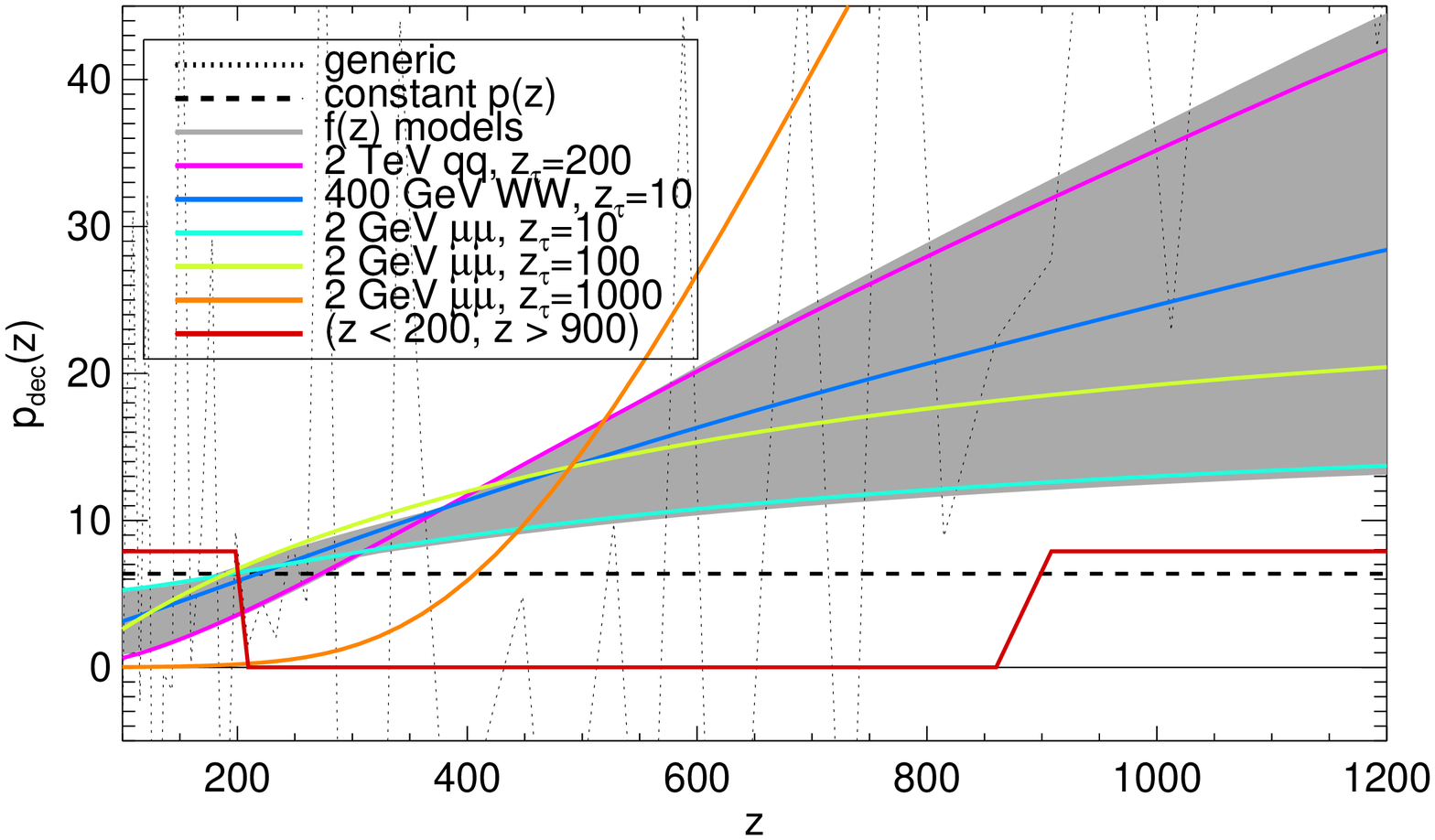}
    \end{minipage}
\centering
\caption{\label{fig:fzscan_planck} {\bf {(a)}} The sensitivity for \PLANCK (single-band), after marginalization, for various models subject to constraints from \WMAP7 single-band at 2$\sigma$. The left figure assumes annihilation-like energy deposition and the right figure assumes decay-like energy deposition. The top panels show: (1) assuming $p(z) \propto e_i(z)$ for each PC, (2) the generic case where all PC coefficients have equal magnitudes $|\varepsilon_i| = \varepsilon = 2/\sqrt{\sum_i \lambda_i^{{WMAP}}}$, (3) constant $p(z)$, and (4) taking $p(z) \propto f(z)$, with $f(z)$ from the models in \cite{Slatyer:2009yq}. For the left figure, the hatched region indicates the range of results from changing the ionization history calculator and including or neglecting the effects of helium and Lyman-$\alpha$ photons, described in \S \ref{app:recfast_vs_cosmorec}-\ref{app:he_lya}. The bottom panels show some sample $z_{\tau}$ models for asymmetric annihilating dark matter (\emph{left}) and decaying species (\emph{right}), as discussed in \S \ref{sec:sensitivity} (the labels describe the initial particle mass, and the SM final state for annihilation or decay), and an extreme case where $p(z)=0$ for $200 < z < 900$ and constant outside that range.  {\bf{(b)}} Same as (a), but for a CVL experiment. {\bf{(c)}} The models in (a) and (b) for annihilation-like (left) and decay-like (right). }
\end{figure*}

%%%%%%%%%%%%%%%%%%%%%%%%%%%%%%%%%%%%%%%%%%%%

For an energy deposition history where the sizes of the coefficients, $|\varepsilon_i|$, are all similar, the respective detectability of the PCs are given simply by their eigenvalues. Literally taking all the coefficients to be the same does not give a physical energy deposition history (since the later eigenvectors are highly oscillatory), but it is in some sense a ``generic'' scenario: none of the PCs have coefficients that are fine-tuned to be small, so slight changes to $p(z)$ or the basis of PCs are unlikely to drastically change the detectability of the different components.

We define detectability of the PCs with respect to this ``generic'' case; of course, detectability of any particular model depends on the relative sizes of coefficients. We consider a number of physical examples below to illustrate that, in some sense, the generic case is a reasonable average over a wide class of models of interest.

As discussed previously, \cite{Slatyer:2009yq} derived a set of energy deposition profiles corresponding to a range of DM annihilation models. These models provide a convenient set of example energy deposition histories, although they all have very similar effects on the CMB (see \S \ref{subsec:dmpca}). We adapt the code developed in \cite{Slatyer:2009yq} and discussed in detail there to obtain similar physical $f(z)$ curves for the case of decaying dark matter with a long lifetime.

While the DM itself must have a lifetime considerably longer than the age of the universe, there could be other metastable species which decay during the redshift range we study ($z\sim 10-1300$), or excited states of the dark matter which decay to the ground state + Standard Model particles (e.g. \cite{Finkbeiner:2008gw, Batell:2009vb, Finkbeiner:2009mi, Cline:2010kv, Bell:2010qt} and references therein). In this case the decay rate would cut off exponentially for $z < z(\tau)$, although heating and ionization of the gas could continue for some time after that: we can again obtain detailed $p(z)$ curves for different decay lifetimes using the methods of \cite{Slatyer:2009yq}.
Models of this type provide a simple class of examples suitable for use with the PCs derived for the case of decaying DM, since the underlying $dE / dt \propto (1+z)^3$ redshift dependence is the same (although for models with lifetimes short enough that the energy deposition has ceased shortly after recombination, the PCs derived for the annihilation-like case may work better).
 
For the annihilating case, asymmetric dark matter scenarios can
furnish a similar set of examples \cite{Hooper:2004dc, Cohen:2009fz, Kaplan:2009ag,
Dutta:2010va, Cohen:2010kn, Falkowski:2011xh}. In such scenarios the
DM sector possesses an asymmetry analogous to that in the baryon
sector, and it is this asymmetry which sets the DM relic density
rather than the annihilation cross section.  In the minimal case there
is thus no \emph{requirement} for an annihilation signal in the
present day or during the epoch of recombination, but it is
nonetheless possible to have a large late-time annihilation signal, by
repopulation of the depleted component at late times, or by
oscillations from the more-abundant to the less-abundant component
\cite{Cohen:2009fz, Cai:2009ia, Falkowski:2011xh}. As a simple example, we consider
models where another species decays to repopulate the less-abundant DM
state \cite{Falkowski:2011xh}, thus causing the annihilation to
``switch on'' as $1 - e^{-t/\tau}$ at a characteristic timescale $\tau$ (with $z_\tau$ being the corresponding redshift). We compute the
$p(z)$ curves for a range of $\tau$. Finally, for both
annihilation and decay we consider the constant $p(z)$ case, studied
in \cite{Chen:2003gz} (for decay) and \cite{Padmanabhan:2005es,
Galli:2009zc} (for annihilation), to facilitate comparison with the
literature.

Figure \ref{fig:fzscan_planck} shows the detectability of the
principal components in \PLANCK and the ideal CVL experiment for these
annihilating and decaying models, with the energy deposition normalized
to lie at the $95\%$ limit from \WMAP7. In the ``generic'' case, we
set the sizes of the coefficients of the \PLANCK (or CVL) PCs to be
$|\varepsilon_i| = \varepsilon = 2/\sqrt{\sum_i
\lambda_i^{{WMAP}}}$. The actual \WMAP7 signal-to-noise for the model
is
\begin{equation}
  \frac{S}{N} = \left( \sum_i \lambda_i^{WMAP} \left[ \sum_j \varepsilon_j e^{Planck}_j \cdot e^{WMAP}_i \right]^2 \right)^{1/2} \nonumber
\end{equation}
and thus depends on the signs of $\varepsilon_i$, but the generic case
is meant to indicate the typical detectability for a class of models,
so we instead use the \WMAP7 constraints to set an overall scale for
$|\varepsilon_i|$.

We also show the detectability for each PC if $p(z) \propto e_i(z)$,
or assuming the energy deposition history has zero overlap with all other
PCs\footnote{If the PCs were the same for the different experiments,
this would give an upper bound on the detectability of the $i$th PC,
given \WMAP7 $2\sigma$ constraints. However, the PCs for different
experiments are not orthogonal, $e_i^{Planck} \cdot e_j^{WMAP} \neq
\delta_{ij}$. A strict upper bound for the $S/N$ of the $i$th \PLANCK
PC is given by $(S/N)_i^{Planck} \le 2 \sqrt{\lambda_i^{Planck}}
\sum_j | e_i^{Planck} \cdot e_j^{WMAP}/\sqrt{\lambda_j^{WMAP}}|$, with
the analogous result for a CVL experiment. However, this quantity is
not very useful as an upper bound; for example, if $p(z)$ is
proportional to a high \WMAP PC, the normalization of $p(z)$ is
essentially unconstrained, but the detectability for \PLANCK may be
very significant if there is even a small overlap with the first
\PLANCK PC.}. As mentioned previously, this is not a physical
assumption (requiring an ``energy deposition'' oscillating rapidly
between positive and negative values): in such a case the effect on
the $C_\ell$'s is so small that the normalization of the ``energy
deposition'' could be very large and still consistent with
\WMAPc. Consequently, arbitrarily high PCs can be measured \emph{if}
they are the sole contributors to the energy deposition history.

We see that models with decay-like redshift dependence and those with annihilation-like redshift dependence tend to have roughly the same number of measurable parameters. 
In both cases, generally 2-3 components are potentially measurable in \PLANCK and up to 5-7 for a CVL experiment.

As a side note, the improvement of these constraints between \WMAP7 and future experiments is in large part due to (anticipated) better measurements of the polarization. In the absence of polarization data (i.e. using the TT spectrum only), we would expect the constraints to weaken by a factor of $\sim 3$ for \WMAP7, $\sim 7$ for \PLANCK, and $\sim 14$ for a CVL experiment. Here we have taken the square root of the eigenvalue of the first principal component as a proxy for sensitivity, which will be approximately true for models with a non-negligible overlap with the first PC. 

%%%%%%%%%%%%%%%%%%%%%%%%%%%%%%%%%%%%%%%%%%%%

\begin{figure*}[thb]
\includegraphics[width=0.47\textwidth]{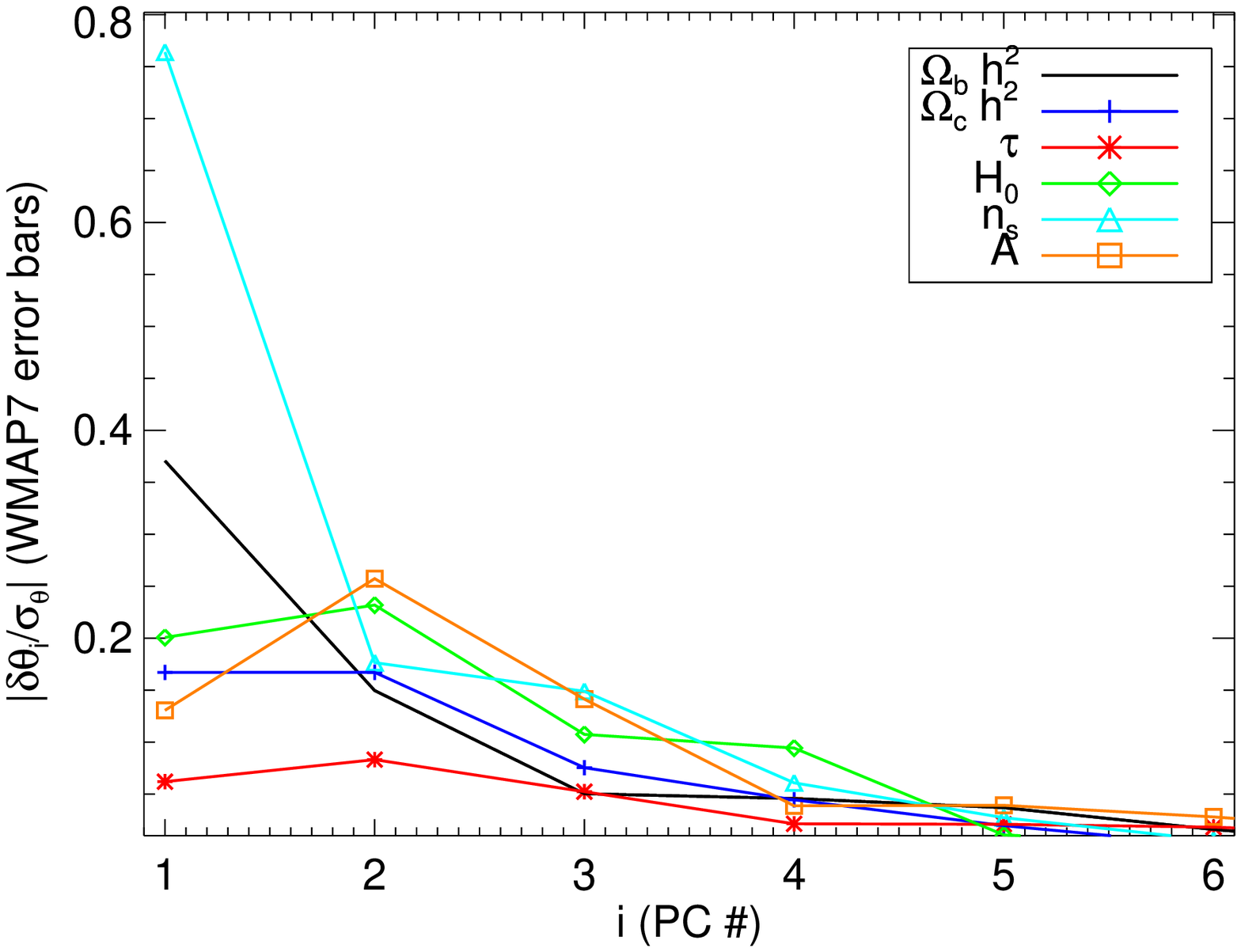}
\includegraphics[width=0.47\textwidth]{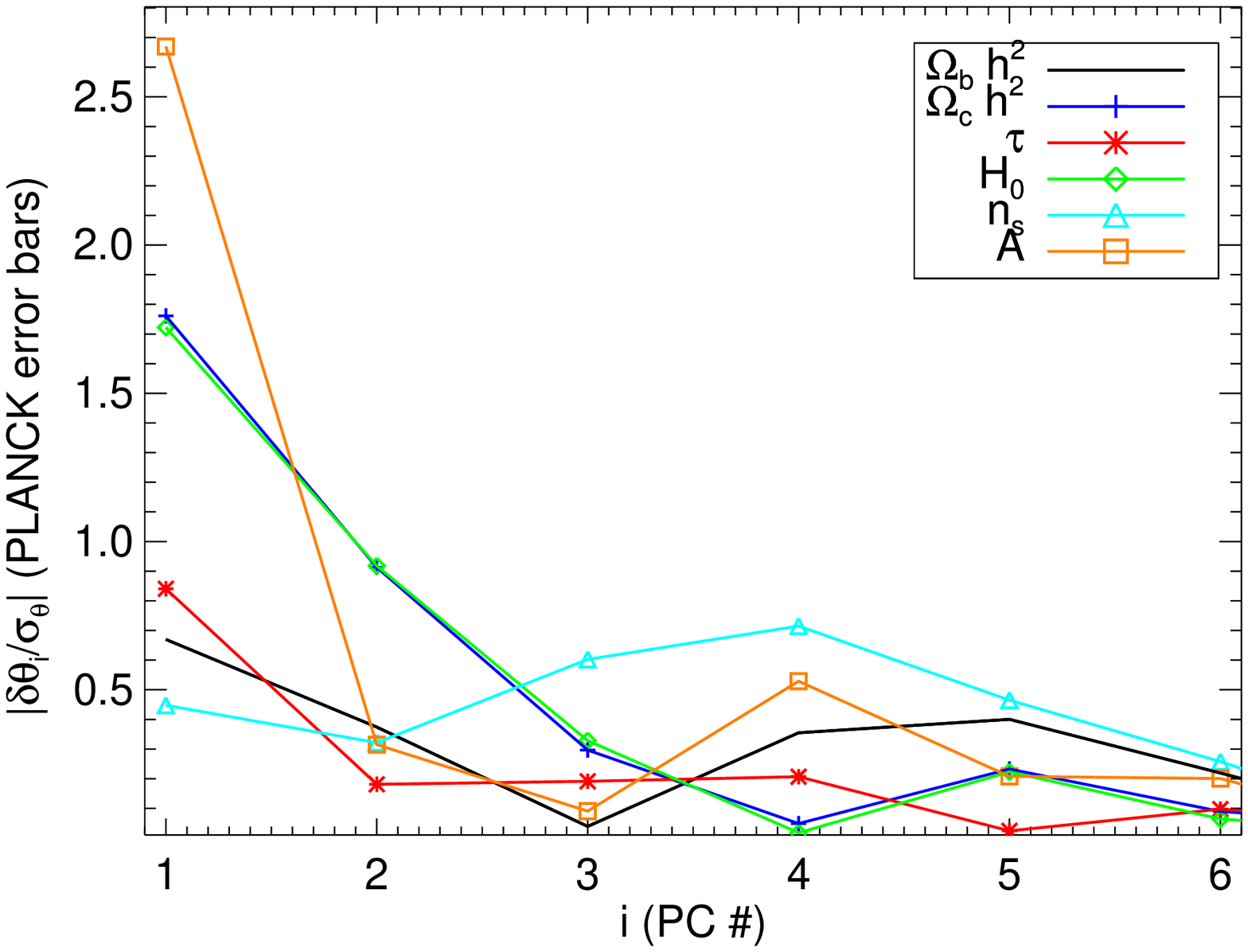}
\caption{For the $i$th PC, the contribution to the bias to
cosmological parameters in \WMAP7 (\emph{left panel}) and \PLANCK
(\emph{right panel}), relative to the error bars forecast from the
Fisher matrix. The normalization is that of the ``generic'' case (see
discussion in \S\ref{sec:sensitivity} or Figure
\ref{fig:fzscan_planck}), where each PC coefficient has the same
absolute value and the overall normalization is the maximum allowed by
\WMAP7 at 2$\sigma$. The total bias for the parameter $\theta$ is
$\sum_i \delta \theta_i$.  }
\label{fig:bias_planck}
\end{figure*}

%%%%%%%%%%%%%%%%%%%%%%%%%%%%%%%%%%%%%%%%%%%%

\subsection{Biases to the cosmological parameters}
\label{sec:bias}

If energy deposition is present but neglected, it can bias the measurement of
the cosmological parameters by a significant amount. For \WMAPc, the partial
degeneracy between varying $n_s$ and the effects of energy deposition means
that the dominant bias is a $1\sigma$ negative shift to $n_s$. The improved polarization sensitivity of \PLANCK largely lifts the degeneracy with $n_s$, but due to the smaller error bars of \PLANCK other parameters develop non-negligible biases: at the maximum
energy deposition allowed by \WMAP7 at $2\sigma$, \PLANCK parameter estimates
are generically biased at $>1\sigma$ for $\omega_c, H_0$, and $A_s$.

Calculation of the biases is exactly complementary to calculating the marginalized Fisher matrix. While the marginalization can be understood as projecting out the degeneracies with the cosmological parameters, the biases are given precisely by the effect of energy deposition in those degenerate directions. To be precise, suppose that some eigenvector $e_j$ has true coefficient $\varepsilon_j \ne 0$ and we falsely assume $\varepsilon_j$ to be zero: then each of the cosmological parameters $\theta_i$ will be shifted by an amount $\delta \theta_i$. The matrix of derivatives $\partial \theta_i / \partial \varepsilon_j$, $i=1..n_c$, $j=1..N$, is given simply by $\sum_k \left(F_c^{-1} F_v^T \right)_{ik} (e_j)_k$.

Thus we can partition the biases into the bias per PC, which is shown in Figure~\ref{fig:bias_planck} for \WMAP7 and \PLANCKc. For a generic energy deposition history, the total bias is dominated by the bias from the first few PCs, consistent with the fact that later PCs are undetectable and can essentially be neglected in any fit to the data. As expected from \cite{Galli:2010it}, the largest bias for \WMAP7 is to $n_s$.

%%%%%%%%%%%%%%%%%%%%%%%%%%%%%%%%%%%%%%%%%%%%

\section{A universal $p_\mathrm{ann}(z)$ for WIMP annihilation}
\label{subsec:dmpca}

\begin{figure}
\includegraphics[width=0.45\textwidth]{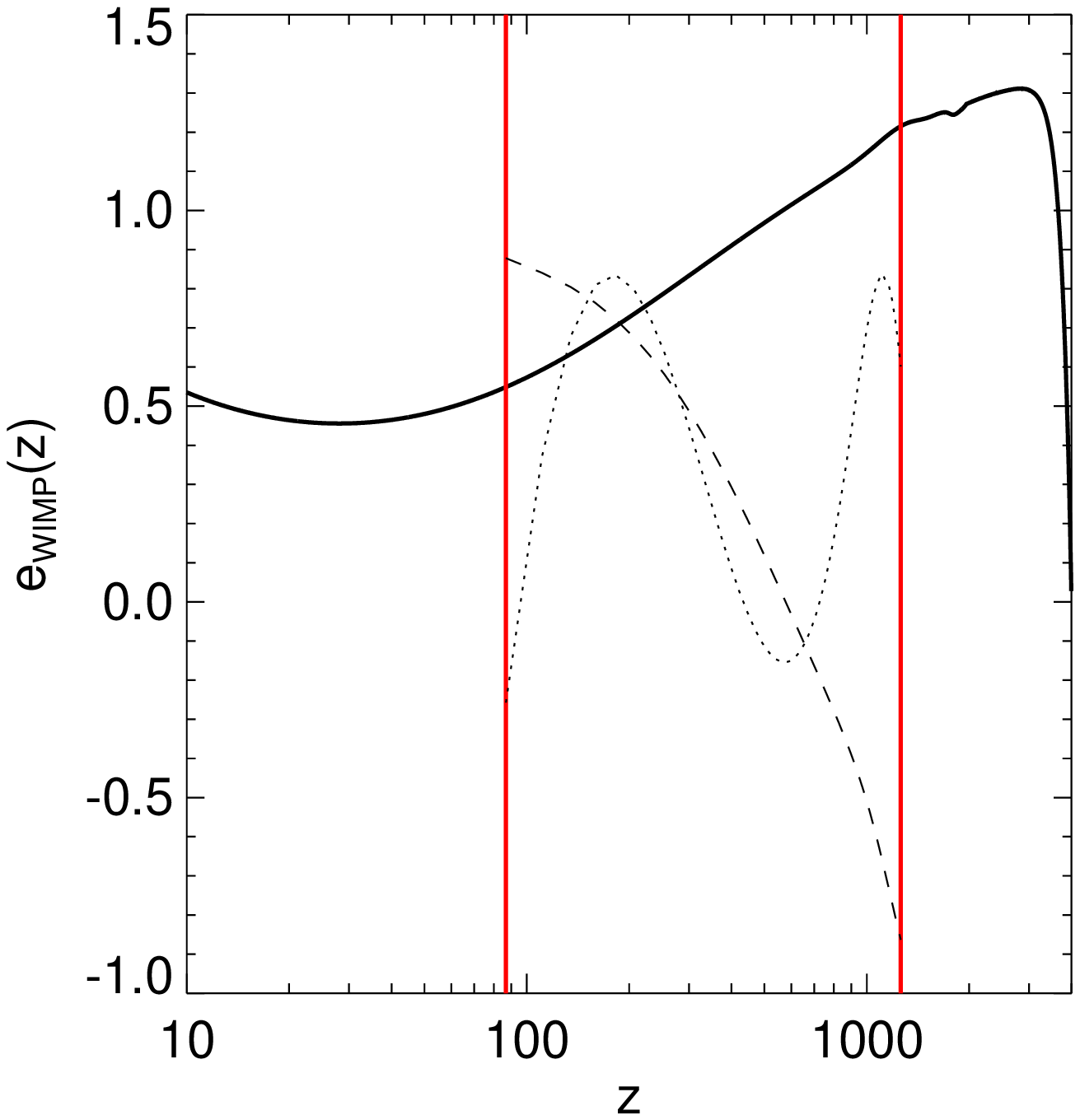}
\caption{The universal $e_\mathrm{WIMP}$ curve (solid black line), normalized as discussed in the text. Note that the principal component analysis is only performed in the redshift range between the vertical red solid lines; outside these lines, we still plot the linear combination of the input energy deposition histories that has been identified as the first principal component, to serve as a canonical energy deposition history for WIMP annihilation, but the $C_\ell$'s are not sensitive to the details of this energy deposition. In the PCA region we also plot the second (dashed) and third (dotted) principal components, with arbitrary normalization.}
\label{fig:universalf}
\end{figure}

Solutions for the redshift dependence of the efficiency function $f(z)$ (and hence the energy deposition history $p_\mathrm{ann}(z)$), for 41 different combinations of dark matter mass and annihilation channel, were presented in \cite{Slatyer:2009yq}. We can use these 41 energy deposition histories, rather than $\delta$-functions in $z$, as the input states for a principal component analysis, specialized for the particular case of conventional WIMP annihilation. Even after marginalization over the other cosmological parameters, we find that in this case the first eigenvalue completely dominates the later ones, accounting for $99.97\%$ of the total variance in \WMAP7, \PLANCK and the CVL forecast: thus, to a very good approximation, for any of the DM models studied in \cite{Slatyer:2009yq} (or any linear combination of the final states studied there), the effect on the $C_\ell$'s is determined entirely by the dot product of $p_\mathrm{ann}(z)$ with the first PC, with the $\ell$-dependence given by mapping the first PC to $C_\ell$-space.

This conclusion agrees with the statements in \cite{Slatyer:2009yq, Hutsi:2011vx} that the effect of DM annihilation can be captured by a single parameter. To put it another way, given equal coefficients for the first two principal components (which is already rather conservative, since the $f(z)$ curves studied are generally very similar to the first PC), the signal corresponding to the first PC would be roughly $60 \times$ larger than the signal corresponding to the second PC: the existence of energy deposition would have to be detected at 60$\sigma$ for even a $1 \sigma$ measurement of the second component to be possible. Measurements of the later components would be far more difficult still: the sum of the first two eigenvalues accounts for $1 - 4.8 \times 10^{-7}$, $1 - 6.0 \times 10^{-7}$, and $1 - 8.3 \times 10^{-7}$ the total variance in the \WMAP7, \PLANCK and CVL cases respectively. The effective $f$-value of various WIMP annihilation models is then just given by the dot product of their $f(z)$ curves with this first principal component. We provide effective $f$-values for all models considered in \cite{Slatyer:2009yq} on our website.

Let us denote this first principal component by $e_\mathrm{WIMP}(z)$. We note that the differences between the $e_\mathrm{WIMP}(z)$ curves corresponding to different experiments (\WMAP, \PLANCK and a CVL experiment) are extremely small, at the sub-percent level for all redshifts, and so it is reasonable to speak of a single such curve. We can choose to normalize $e_\mathrm{WIMP}(z)$ so that when it is multiplied by a given (dimensionful) factor $\varepsilon$ to obtain an energy deposition history $p_\mathrm{ann}(z) = \varepsilon e_\mathrm{WIMP}(z)$, the significance of the resulting signal in a particular experiment is the same as that of a constant-$p_\mathrm{ann}$ energy deposition with $p_\mathrm{ann} = \varepsilon$. This normalization simplifies comparisons with the earlier literature, in that the constraint on $\varepsilon$ is precisely the same as the familiar limit on constant $p_\mathrm{ann}$. The choice of experiment affects the normalization at the percent level; as a default, we will use the normalization appropriate for \PLANCKc. We plot the resulting $e_\mathrm{WIMP}(z)$ curve in Figure \ref{fig:universalf}; this curve and the corresponding $C_\ell$ shifts are also available online in tabulated form (see Appendix \ref{app:website}).

We could of course also write $e_\mathrm{WIMP}(z)$ in terms of the previously calculated principal components, just like any other energy deposition history. For example, for the \PLANCK PCs derived above, an energy deposition $\varepsilon \times e_\mathrm{WIMP}(z) = \sum_{i} \varepsilon_i e_i$ corresponds to $\{ \varepsilon_1, \varepsilon_2, \varepsilon_3, ...\} = \varepsilon \times \{ 4.64, -0.396, 3.11, ... \}$. In particular, this implies that the shapes of the $\delta C_\ell$'s induced by WIMP annihilation are quite close to those shown for the first PC in Figures \ref{fig:deltaclpcs}-\ref{fig:deltaclpcsperp}.

\begin{figure}
\includegraphics[width=0.45\textwidth]{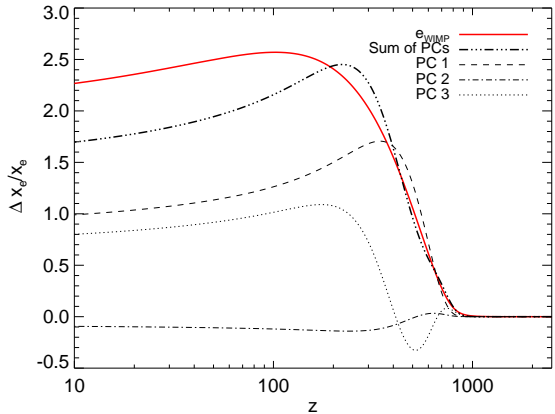}
\caption{Fractional change to the ionization fraction $x_e$ in the presence of energy deposition, for the universal $e_\mathrm{WIMP}$ curve (\emph{red, solid}). We also write the $e_\mathrm{WIMP}(z)$ curve as a linear combination of the \PLANCK principal components, and show (in \emph{black}) the effect on the ionization history of the first three principal components individually (weighted by their contribution to $e_\mathrm{WIMP}$), and their (weighted) sum.  The curves shown are extrapolated from the linear (small energy deposition) regime, with normalization factor $\varepsilon = 2 \times 10^{-27}$ cm$^3$/s/GeV for the $e_\mathrm{WIMP}$ curve.}
\label{fig:frac_delta_xe_eWIMP}
\end{figure}

In Figure \ref{fig:frac_delta_xe_eWIMP} we show, for reference, the modification to the ionization history associated with the $e_\mathrm{WIMP}$ curve, and the contributions from the first three principal components. We see that the first three PCs provide a good description of the ionization history modifications at $300 \lesssim z \lesssim 1000$, and in both cases the ionization fraction is essentially unaffected for $z \gtrsim 1000$; we can infer that the effect on the CMB of the discrepancy at lower redshifts is small.

%%%%%%%%%%%%%%%%%%%%%%%%%%%%%%%%%%%%%%%%%%%%

\section{CosmoMC Results \label{sec:cosmomc}}

%%%%%%%%%%%%%%%%%%%%%%%%%%%%%%%%%%%%%%%%%%%%

Everything we have done so far assumes both linearity and that the Fisher matrix is an adequate description of the likelihood function. We now present results of a full likelihood analysis using the \texttt{CosmoMC} Markov chain Monte Carlo code, in particular examining the biases to the cosmological parameters and the detectability of the PCs. Throughout this section we use \texttt{RECFAST} 1.5 as our ionization history calculator, since we have established that the PCs are unaffected by this choice (see \S \ref{app:recfast_vs_cosmorec}) and the interface to \texttt{CosmoMC} is better established. 

We sample the six cosmological parameters  $\omega_b$, $\omega_c$, $n_s$, $\ln 10^{10} A_s$(k = 0.002/Mpc), $\tau$ and $H_0$, all with flat priors. We consider purely adiabatic initial conditions. We parameterize the energy deposition due to dark matter annihilation using the marginalized principal components in redshift space $e_i$ presented in \S \ref{sec:pca}. We thus include 0, 1, 3, 5 or 7 additional parameters corresponding to the coefficients of the principal components with highest significance, as determined from the Fisher Matrix analysis\footnote{We use odd numbers of principal components to illustrate the effects of including more PCs because for near-constant $p_\mathrm{ann}$, the even-numbered principal components tend to have very small coefficients, and so including them does not substantially change the results.}. We impose flat priors on these parameters. Our treatment of the energy deposition is the same as described in the previous sections; we do \emph{not} include Lyman-$\alpha$ as a default (see \S \ref{app:he_lya}).

The MCMC convergence diagnostic tests are performed on 4 chains using the Gelman and Rubin (variance of chain
mean)/(mean of chain variances) R - 1 statistic for each parameter. Our constraints and the $1 - D$ and $2 - D$ likelihood contour plots are obtained after marginalization over the remaining nuisance parameters, again using the programs included in the \texttt{CosmoMC} package. We use a cosmic age top-hat prior of 10 Gyr $\leq t_0 \leq$ 20 Gyr.

We first determine the constraints on parameters using the seven-year WMAP data \cite{Komatsu:2010fb} (temperature and polarization) with the routine for computing the likelihood supplied by the WMAP team. For this case only, we also marginalize over a possible contamination from a Sunyaev-Zeldovich component (see e.g. \cite{Larson:2010gs}).

We then generate simulated data for \PLANCK and a CVL experiment using a fiducial cosmological model given by the best fit WMAP7 model. We simulate the data assuming in one case no energy deposition, and in another case an energy deposition history with constant $p_\mathrm{ann}=1.78\times 10^{-27}$ cm$^3$/s/GeV.
We model experimental noise as described in Equation \ref{eq:noise}. For the \PLANCK experiment, we use the specifications reported in Table \ref{tab:exptproperties} for the 147 GHz channel only. For the CVL experiment, we limit the maximum resolution of the experiment to $\ell_{max}=2500$. In each case, we use the PCs developed for that particular experiment, as described in \S \ref{sec:pca}.

%%%%%%%%%%%%%%%%%%%%%%%%%%%%%%%%%%%%%%%%

\subsection{Constraints and forecasts}

\begin{table*}[t]
\begin{tabular}{cc|r}
 \hline Number of &PC&WMAP7\\
PCs used&&95\%c.l.\\
\hline&&\\
1&PC&$<1.2\times 10^{-26}$cm$^3$/s/GeV
\\&&\\ \hline &&\\
1&$e_\mathrm{WIMP}(z)$ & $< 2.43 \times 10^{-27}$cm$^3$/s/GeV\\
&&\\ \hline
\end{tabular}
  \caption{\label{tab:cosmomcresWMAP}Upper limits on the first principal component amplitude using \WMAP7 data. In the last line we also show the constraints on the $e_\mathrm{WIMP}(z)$ principal component  presented in \S \ref{subsec:dmpca}. The uncertainties reported are upper limits at the 95\% c.l. We show the constraints obtained when only the first principal component is varied, together with the cosmological parameters. We assume in this case a flat positive prior on the amplitude of the principal component.}
\end{table*}

\begin{table*}[t]
\begin{tabular}{cc|rr|rr}
 \hline Number of &PC&\multicolumn{2}{c|}{Planck}&\multicolumn{2}{c}{CVL}\\
PCs used&&\multicolumn{1}{c}{$p_\mathrm{ann}=0$}&$p_\mathrm{ann}=1.78\times 10^{-27}$ cm$^3$/s/GeV&\multicolumn{1}{c}{$p_\mathrm{ann}=0$}&$p_\mathrm{ann}=1.78\times 10^{-27}$cm$^3$/s/GeV\\
\hline&&&&&\\
1&PC 1&$(0.2\pm1.1)\times10^{-27}$ &$(8.9\pm1.6)\times10^{-27}$&$(0.1\pm  4.9) \times 10^{-28}$&$(9.04\pm  0.81)\times10^{-27}$
\\&&&&&\\
3&PC 1&$(0.4\pm1.1)\times10^{-27}$&$(8.8\pm1.5)\times10^{-27}$&$(0.6\pm5.0)\times10^{-28}$ &$(8.71\pm  0.81)\times10^{-27}$ \\
3&PC 2&$(0.4  \pm 2.4)\times 10^{-27}$&$(0.7\pm 3.2)\times10^{-27}$&$(0.1\pm1.1)\times 10^{-27}$ &$(0.5\pm  1.5)\times 10^{-27}$\\
3&PC 3&$(1.7\pm 4.1)\times10^{-27}$&$(7.1\pm 5.8 )\times10^{-27}$&$(-0.3\pm 1.7)\times 10^{-27}$ &$(5.5\pm  2.5)\times 10^{-27}$ \\&&&&&\\
5&PC 1&-&-&$(0.9\pm 5.0)\times 10^{-28}$&$(8.87  \pm0.83)\times 10^{-27}$\\
5&PC 2&-&-&$(0.1\pm1.1)\times 10^{-27}$&$(0.4 \pm  1.5)\times 10^{-27}$\\
5&PC 3&-&-&$(-0.1\pm 1.8)\times 10^{-27}$&$(5.4 \pm  2.7)\times 10^{-27}$\\
5&PC 4&-&-&$(0.3\pm2.5)\times 10^{-27}$&$(1.9  \pm 2.8)\times 10^{-27}$\\
5&PC 5&-&-&$(0.2\pm3.4)\times 10^{-27}$&$(3.1  \pm 4.1)\times 10^{-27}$\\\hline&&&&&\\
1&$e_\mathrm{WIMP}(z)$ &$(0.3\pm2.2)\times10^{-28}$&$(1.84\pm0.31)\times 10^{-27}$ &$(-0.1 \pm 1.0)\times10^{-28}$&$(1.81\pm 0.16)\times 10^{-27}$\\&&&&&\\\hline

\end{tabular}
  \caption{\label{tab:cosmomcresPLANCKCVL}Constraints on the principal component amplitudes using simulated data for Planck and for a Cosmic Variance Limited experiment, assuming a fiducial energy deposition history of $p_\mathrm{ann}=1.78\times 10^{-27}$cm$^3$/s/GeV or no energy deposition ($p_\mathrm{ann}=0.$). In the last line we also show the constraints on the $e_\mathrm{WIMP}(z)$ principal component  presented in \S \ref{subsec:dmpca}. The uncertainties reported are at the 68\% c.l. We show the constraints obtained when different numbers of principal components are varied at the same time, together with the 6 $\Lambda CDM$ cosmological parameters. The results are reported in cm$^3$/s/GeV.}
\end{table*}

\begin{figure}[tb]
\includegraphics[width=0.5\textwidth]{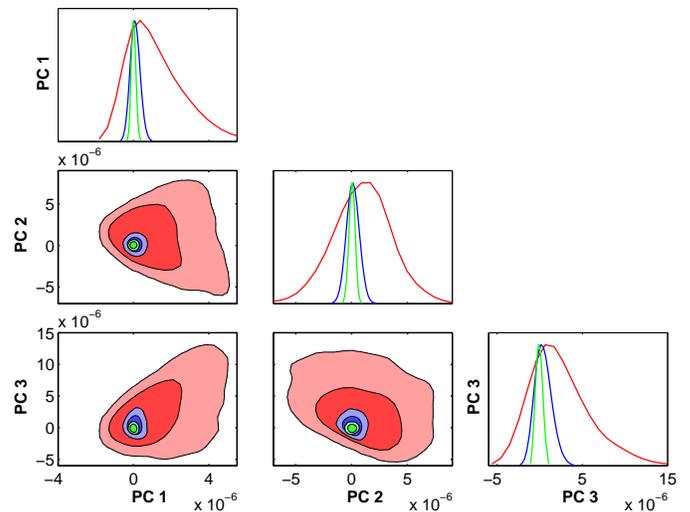}
\caption{Constraints from the seven-year \WMAP data (red), and from simulated data for Planck (blue) and a cosmic variance limited experiment (green). The plot shows marginalized one-dimensional distributions and two-dimensional 68\% and 95\% limits. The mock data for Planck and the CVL experiment assumed no dark matter annihilation. Three Principal Components were used in each run to model the energy deposition from dark matter annihilation. The units of the PC coefficients here are in m$^3$/s/kg, with $1\times 10^{-6}$m$^3$/s/kg = $1.8 \times 10^{-27}$ cm$^3$/s/GeV.
}
\label{fig:improvement}
\end{figure}

In Table \ref{tab:cosmomcresWMAP} we give the \WMAP7 upper limit at 95\% c.l. obtained on the amplitude of the first principal component (for generic energy deposition histories) and on the amplitude of the principal component $e_\mathrm{WIMP}(z)$ (the universal WIMP energy deposition history described in \S \ref{subsec:dmpca}). For these constraints we imposed a positive flat prior on the amplitude of the principal component.  This physical assumption is convenient in order to avoid a region of parameter space where the likelihood function is abruptly cut at some negative value of the principal component amplitude, where the recombination history calculation breaks down.

Figure \ref{fig:improvement} shows the 1-D and 2-D contour plots from \WMAP7 for the first three PCs, and the forecast improvements for \PLANCK and a CVL experiment. The results shown are obtained varying the standard cosmological parameters together with the amplitudes of the first 3 principal components for each experiment. 

From these plots, it is evident that the likelihoods for the PC amplitudes obtained from \WMAP7 are highly non-Gaussian. One reason is that the \WMAP7 data allow sets of coefficients for the principal components such that the corresponding energy deposition is negative and unphysical for some redshift. In particular, 
if the energy deposition is negative and sufficiently large in magnitude, this can cause the ionization history calculation to break down. If the coefficients of the PCs are all large, the condition that some linear combination of the PCs be non-negative (or where negative, sufficiently small) imposes quite a non-trivial constraint on the coefficients, which is reflected in peculiar-looking boundaries for the favored regions. This effect is clearly visible in the \WMAP results in Figure \ref{fig:improvement}; for \PLANCK the energy deposition history is much more constrained, and so this problem does not arise to nearly the same degree.

Another issue is that adding more principal components to the fit does \emph{not} always lead to a better reconstruction of the cosmological parameters and energy deposition history. By construction, higher PCs are less constrained by the data, and so the favored regions for their coefficients extend to much higher values, corresponding to very large energy deposition at particular redshifts. This means both that the previous problem of unphysical energy deposition histories reasserts itself, and that the effects of the energy deposition become nonlinear. 

As a consequence, the property of orthogonality between principal components does not hold anymore, and unexpected degeneracies between the PC parameters can arise: Figure \ref{fig:improvement} includes a 2-D contour plot showing the degeneracy between PC 2 and PC 3 for \WMAP7. This problem affects the \PLANCK  and CVL cases only when too many PCs that are completely unconstrained by the data are included in the MCMC runs. Thus there is an optimal number of PCs to include in the reconstruction for each experiment: beyond this point, adding further PCs to the reconstruction only obfuscates the results.

\begin{figure*}[tb]
\includegraphics[width=0.45\textwidth]{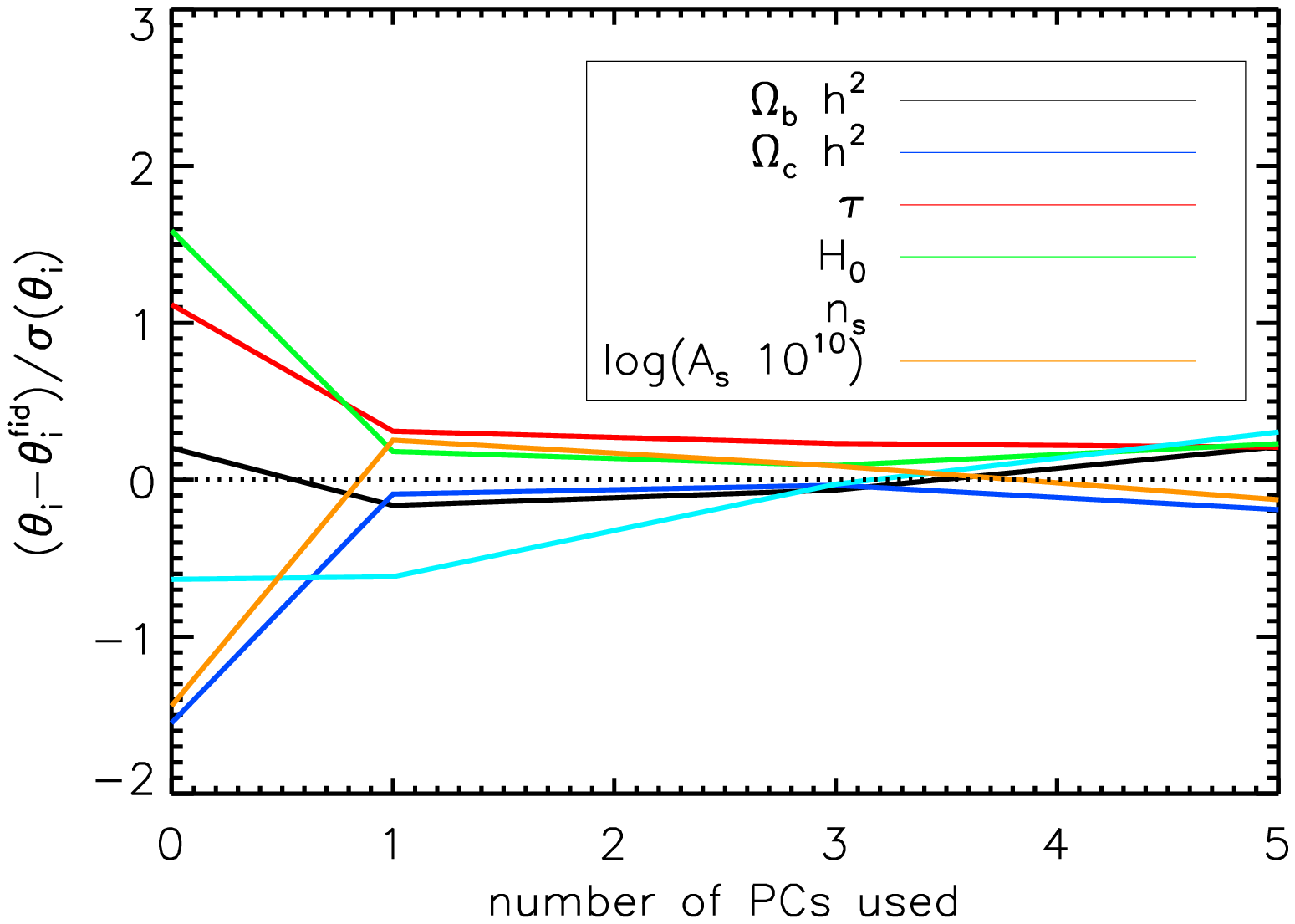}
\includegraphics[width=0.45\textwidth]{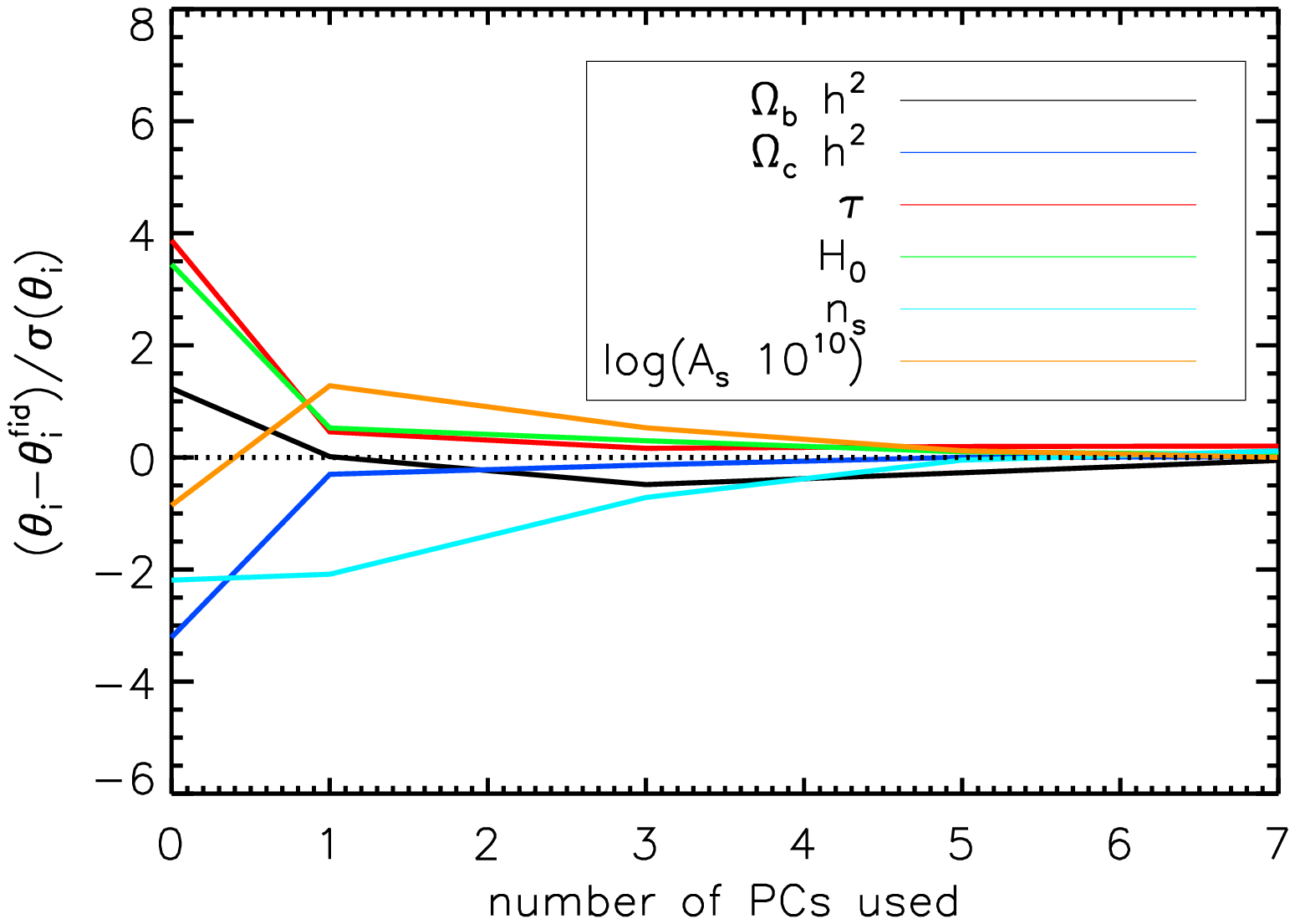}
\caption{Bias on the cosmological parameters using models with $\Lambda CDM$+ $n$ principal components for \PLANCK (\emph{left panel}) and CVL (\emph{right panel}) simulated data. Here, the bias is defined as $(\theta_i-\theta_i^{fid})/\sigma{\theta_i}$, where $\theta_i$ is the value of the mean value of the parameter from \texttt{CosmoMC}, $\theta_i^{fid}$ is the \WMAP7 marginalized value used as fiducial for the mock data, and $\sigma(\theta_i)$ is $1-\sigma$ error bound from the runs.
 Both sets of mock data assumed a constant-$p_\mathrm{ann}(z)$ energy deposition history with $\pann=1\times 10^{-6}$m$^3$/s/kg = $1.8 \times 10^{-27}$ cm$^3$/s/GeV. Note that this plot should \emph{not} be directly compared to Figure \ref{fig:bias_planck} (which shows the bias per unit-normalized PC, rather than the total remaining bias after a certain number of PCs have been reconstructed from the fit).
}
\label{fig:cosmomcbias}
\end{figure*}

In Figure \ref{fig:cosmomcbias}, we show how the bias to the cosmological parameters is reduced with the inclusion of more PCs, for both \PLANCK and a CVL experiment. We find that the best results are obtained when the number of included PCs equals the maximum number of ``measurable'' PCs (in the sense of \S \ref{sec:sensitivity}), i.e.  three PCs for \PLANCK and five for a CVL experiment\footnote{Note that the number of ``measurable'' PCs may \emph{not} reflect the actual number of PCs with S/N $> 1$ for any specific energy deposition history, which depends on the values of the $\varepsilon_i$ coefficients for that history; it is simply an estimate of the number of PCs that could feasibly be reconstructed from the data, from scanning over a range of models, and seems to also well describe the number of PCs with small enough error bars that nonlinearities do not cause problems.}. Later PCs cannot be reconstructed from the data and their inclusion does not improve the residual bias.

Table \ref{tab:cosmomcresPLANCKCVL} shows the forecast constraints (with uncertainties at 68\% c.l.) on the principal components amplitudes using simulated data for \PLANCK and for the CVL case, assuming a fiducial model with no energy deposition  ($p_\mathrm{ann}=0$) or with energy deposition described by constant $p_\mathrm{ann}=1.78\times 10^{-27}$cm$^3$/s/GeV. In these cases we let the amplitudes of the principal components assume both positive and negative values. In the table we report the results obtained when including different numbers of  principal components. The uncertainties on the amplitudes of the PCs (e.g. on PC 1) do not substantially change with the number of principal components included in the run.  Table \ref{tab:cosmomcresPLANCKCVL} also presents the constraints obtained using the  principal component $e_\mathrm{WIMP}(z)$ described in \S \ref{subsec:dmpca}.  

\begin{figure*}
\includegraphics[width=\textwidth]{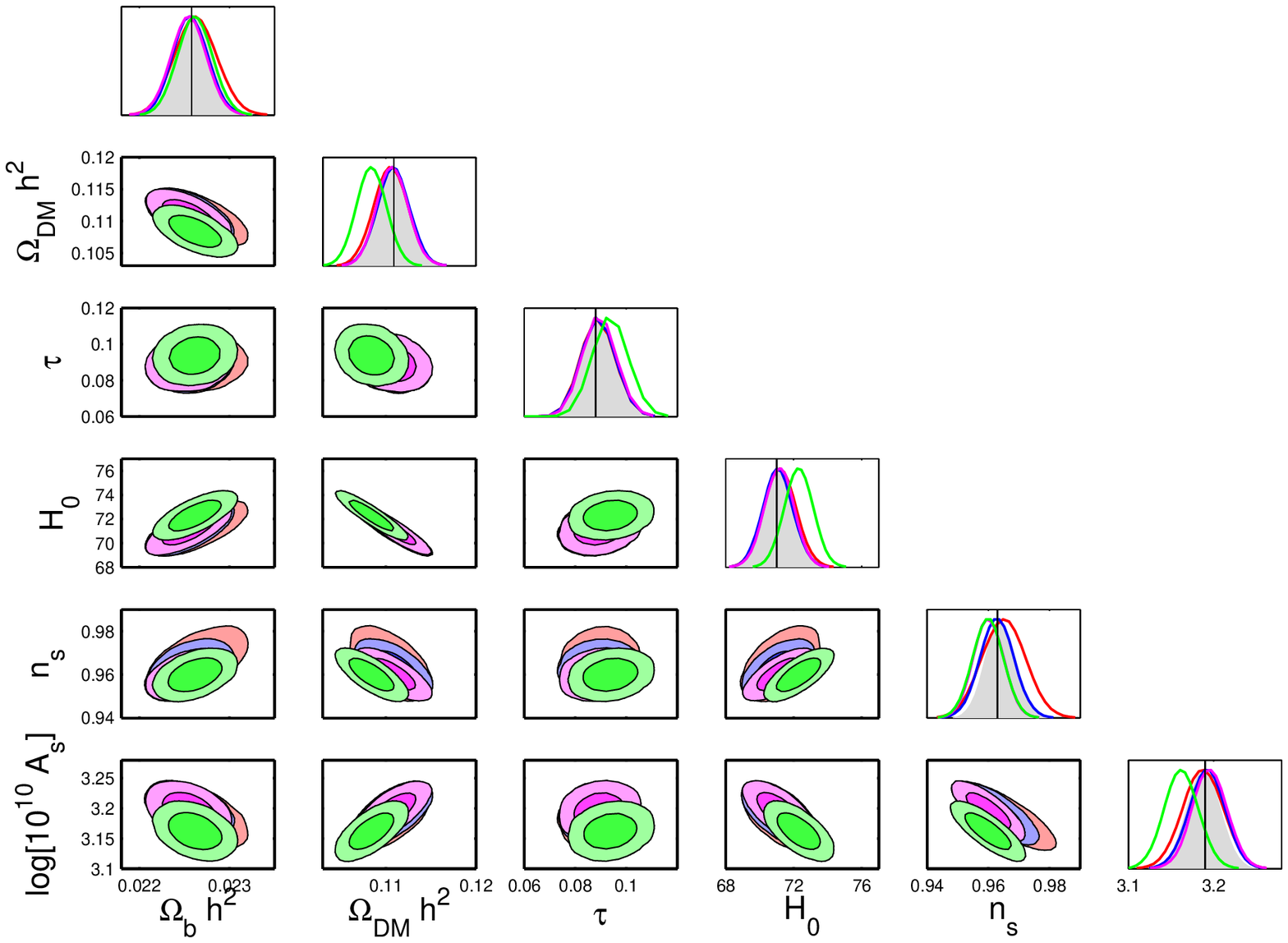}
\caption{Constraints from simulated data for \PLANCK on $\Lambda CDM$ parameters + 0 principal components (green), $\Lambda CDM$+ 1 PCs (magenta),  $\Lambda CDM$+ 3 PCs (blue), $\Lambda CDM$+ 5 PCs (red). 
 The plot shows marginalized one-dimensional distributions and two-dimensional 68\% and 95\% limits. The mock data for \PLANCK assumed for the solid lines includes energy deposition with constant $p_\mathrm{ann}=1\times 10^{-6}$m$^3$/s/kg = $1.8 \times 10^{-27}$ cm$^3$/s/GeV. The grey area shows the case of a mock data with no energy injection and a model $\Lambda CDM$+ 0 principal components. Only the cosmological parameters are shown.}
\label{fig:planckcosmo}
\end{figure*}

\begin{figure*}
\includegraphics[width=\textwidth]{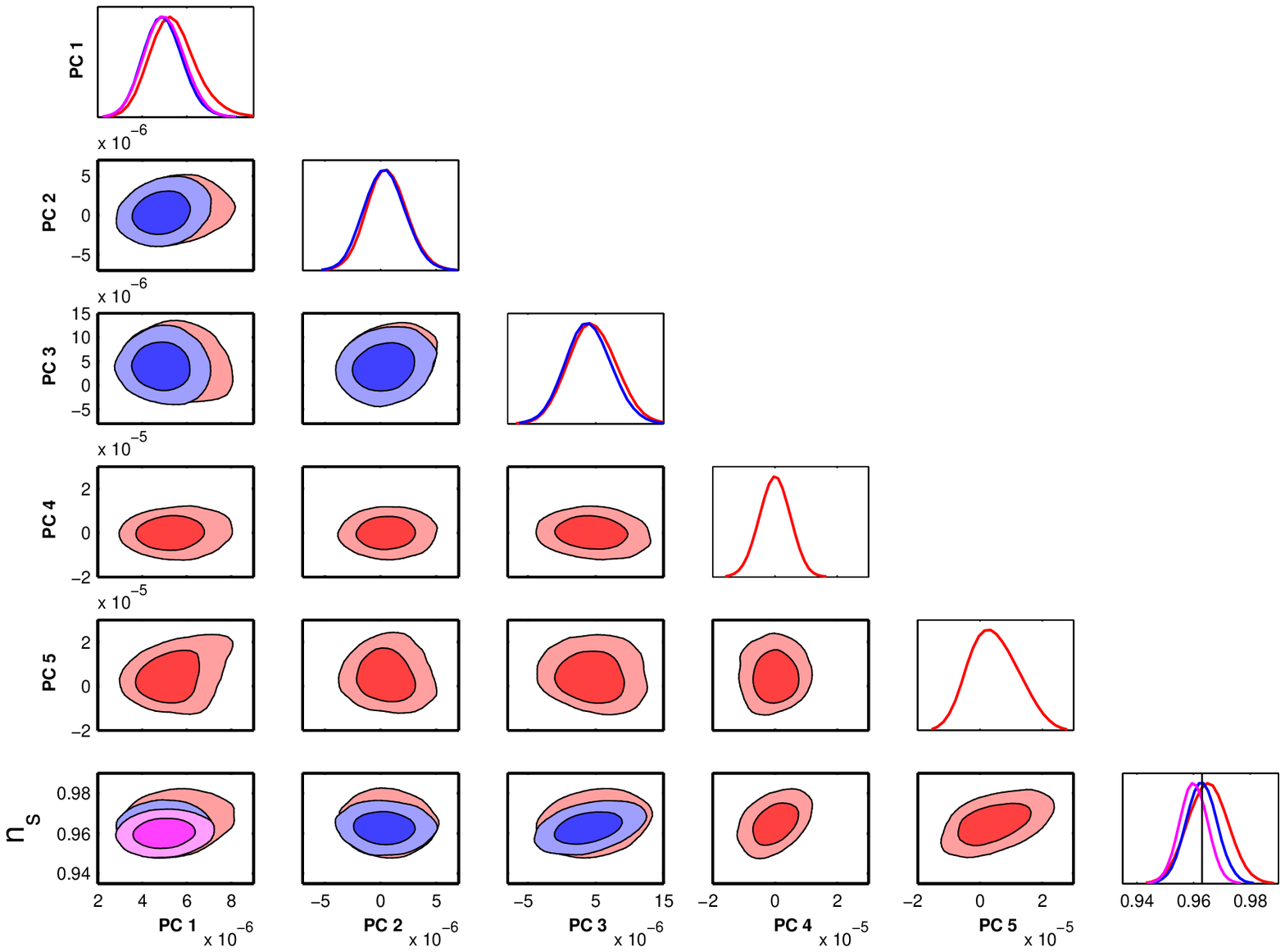}
\caption{Constraints from simulated data for \PLANCK  on $\Lambda CDM$+ 1 PCs (magenta),  $\Lambda CDM$+ 3 PCs (blue), $\Lambda CDM$+ 5 PCs (red). 
 The plot shows marginalized one-dimensional distributions and two-dimensional 68\% and 95\% limits. The mock data for \PLANCK assumed for the solid lines includes energy deposition with constant $p_\mathrm{ann}=1\times 10^{-6}$m$^3$/s/kg = $1.8 \times 10^{-27}$ cm$^3$/s/GeV, with the PC coefficients shown in units of m$^3$/s/kg.  Only the principal components and $n_s$ are shown.}
\label{fig:planckpcs}
\end{figure*}

Figures \ref{fig:planckcosmo}-\ref{fig:planckpcs} show the constraints on the various parameters from simulated \PLANCK data with a toy-model energy deposition history corresponding to constant $p_\mathrm{ann}(z) = 1.78\times 10^{-27}$ cm$^3$/s/GeV. As mentioned previously, while the coefficients of the early PCs are reconstructed well, we see that including additional PCs beyond the three expected to be measurable degrades the measurement of $n_s$ and to a lesser extend $A_s$.

\begin{figure*}
\includegraphics[width=\textwidth]{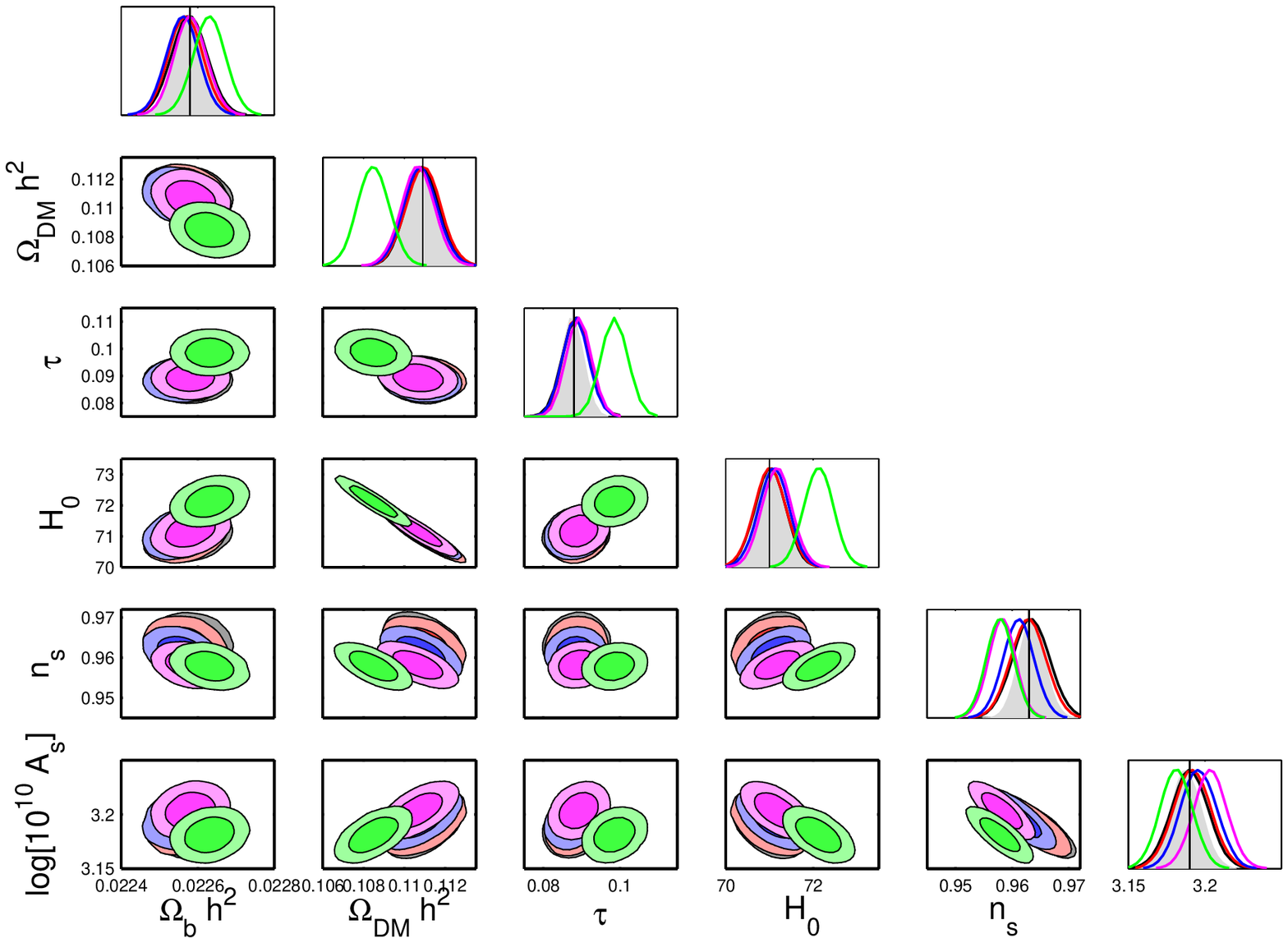}
\caption{Constraints from simulated data for a cosmic variance limited experiment on $\Lambda CDM$+ 0 principal components (green), $\Lambda CDM$+ 1 PCs (magenta), $\Lambda CDM$+ 3 PCs (blue), $\Lambda CDM$+ 5 PCs (red), $\Lambda CDM$+ 7 PCs (black). 
 The plot shows marginalized one-dimensional distributions and two-dimensional 68\% and 95\% limits. The mock data for the CVL experiment assumed for the solid lines includes an energy deposition history with constant $p_\mathrm{ann}=1\times 10^{-6}$m$^3$/s/kg = $1.8 \times 10^{-27}$ cm$^3$/s/GeV. The grey area shows the case of a mock data with no energy injection and a model $\Lambda CDM$+ 0 principal components. Only the cosmological parameters are shown.
}
\label{fig:cvlcosmo}
\end{figure*}

\begin{figure*}
\includegraphics[width=\textwidth]{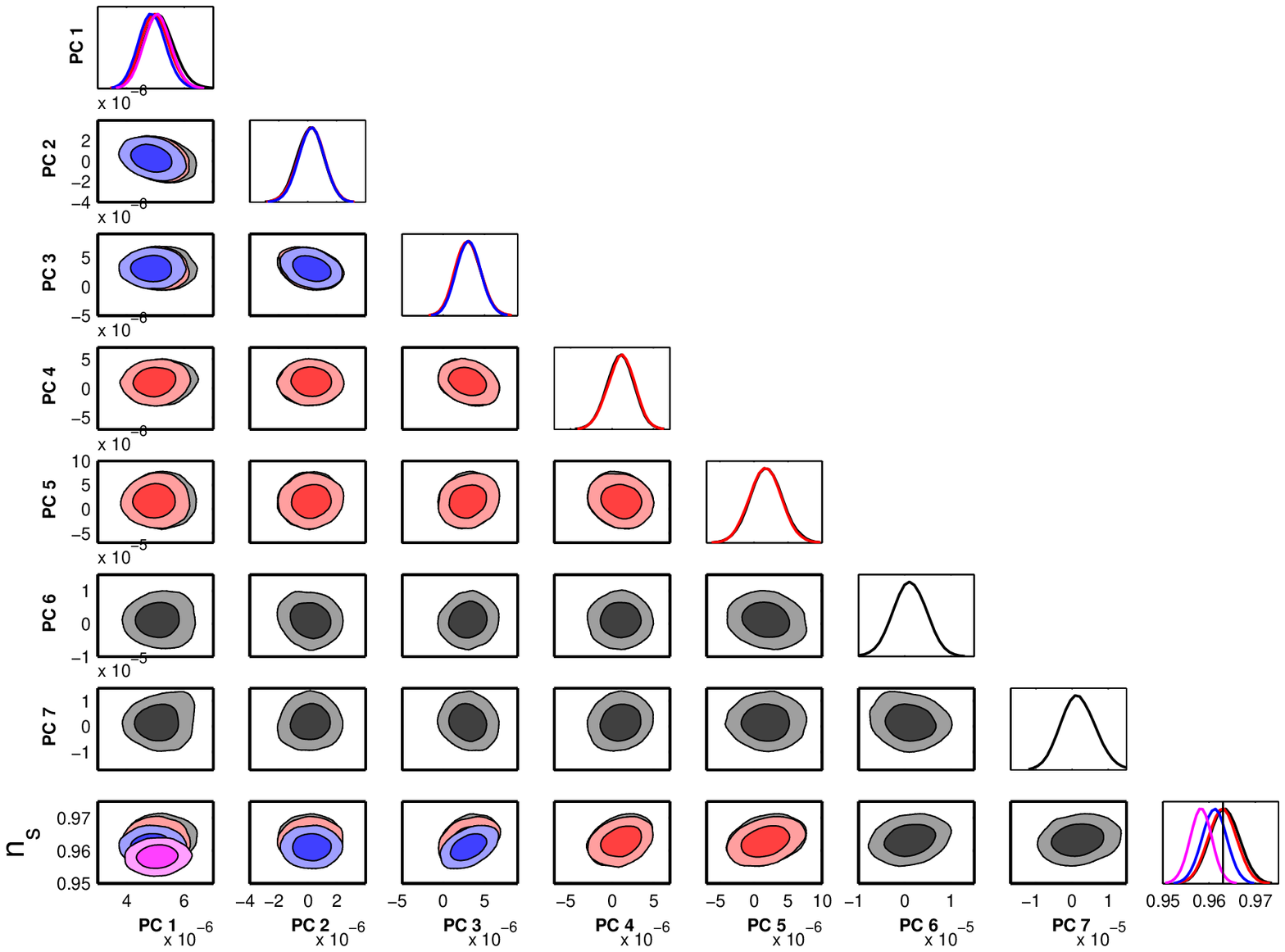}
\caption{Constraints from simulated data for a cosmic variance limited experiment on $\Lambda CDM$+ 1 PCs (magenta), $\Lambda CDM$+ 3 PCs (blue), $\Lambda CDM$+ 5 PCs (red), $\Lambda CDM$+ 7 PCs (black). 
 The plot shows marginalized one-dimensional distributions and two-dimensional 68\% and 95\% limits. The mock data for the CVL experiment assumed for the solid lines includes an energy deposition history with constant $p_\mathrm{ann}=1\times 10^{-6}$m$^3$/s/kg = $1.8 \times 10^{-27}$ cm$^3$/s/GeV, with the PC coefficients shown in units of m$^3$/s/kg.  Only the principal components and $n_s$ are shown.}
\label{fig:cvlpcs}
\end{figure*}

Figures \ref{fig:cvlcosmo}-\ref{fig:cvlpcs} show the constraints on the various parameters from a simulated CVL experiment with the same energy deposition history as in the \PLANCK case. We now see that including at least five PCs is necessary to remove the bias to the cosmological parameters, especially $n_s$ and $A_s$ but going from five to seven PCs neither greatly improves nor degrades the reconstruction, and is only significant at all in the case of $n_s$.

%%%%%%%%%%%%%%%%%%%%%%

%%%%%%%%%%%%%%%%%%%%%%%

\subsection{Application of constraints}
 
Once the constraints on the individual principal components are obtained, they can be used to set general bounds.
Given an arbitrary energy deposition history, it can be decomposed into the principal component basis we have supplied online (see Appendix~\ref{app:website}), and the coefficients in that basis compared to the limits presented here for \WMAP7. 

As a simple example, suppose we wish to set limits on an energy deposition
history with positive constant $p_\mathrm{ann}$, by projecting onto the
constraints obtained for the first principal component using \WMAP7
data. Recall that we 
impose a positive prior on the amplitude of the principal component.
The upper limit on the first principal component using \WMAP7 data is then \footnote{This value depends on the normalization of the principal components used. Here we employ the PCs supplied online (see Appendix \ref{app:website}), which are orthonormal when expressed as a vector sampled at 50 redshifts.} $\varepsilon_1<1.2\times10^{-26}$cm$^3$/s/GeV at 95\% confidence. From Equation \ref{eq:decomposedpz} and the discussion in \S \ref{sec:fisher}, we see that for a constant-$p_\mathrm{ann}$ energy deposition history we have 
$\varepsilon_i = p_\mathrm{ann}\sum_{j=1}^N e_i(z_j)$:  
the resulting derived upper limit on an energy deposition history with constant positive $p_\mathrm{ann}$ is $ p_\mathrm{ann}<2.8\times 10^{-27}$cm$^3$/s/GeV at 95\% confidence. This constraint can be compared with the one obtained directly by sampling the cosmological parameters with a positive constant $p_\mathrm{ann}$, which gives an upper limit of $p_\mathrm{ann}<2.7\times 10^{-27}$cm$^3$/s/GeV at 95\% confidence.
In a more general case when more than one principal component is considered in the analysis, the limit on $p_\mathrm{ann}$ can be determined with a  $\chi^2$ analysis as in Equation \ref{deltachi}, writing the theoretically predicted amplitude $\bar{\varepsilon}_i$ as a function of $p_\mathrm{ann}$.

For WIMP models, the \WMAP7 limit for $e_\mathrm{WIMP}(z)$  in Table~\ref{tab:cosmomcresWMAP} can be applied directly given the effective $f$-value; these were described in \S\ref{subsec:dmpca} and are available online for a range of models (see Appendix \ref{app:website}). As another consistency check, we can apply the \WMAP7 limit on the first principal component to the WIMP case, $p_\mathrm{ann}(z) = \varepsilon e_\mathrm{WIMP}(z)$. For \WMAP7, the coefficient $\varepsilon_1$ is given by $4.41 \varepsilon$, giving a bound of $\varepsilon < 2.7 \times  10^{-27}$cm$^3$/s/GeV. The true bound on $\varepsilon$ is roughly 10$\%$ stronger.

Once \PLANCK data are available, the same analysis can be redone with real data, and a very broad range of models can then be confronted with the resulting PC-based limits (the only exceptions being models which have negligible overlaps with the first few PCs and thus evade the constraints in this form; studying limits on such models will still require a separate analysis).

Using mock \PLANCK data, we have confirmed that the constraints on the first three principal components can be used to  recover the correct limit on a particular energy deposition history -- that is, the limit that we would obtain by directly varying the energy deposition amplitude for that specific model. Again taking the constant-$p_\mathrm{ann}$ case as a simple example, but now using \PLANCK mock data with an energy deposition history with constant $p_\mathrm{ann}=1.78\times 10^{-27}$cm$^3$/s/GeV, the amplitudes for the first 3 principal components are:
\begin{eqnarray}
 \varepsilon_1&=&(8.8\pm{1.5})\times10^{-27}\mathrm{cm}^3\mathrm{/s/GeV}\\
\varepsilon_2&=&(6.8\pm 31.5 )\times10^{-28}\mathrm{cm}^3\mathrm{/s/GeV}, \\ 
\varepsilon_3&=&(7.1\pm 5.8)\times10^{-27}\mathrm{cm}^3\mathrm{/s/GeV},
\end{eqnarray}
where the errors are at 68\% confidence. From a $\chi^2$ analysis of these results, the recovered energy deposition is given by $p_\mathrm{ann}=(1.90\pm 0.50)\times 10^{-27}$cm$^3$/s/GeV at 68\% c.l.. This constraint can be compared with the one obtained directly by sampling the cosmological parameters with a constant $p_\mathrm{ann}$, which gives an upper limit of $p_\mathrm{ann}=(1.90\pm0.32)\times 10^{-27}$cm$^3$/s/GeV at 68\% c.l.. These checks confirm the validity and usefulness of the principal component decomposition.

%%%%%%%%%%%%%%%%%%%%%%
%%%%%%%%%%%%%%%%

\subsection{Bias on cosmological parameters and agreement with the Fisher matrix}

\begin{figure*}
\includegraphics[width=0.34\textwidth]{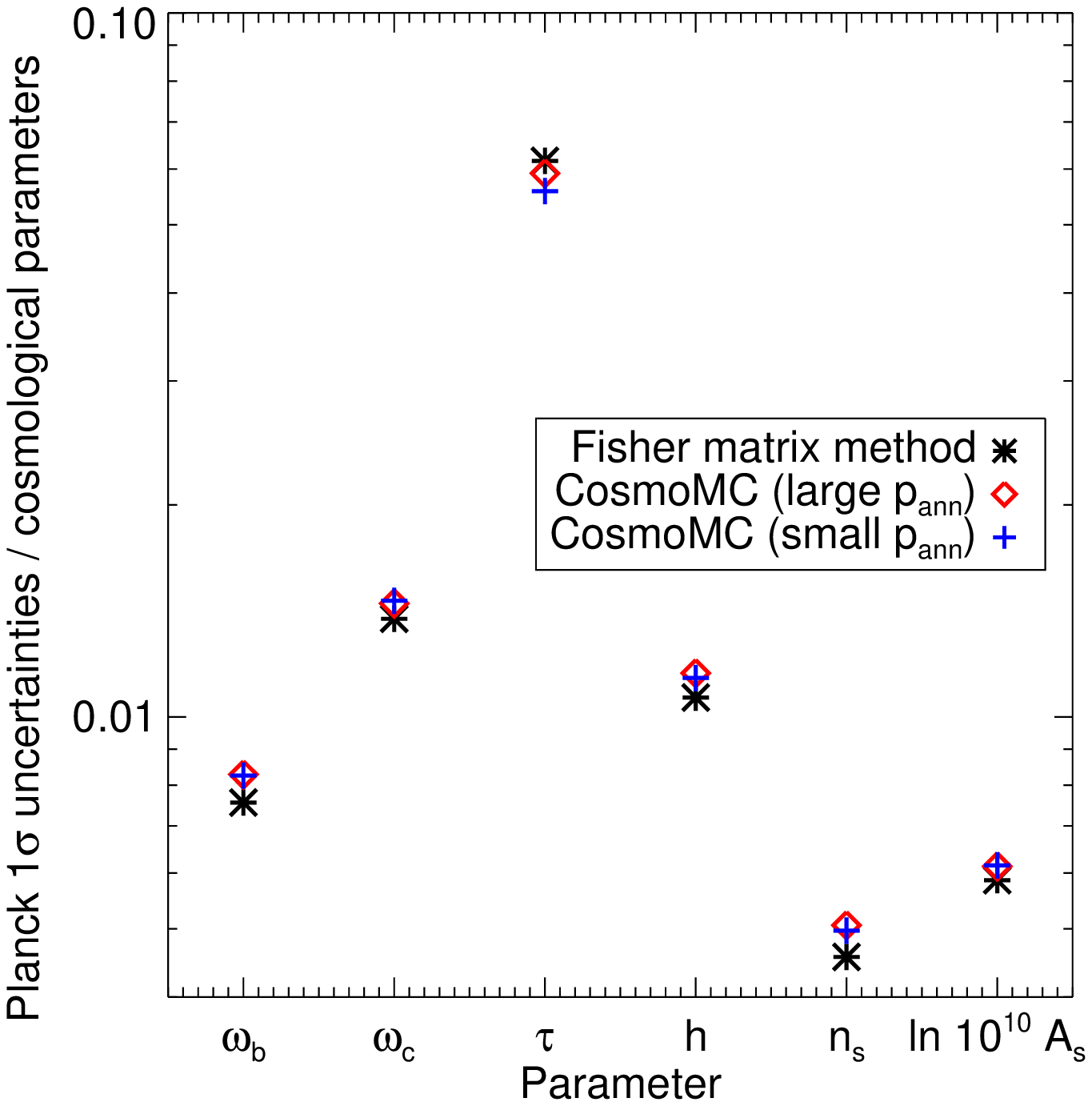}
\includegraphics[width=0.34\textwidth]{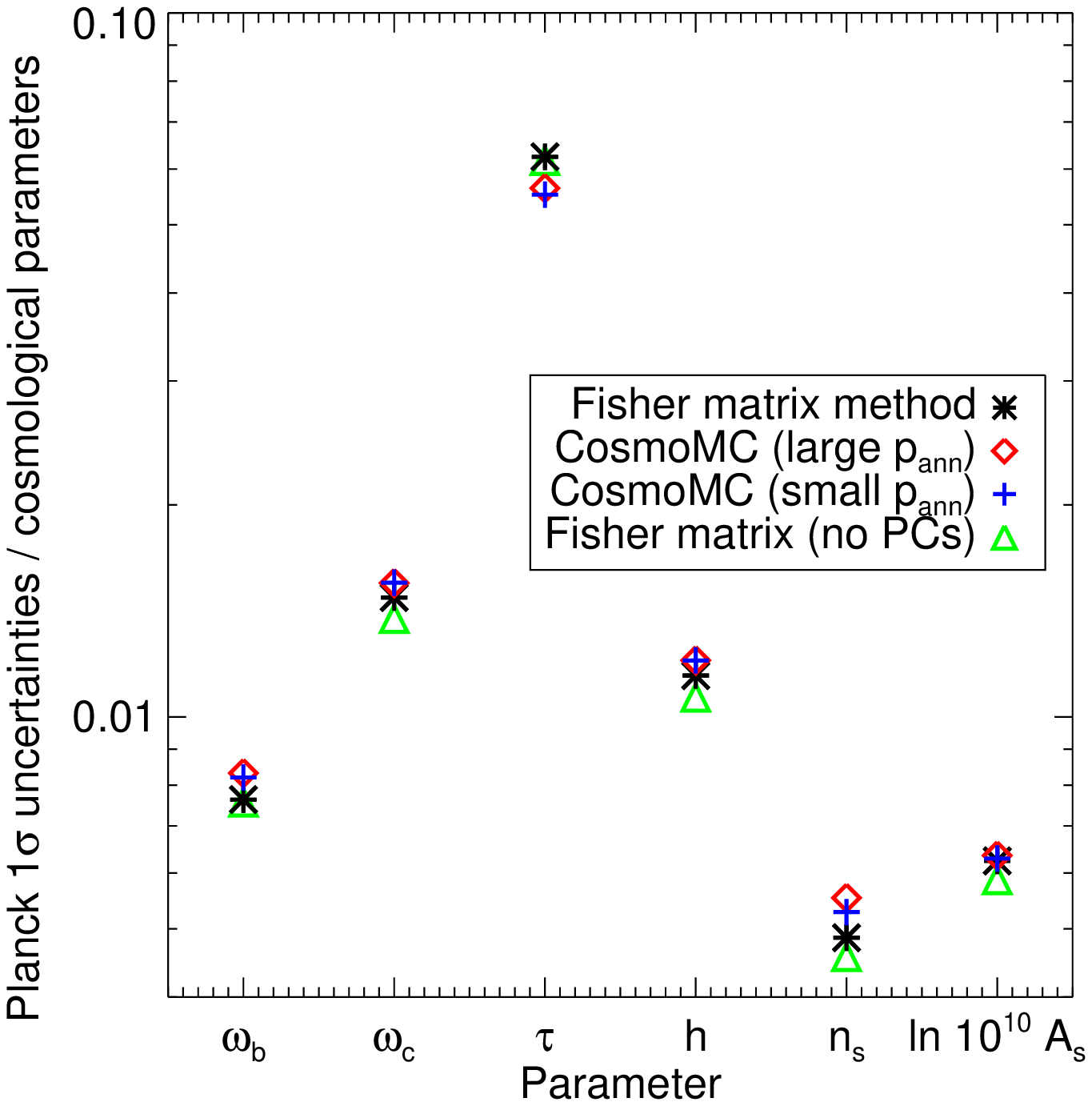}
\includegraphics[width=0.284\textwidth]{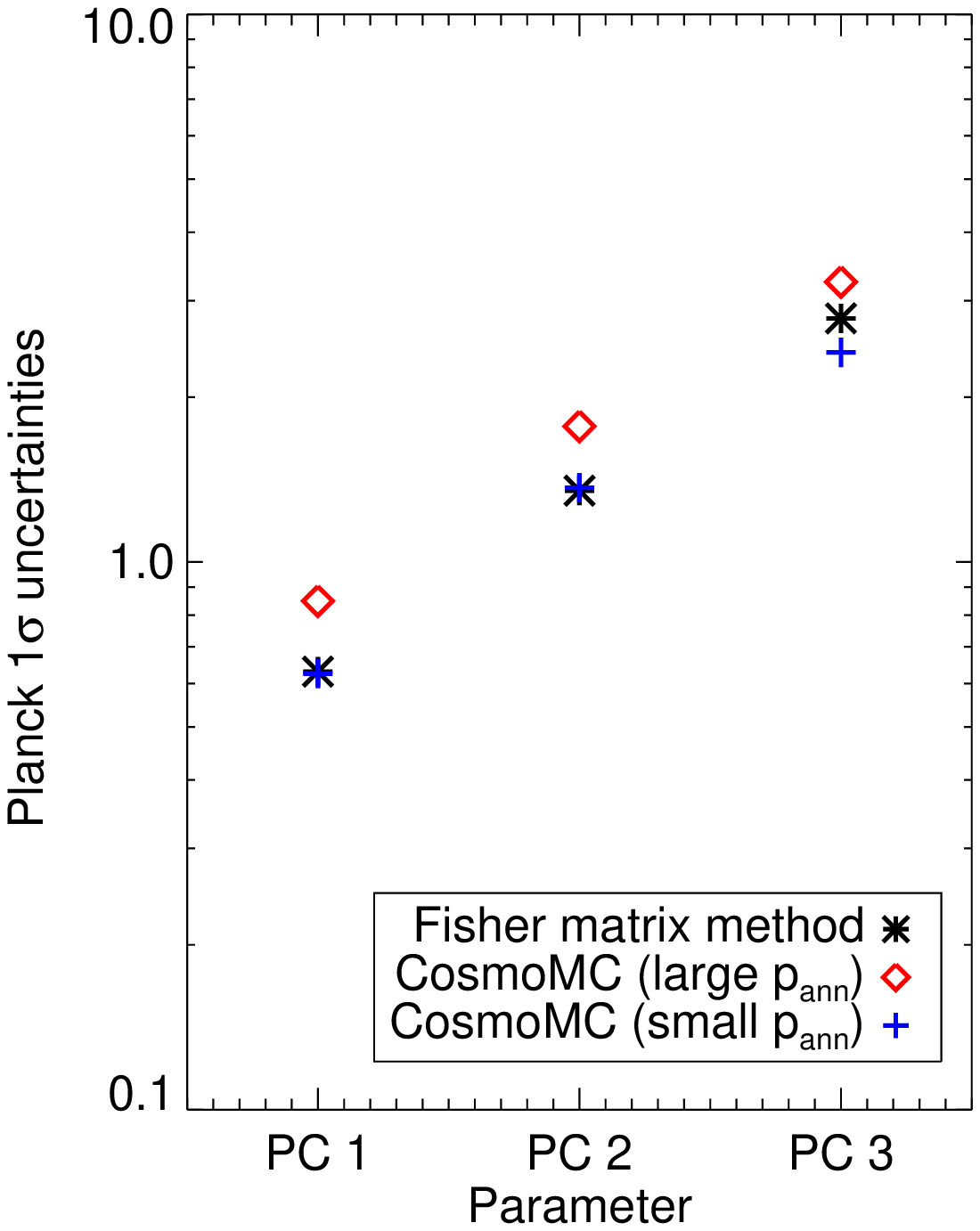}
\caption{Error bars for the cosmological parameters and coefficients of the principal components, in mock \PLANCK data, simulated assuming a constant-$p_\mathrm{ann}(z)$ energy deposition history. The points marked ``large $p_\mathrm{ann}$'' have $p_\mathrm{ann}=1\times 10^{-6}$m$^3$/s/kg = $1.8 \times 10^{-27}$ cm$^3$/s/GeV, while for the points marked ``small $p_\mathrm{ann}$'' the value is a factor of 10 lower. In the \emph{left panel} the fit to the data is performed using only the standard six cosmological parameters, whereas in the \emph{center panel} and \emph{right panel} three PCs are also included in the fit. For ease of comparison, the green triangles in the center panel reproduce the black asterisks in the left panel. For the PCs, the dimensionless uncertainties plotted here should be multiplied by $p_\mathrm{ann}$ to obtain the uncertainties on the $\varepsilon_i$ coefficients. The uncertainties on the cosmological parameters have been divided by the central values of the respective parameters.}
\label{fig:cosmomcsigma}
\end{figure*}

\begin{figure}
\includegraphics[width=0.4\textwidth]{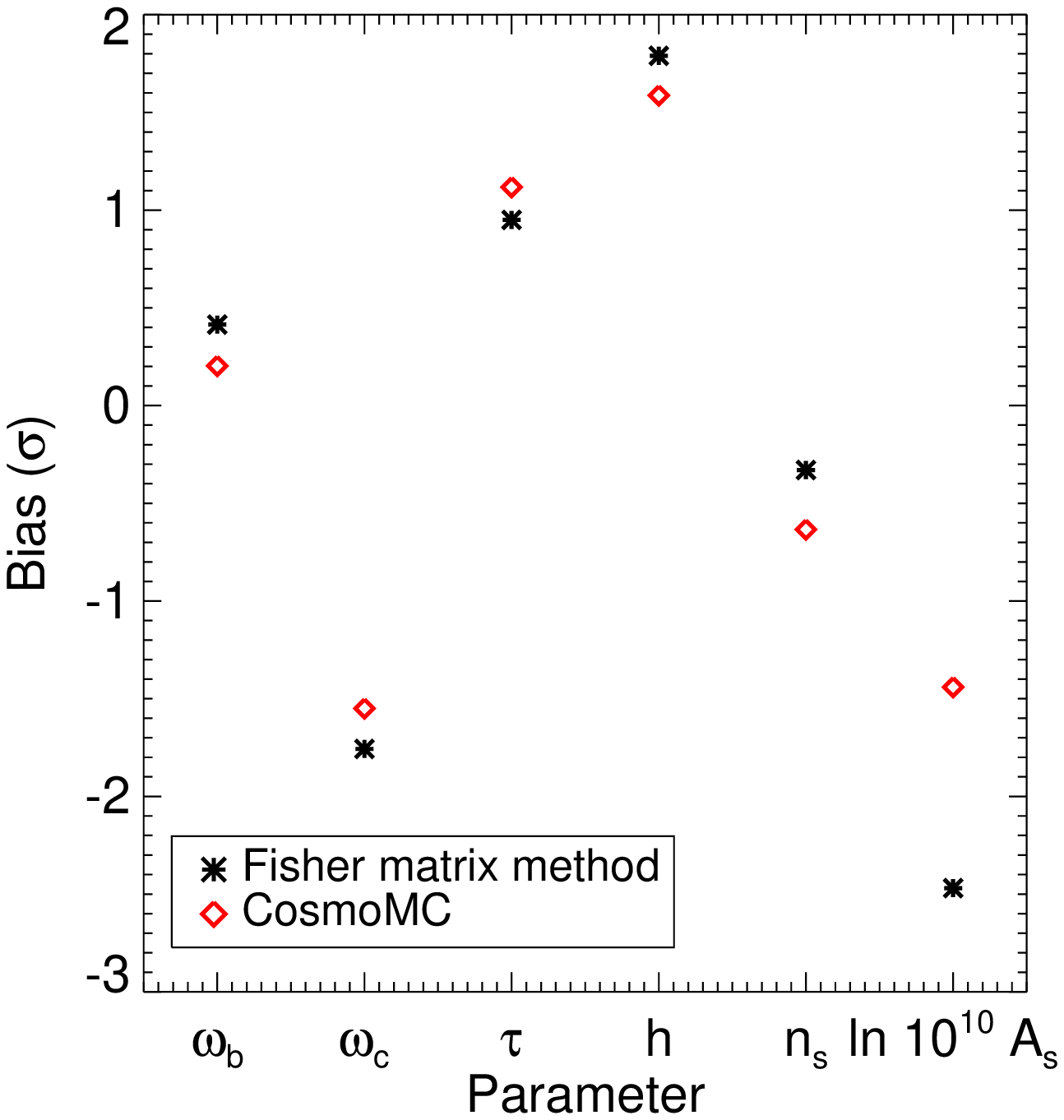}
\caption{Biases to the cosmological parameters in mock \PLANCK data, simulated assuming a constant-$p_\mathrm{ann}(z)$ energy deposition history with $p_\mathrm{ann}=1\times 10^{-6}$m$^3$/s/kg = $1.8 \times 10^{-27}$ cm$^3$/s/GeV. The fit to the data is performed using only the standard six cosmological parameters.}
\label{fig:cosmomcbiascomparison}
\end{figure}

In Figures \ref{fig:cosmomcsigma}-\ref{fig:cosmomcbiascomparison}, we compare the error bars on the cosmological parameters and principal components, and the bias to the cosmological parameters due to assuming no energy deposition, for the Fisher matrix method and \texttt{CosmoMC}. We find in general that the results of the Fisher matrix method are in good agreement with the full likelihood analysis, accurately predicting the error bars on the reconstructed values of the PCs and the cosmological parameters. For example, for the case of the \PLANCK fit with three PCs, the Fisher matrix predictions for the errors on the reconstructed PCs are within $5\%$ of the true values for the first two PCs, and $\sim 15\%$ different for the third PC, if the true energy deposition is small enough to lie in the linear regime. The biases to the cosmological parameters, when no PCs are included in the fit but the ``true'' energy deposition history is one of constant positive $p_\mathrm{ann}$, are not quite as well matched; the Fisher matrix method adequately captures the directions and approximate sizes of the various biases, but significantly overpredicts the bias to $A_s$ in particular, in the example presented here.  To obtain precise measurements of the biases, a \texttt{CosmoMC} analysis like the one we have performed is essential. 

\section{Conclusion}

Principal component analysis provides a simple and effective
parameterization for the effect of arbitrary energy deposition histories on
anisotropies in the cosmic microwave background. We find that for DM
annihilation-like energy deposition histories the first principal component,
describing the bulk of the effect, is peaked around $z\sim 500-600$,
at somewhat lower redshift than previously expected; the later
principal components provide corrections to this basic weighting
function.

The principal components, derived from a Fisher matrix approach, are
stable against a wide variety of perturbations to the analysis,
including choice of code calculating the ionization history, additions
to the usual set of cosmological parameters, the inclusion or
exclusion of ionization on helium, the range of included multipoles,
and the choice of binning. (For further discussion, see
Appendix~\ref{app:validation}.) The one significant potential change
to the PCs arises from how deposited energy is attributed to
additional Lyman-$\alpha$ photons: we have showed the effect of on one
hand neglecting this channel, and on the other of assuming that all
the energy attributed to ``excitations'' is converted into
Lyman-$\alpha$, which should bracket the true result. We eagerly await
a more careful analysis of this problem.

Within the Fisher matrix formalism, it is straightforward to take into account degeneracies with the standard cosmological parameters. We have presented predictions for the (significant!) biases that would arise in \PLANCK as a result of falsely assuming energy deposition to be zero, for each of the principal components. We have confirmed the previously noted degeneracy between energy deposition and $n_s$, and to a lesser degree with $A_s$, $\omega_b$ and $\omega_c$, in \WMAP data; since our analysis decomposes the biases according to the principal components that generate them, it is now trivial to compute the biases to the cosmological parameters for any arbitrary energy deposition history, in \WMAP or in mock \PLANCK data.

For a wide range of energy deposition histories, spanning models of dark matter annihilation and decaying species where annihilation or decay can begin or end abruptly on characteristic timescales shorter than the age of the universe, the coefficients of up to \emph{three} principal components are potentially measurable by \PLANCKc, for energy deposition histories satisfying $95\%$ confidence limits from \WMAPc, opening up the exciting possibility of distinguishing different models of energy deposition. For a CVL experiment, up to \emph{five} coefficients  are measurable. 

For the ``standard'' WIMP annihilation case, principal component
analysis on a large set of WIMP models yields a single principal
component $e_{\rm WIMP}(z)$ that describes the effect on the
$C_\ell$'s of all the models very well; any model is then
parameterized simply by the coefficient of $e_{\rm WIMP}(z)$ (or
equivalently, effective $f$).  Our analysis confirms previous
statements in the literature, and we have provided this ``universal
$f(z)$'' curve for future WIMP annihilation studies.

We performed an accurate MCMC analysis of current \WMAP7 data to impose constraints on the measurable principal component amplitudes, and to forecast constraints for future experiments such as \PLANCK or a CVL experiment. We find good agreement with the Fisher matrix analysis, although the MCMC analysis is required to accurately predict the biases on the cosmological parameters. We have illustrated how it is possible to recover the constraints on an arbitrary energy deposition history from the constraints on the amplitudes of the principal components. The reconstructed constraints are in very good agreement with the constraints obtained by directly sampling a specific energy deposition history, confirming the validity and usefulness of the principal component decomposition.

We wish to acknowledge helpful conversations with Jens Chluba, Joanna Dunkley, Fabio Iocco, Eric Linder, Alessandro Melchiorri, Nikhil Padmanabhan and Joshua Ruderman.  We are
grateful to Antony Lewis for his advice on increasing the numerical accuracy
of CAMB.  DPF and TL are partially supported by NASA Theory grant NNX10AD85G. TRS is supported by the NSF under grant NSF AST-0807444 and the DOE under grant DE-FG02-90ER40542.
This research made use of the NASA Astrophysics Data System (ADS) and the IDL
Astronomy User's Library at Goddard \footnote{Available at
\texttt{http://idlastro.gsfc.nasa.gov}}.

\onecolumngrid

\clearpage

%%%%%%%%%%%%%%%%%%%%%%%%%%%%%%%%%%%%%%%%%%%%

\begin{appendix}

\section{Validation of the PCA method}
\label{app:validation}

In this appendix we discuss a number of specific issues that might affect the results of the principal component analysis, and motivate our default choices.

\subsection{Choice of binning}

%%%%%%%%%%%%%%%%%%%%%%%%%%%%%%%%%%%%%%%%%%%%

\begin{figure*}
\includegraphics[width=0.45\textwidth]{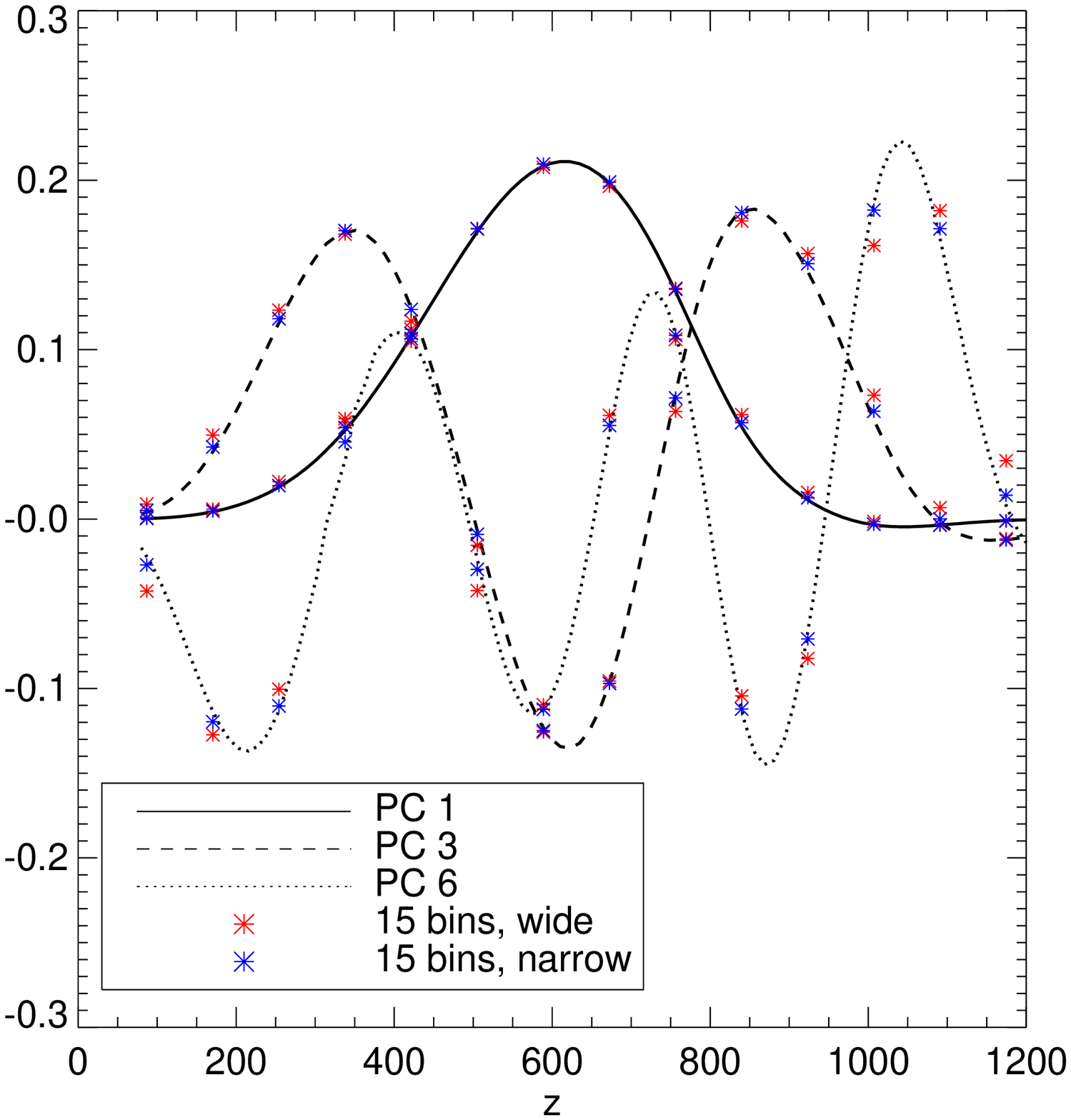}
\includegraphics[width=0.45\textwidth]{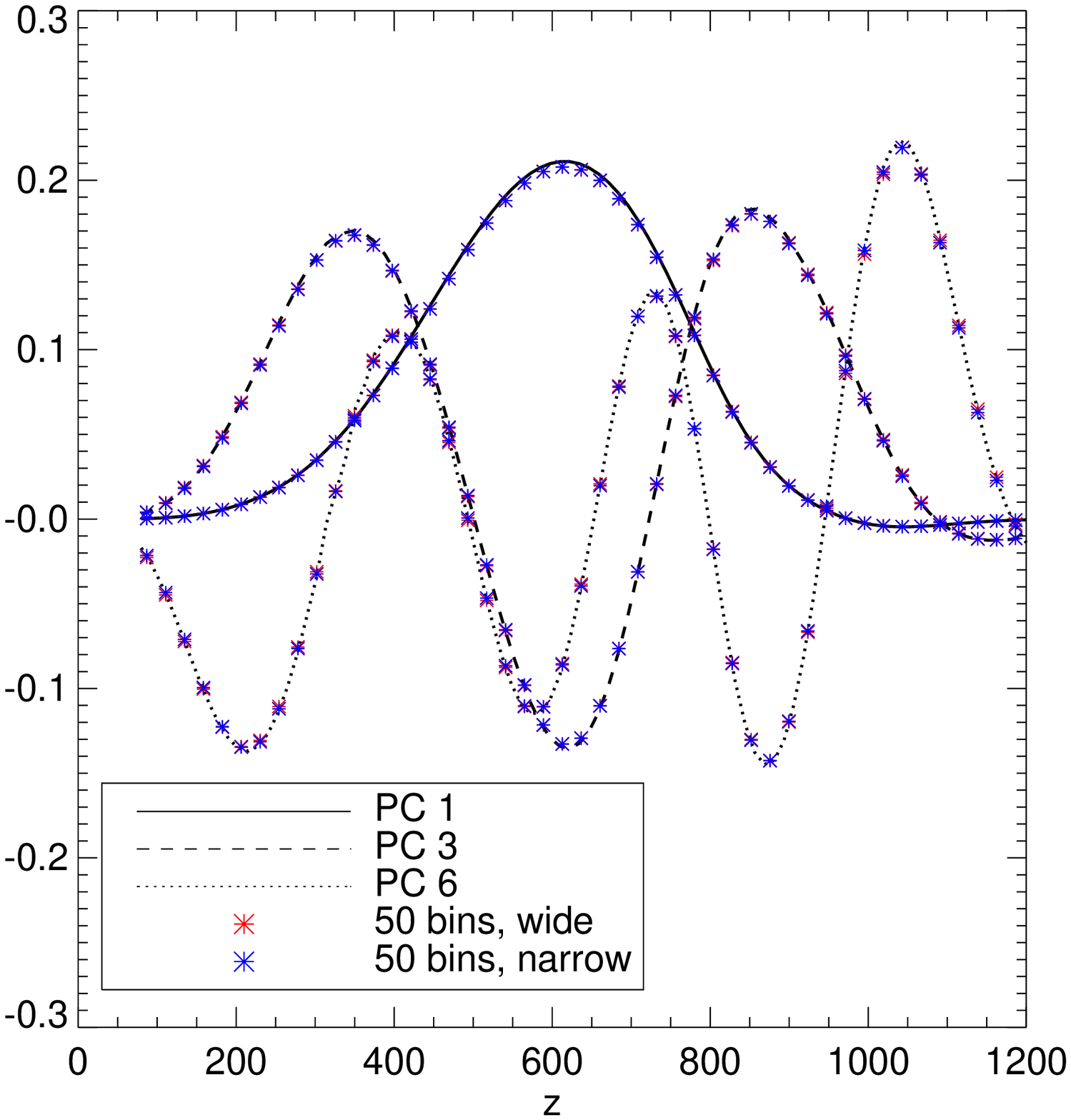}
\caption{An example of the dependence of the principal components on the number of bins and the width of the Gaussians, for the first, third and sixth principal components in \PLANCKc, after marginalization over the cosmological parameters (the later principal components are shown as they are more sensitive to small changes in the binning). Black lines indicate the case with 100 linearly spaced bins and $\sigma =$ bin width/2; in the left (right) panel, red and blue dots indicate respectively the case with 15 (50) linearly spaced bins and $\sigma=$ bin width/2, and the case with 15 (50) linearly spaced bins and $\sigma =$ bin width/4.}
\label{fig:bindependence}
\end{figure*}

%%%%%%%%%%%%%%%%%%%%%%%%%%%%%%%%%%%%%%%%%%%%

As discussed previously, to approximate $\delta$-functions in energy deposition, as a default we employ (for the annihilation-like case) 50 linearly-spaced redshift bins covering the redshift range from $z=80$ to $z=1300$. Figure \ref{fig:bindependence} shows the effect of changing the Gaussian width and the number of bins for some sample PCs; doubling the number of bins or changing the Gaussian width by a factor of 2 has no effect on the PCs, although reducing the number of bins to 15 does affect the PCs slightly, especially at the lowest and highest redshifts.

In \S \ref{sec:fisher} we briefly discussed the choice of log vs linear binning, preferring to use the former for decay-like scenarios and the latter for annihilation-like scenarios. In the annihilation-like case, the choice of linear-spaced bins in redshift was not inevitable; log-spaced bins seem equally natural. The choice of log or linear binning is somewhat subtle, as it affects whether or not two energy deposition histories are considered orthogonal, and how much different redshifts contribute to the norm of a particular energy deposition profile. Consequently, the two choices give rise to different sets of principal components and eigenvalues, and a ``generic'' energy deposition history is somewhat different between the two cases. However, the eigenvalues of the first several PCs are quite similar, and using log binning instead of linear binning does not significantly affect the results of \S \ref{sec:sensitivity}: in both cases a similar (small) number of PCs are sufficient to describe a very broad class of energy deposition histories, for the purpose of future experiments. For individual models, of course, the number of measurable PCs depends on the choice of basis: a $p(z)$ curve which happens to be very well described as a linear combination of the first two PCs in the log-binned case may not be nearly so well described by the first two PCs in the linear-binned case, or vice versa.

In Figure \ref{fig:pcs} we showed the \PLANCK PCs for both linear and log binning in the annihilation case. The first PC is peaked at higher redshift in the log-binned case: this is to be expected, since in the log-binned case the amount of energy deposited per bin for a constant $p(z)$ has an extra $(1+z)$ scaling relative to the linear case (simply due to the wider bins at high redshift). It can be shown that this simple rescaling largely (but not completely) describes the difference between log and linear binning for the first PC, but the later PCs shift by a larger amount and in more complex ways, as each must be orthogonal to all previous PCs.

\subsection{The effect of noise and maximum $\ell$}

\begin{figure*}
\includegraphics[width=0.45\textwidth]{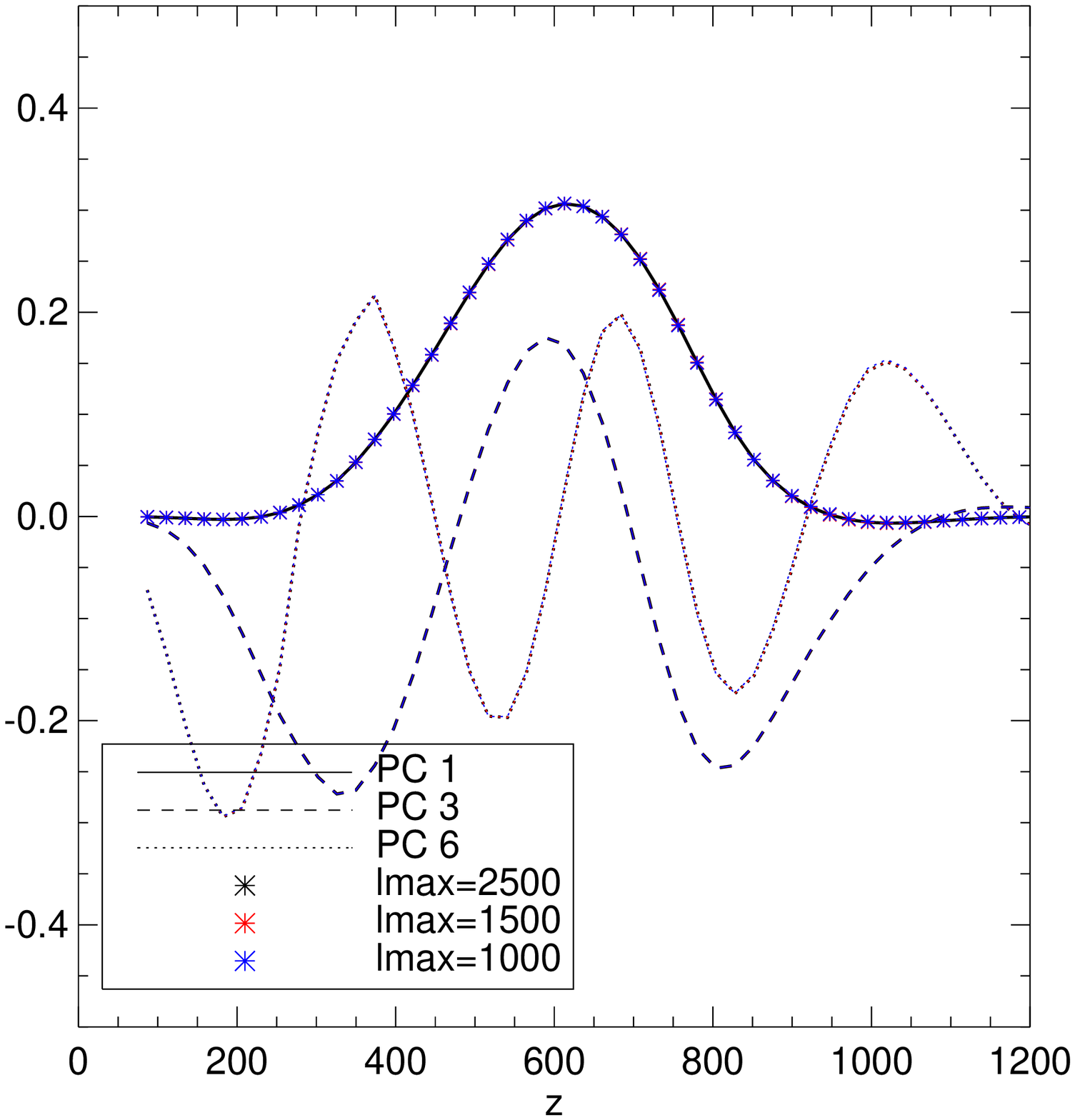}
\includegraphics[width=0.45\textwidth]{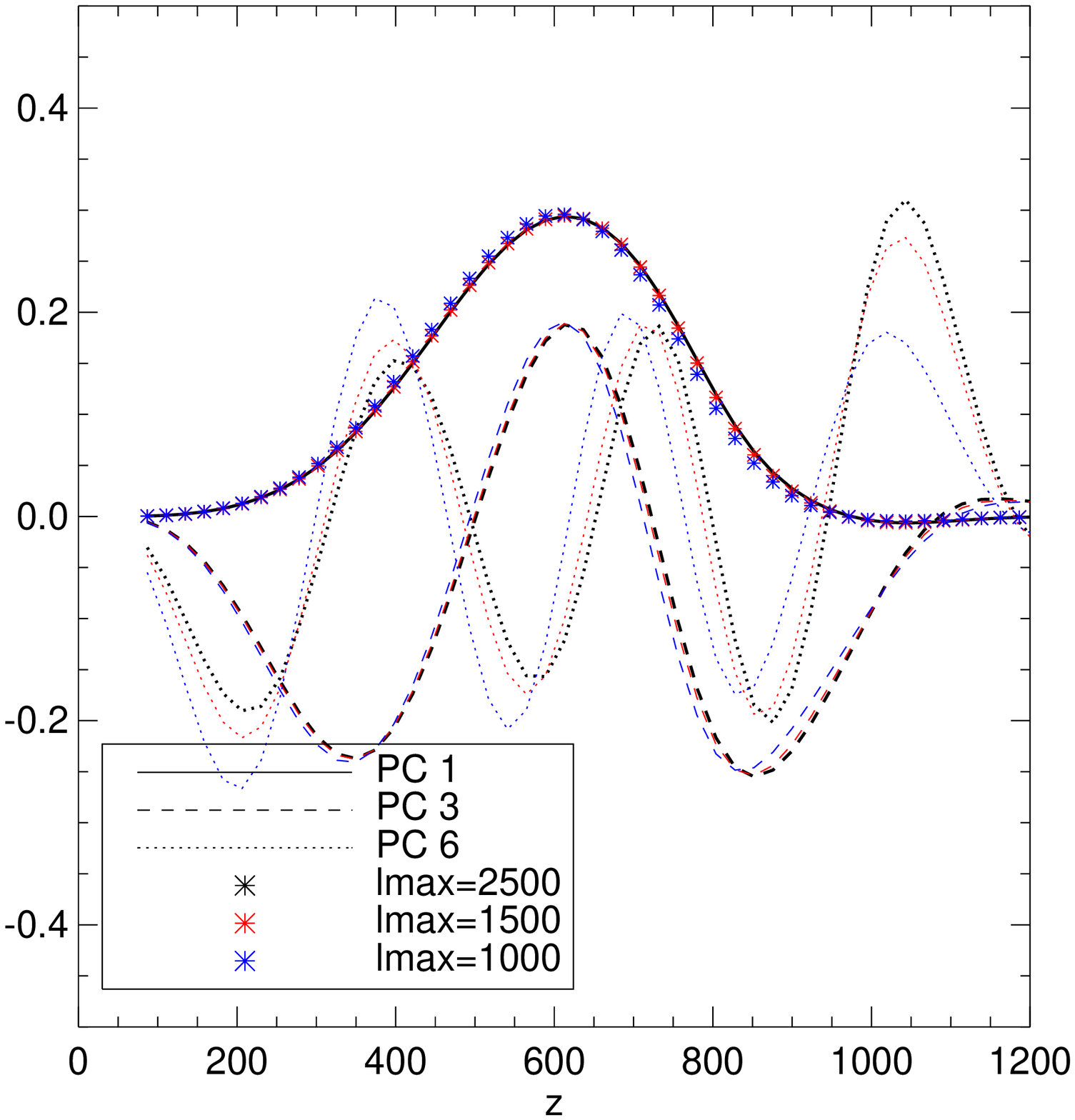}
\caption{The dependence of a selection of PCs on the range of $\ell$ included in the analysis, for \WMAP7 (\emph{left}) and \PLANCK (\emph{right}).}\label{fig:lmaxpcs}
\end{figure*}

As a default, we include $\ell=2..2500$ in our analysis, for both temperature and polarization. $\ell$'s above 1000 do not noticeably affect the principal components for \WMAPc, but have significant effects for \PLANCK (and the CVL case), especially in the higher PCs. Figure \ref{fig:lmaxpcs} shows the effect on a selection of PCs on changing $\ell_\mathrm{max}$. The results also depend on the estimated sensitivity of the experiment. 

\subsection{Choice of ionization history calculator and recombination corrections \label{app:recfast_vs_cosmorec}}

\begin{figure*}
\includegraphics[width=0.3\textwidth]{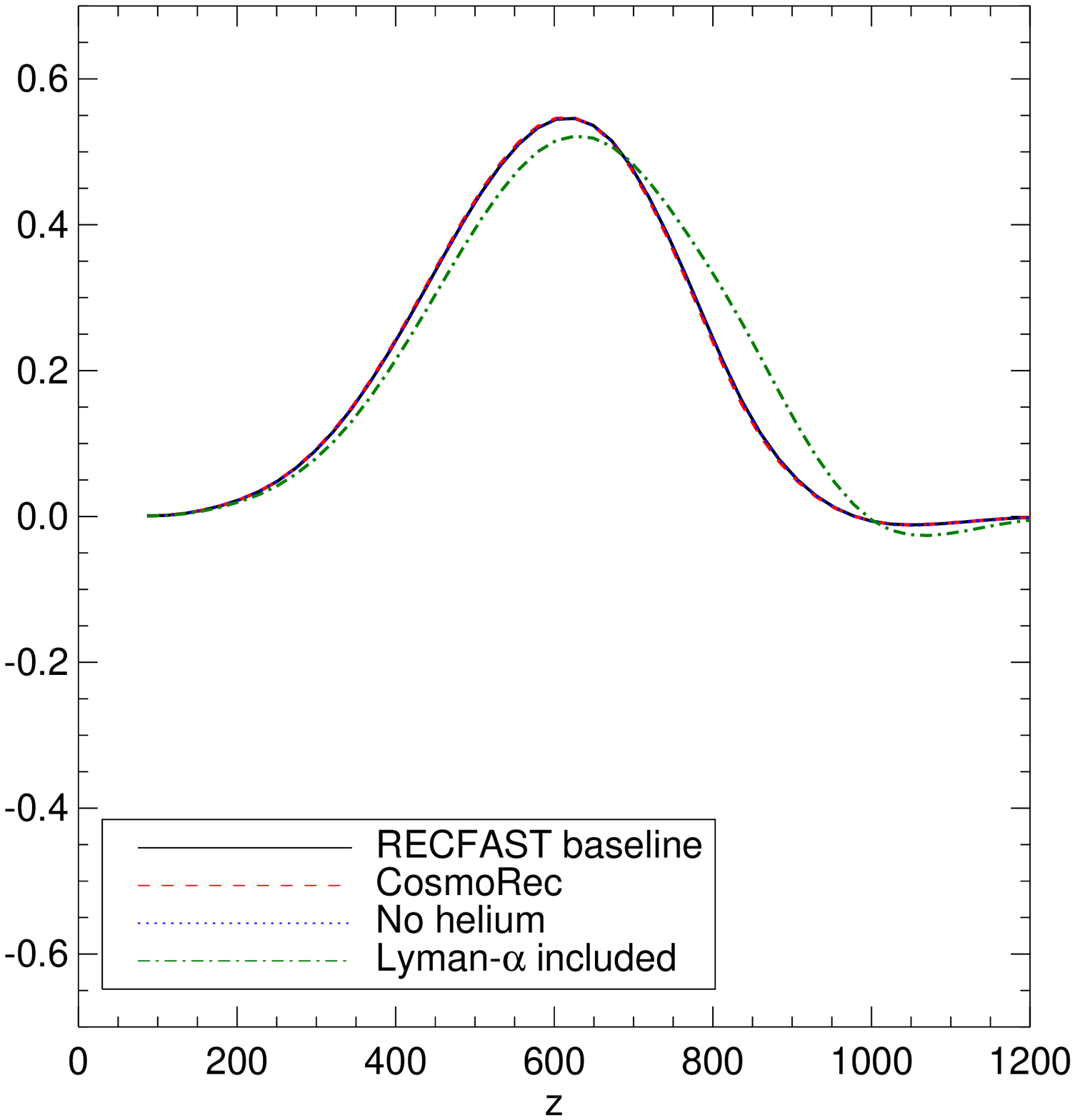}
\includegraphics[width=0.3\textwidth]{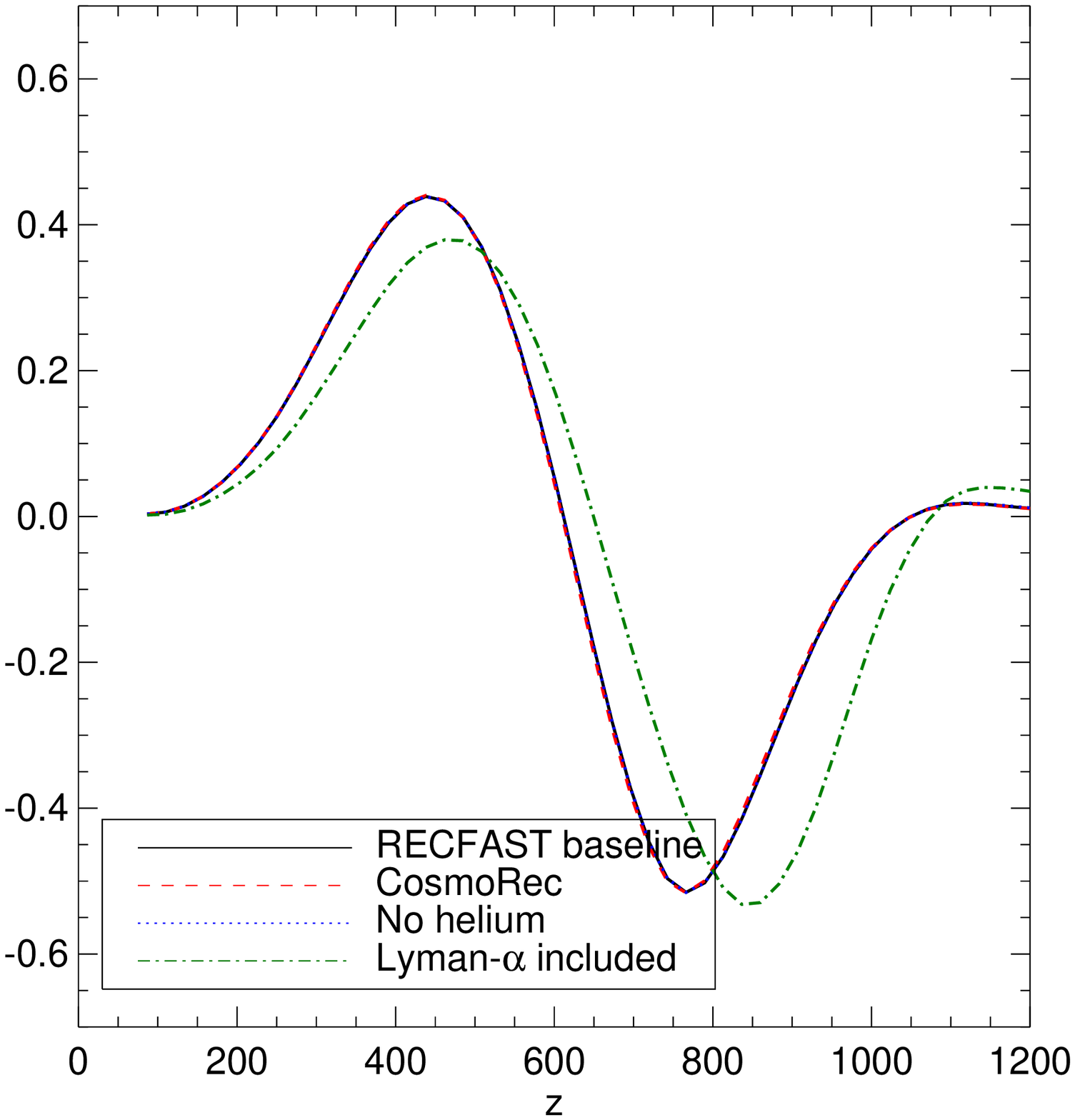}
\includegraphics[width=0.3\textwidth]{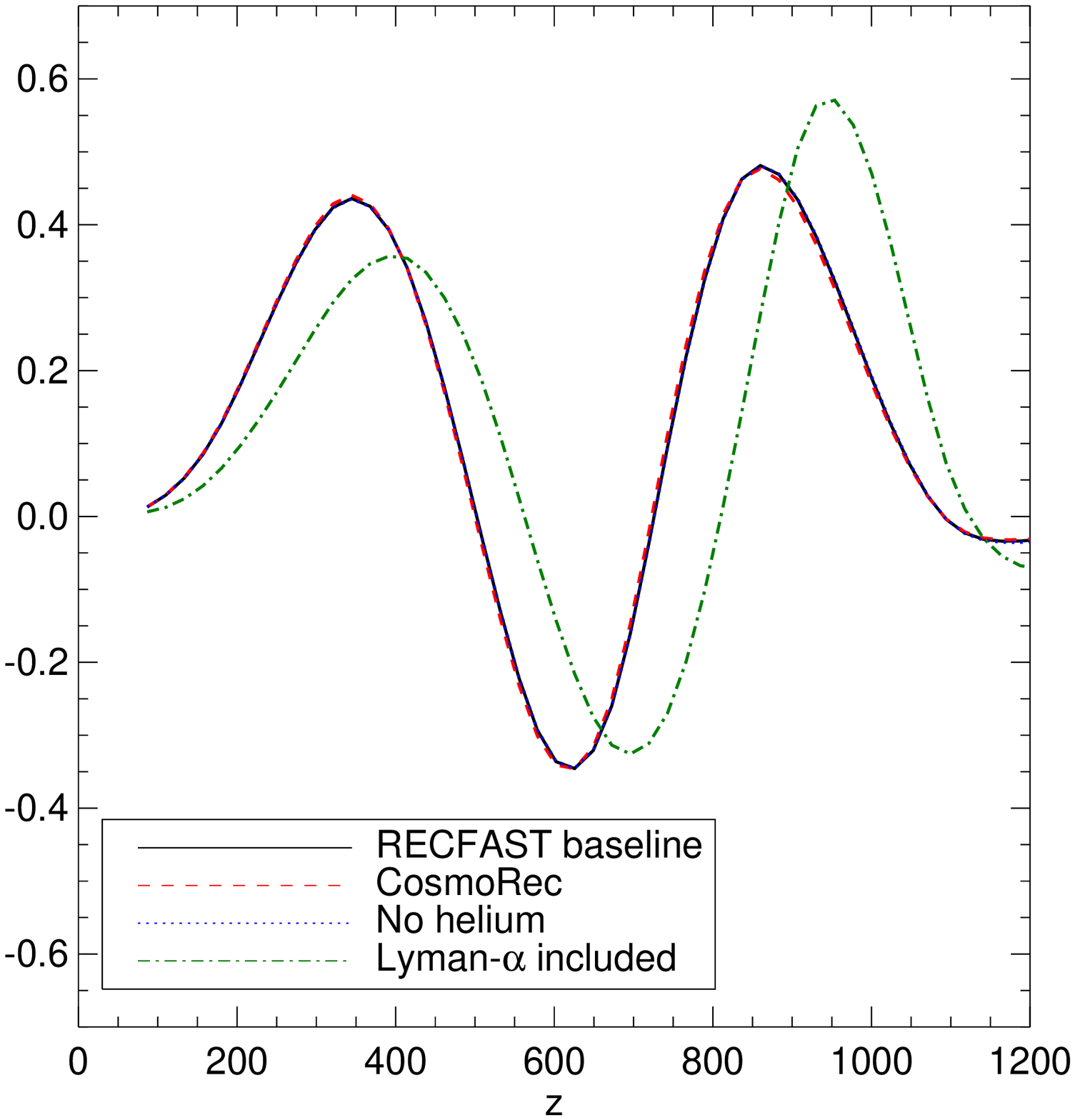} 
\caption{The first three principal components for \PLANCKc, after marginalization, computed using \texttt{RECFAST 1.5} and \texttt{CosmoRec}. In the baseline case (as in \texttt{CosmoRec}), ionization of helium is included but injection of Lyman-$\alpha$ photons is not. We also show the effects of including a contribution to Lyman-$\alpha$ photons, and neglecting helium. The effect of helium ionization on the PCs is negligible because it is approximately a redshift-independent effect.}
\label{fig:recfastcomparison}
\end{figure*}

Recent improvements in the detailed treatment of recombination have motivated new codes to compute the ionization history, in particular \texttt{CosmoRec} \cite{Chluba:2010ca} and \texttt{HyRec} \cite{AliHaimoud:2010dx}. The inclusion of these additional recombination corrections can modify the effect of energy deposition on the ionization history and the CMB, thus shifting the derivatives and modifying the principal components. However, comparing the principal components obtained using \texttt{RECFAST} 1.5 to those obtained using \texttt{CosmoRec}, we find that there is essentially no difference. The overall normalization of the effect, for a given energy deposition history, is slightly smaller (by a few percent) in \texttt{CosmoRec} as compared to \texttt{RECFAST} 1.5, but as shown in Figure \ref{fig:recfastcomparison} the shapes of the PCs are unaffected. This shift in normalization will not change the detectability of the maximum \WMAP-allowed signal in \PLANCKc, and will only very slightly modify the amount of energy deposition required to produce such a signal.

\subsection{Helium and Lyman-$\alpha$}
\label{app:he_lya}

Ionization of helium also contributes to the population of free electrons, and is included automatically in \texttt{CosmoRec}: however, some previous studies have neglected it, focusing only on the H contribution. We include helium by default throughout, but show in Figure \ref{fig:recfastcomparison} the (negligible) effect on the PCs of leaving it out. Similarly to the choice of ionization history calculator, including helium increases the total effect of a given energy deposition history, but does not significantly modify the redshift dependence and hence the PCs.

A more difficult question is the effect of additional Lyman-$\alpha$ photons. The energy from dark matter annihilation that is deposited to the gas is partitioned between excitation, ionization and heating. In the analysis of \cite{Padmanabhan:2005es}, which we have followed, only the latter two processes are included; excitations are assumed to have no significant effect. However, it was pointed out in \cite{Chen:2003gz} that the increased population of Lyman-$\alpha$ photons from excitations -- or equivalently, the higher fraction of H in an excited state -- facilitates ionization and thus indirectly increases the ionization fraction.

We studied the effect of including Lyman-$\alpha$ photons, following \cite{Chen:2003gz, Bean:2003kd} and assuming all the energy partitioned into excitation produces Lyman-$\alpha$ photons. The true picture is probably more complicated, but its study lies beyond the scope of this work, and this choice and our baseline case with no additional Lyman-$\alpha$ photons should bracket the true solution. Figure \ref{fig:recfastcomparison} shows the effect on the PCs of including the Lyman-$\alpha$ contribution. The corresponding effect on the $S/N$ for the various PCs is indicated by the hatched region in Figure \ref{fig:fzscan_planck}; the bound on constant $p_\mathrm{ann}$ is strengthened by $\sim 7-10\%$ (depending on the experiment). 

%%%%%%%%%%%%%%%%%%%%%%%%%%%%%%%%%%%%%%%%%%%%

\subsection{Additional cosmological parameters}

Since the primary effect of dark matter annihilation is an $\ell$-dependent damping of the temperature anisotropies, one might ask if it is degenerate with other cosmological parameters that have a similar effect: for example, including running of the scalar spectral index could better mimic the profile of the damping with respect to $\ell$, increasing the number of massless neutrino species $N_\mathrm{eff}$ also suppresses small-scale anisotropy, and the primordial helium mass fraction $Y_P$ has degeneracies with $N_\mathrm{eff}$ \cite{Hou:2011ec}. Accordingly, we have redone the principal component analysis including each of these parameters separately, as well as all of them simultaneously. 

In all cases, we find only negligible shifts to the principal components. The forecast constraint on constant $p_\mathrm{ann}$ from \PLANCK is weakened by $\sim 7\%$ if running of the spectral index is included, and by $\sim 12\%$ if all three parameters are used. For \WMAP7, the Fisher-matrix-estimated limit weakens by $\sim 11\%$ and $\sim 13\%$ in these two cases.

%%%%%%%%%%%%%%%%%%%%%%%%%%%%%%%%%%%%%%%%%%%%%

\section{Review of marginalization and biases}
\label{app:marg}

\subsection{Marginalization as projection}

Marginalization over some parameters in the Fisher matrix is
equivalent to a projection of the vector space spanned by all the
parameter perturbations. Another way of saying this is that given the
transfer matrix (Equation \ref{eq:transfermatrix}) for the parameters of interest, then the marginalized
Fisher matrix for those parameters is formed by contracting the {\it
projected} transfer matrix with the covariance matrix $\Sigma^{-1}$.

We illustrate these statements for the situation considered in this
paper. The total vector space consists of the $\delta C_\ell$'s
spanned by energy deposition and cosmological parameter perturbations,
and we wish to marginalize over the cosmological parameters. Recall
that the full Fisher matrix is
\begin{equation}
    \left( \begin{array}{cc} 
      F_e & F_v \\
      F_v^T & F_c
      \end{array} \right) = 
    \left( \begin{array}{cc} 
      T_e^T \Sigma^{-1} T_e & T_e^T \Sigma^{-1} T_c \\
      T_c^T \Sigma^{-1} T_e & T_c^T \Sigma^{-1} T_c
      \end{array} \right),
\end{equation}
where $T_e$ and $T_c$ are transfer matrices mapping energy deposition histories
and cosmological parameter perturbations, respectively, to the space
of $\delta C_\ell$'s. For ease of notation we suppress indices. Using
Equation \ref{eq:fisher_inversion}, the marginalized Fisher matrix can
be written as
\begin{align}
  F & = F_e - F_v F_c^{-1} F_v^T
% \nonumber \\   & = T_e^T \Sigma^{-1} \left(1 - T_c \left( F_c \right)^{-1} T_c^T \Sigma^{-1} \right) T_e  
    = \left( P T_e \right)^T \Sigma^{-1} \left( P T_e \right)
\end{align}
where $P = \left(1 - T_c \left( F_c \right)^{-1} T_c^T \Sigma^{-1}
\right)$ satisfies $P^2 = P$ and so is a projection operator. $P$
projects out any component of $\delta C_\ell$ which can be effectively
absorbed by a change in the cosmological parameters. This can also be
seen if we act with $P$ on a generic perturbation in the cosmological
parameters: $P T_c \delta \theta_\alpha =0$. Accordingly, any
perturbation to the fiducial CMB model can be written as
\begin{equation}
  \delta C_\ell = (1-P)\delta C_\ell + P\delta C_\ell 
         = \delta C_\ell^{||}+ \delta C_\ell^{\perp}.
\end{equation} 
where $\perp$ and $||$ mean perpendicular or parallel to the
cosmological parameter perturbations.

Thus $P T_\epsilon$ is a projected transfer matrix, taking energy
deposition histories to the subspace of $\delta C_\ell$'s which are orthogonal to
cosmological parameter perturbations. This projection depends on the
noise parameters in $\Sigma^{-1}$, since the notion of orthogonality
depends on our definition of norm. Intuitively, if an energy deposition
and cosmological parameter perturbation have very similar effects at
low $\ell$ but are different at high $\ell$, then the projection
operator may give a very small $\delta C_\ell^\perp$ in the case of
\WMAP and a larger $\delta C_\ell^\perp$ in the case of \PLANCKc.

The eigenvectors ${e_i}$ of $F$ (or principal components) with the
largest eigenvalues are correspondingly those with the largest
measurable $\delta C_\ell^\perp$. The $e_i$ also map to an orthogonal
vector space of $\delta C_\ell^\perp$'s.  The unprojected $h_i=T_e
e_i$ are in general not orthogonal because $h_i\cdot h_j= (T_e e_i)^T
\Sigma^{-1} (T_e e_j)=e_i^{T}F_e e_j$. However, defining
$h_i^\perp=PT_e e_i$, we have
\begin{equation}
  h_i^\perp\cdot h_j^\perp= (PT_e e_i)^T\Sigma^{-1} (PT_e e_j) 
       = e_i^{T} F e_j =\delta_{ij}\lambda_i.
\end{equation}
Because the $h_i^\perp$ are orthogonal, we can extract the
coefficients of the principal components from a generic $R_\ell \in
\{\delta C_\ell\}$:
\begin{align}
  \varepsilon_i &= \frac{1}{\lambda_i} (P T_e e_i)^T \Sigma^{-1} R_\ell =  \frac{R \cdot h_i^\perp}{\lambda_i}.
\end{align}

\subsection{Biases to cosmological parameters}

The parallel components from energy deposition, $\delta C_\ell^{||} = (1-P)T_e e_i \varepsilon_i$, correspond to the biases to the cosmological parameters. Suppose there is some energy deposition ($\varepsilon_i \neq 0$), but it is not included in the fit to the data. Then the best fit cosmological parameters will absorb any $\delta C_\ell^{||}$ from the energy deposition, and the measurements of the true cosmological parameters would be biased by the amount it takes to produce $\delta C_\ell^{||}$:\begin{equation}
  \delta C_\ell^{||}  = T_c \delta \theta_\alpha = (1-P)T_e e_i \varepsilon_i.
\end{equation}
Multiplying both sides by $(F_c)^{-1} T_c^T \Sigma^{-1}$, we have
\begin{align}
  \delta \theta_\alpha
  = F_c^{-1} F_v^T e_i \varepsilon_i.
\end{align}

\section{Web files description}
\label{app:website}

To facilitate the analysis described above, we provide the PC vectors defined
above on our web page\footnote{http://nebel.rc.fas.harvard.edu/epsilon}.  The
files are provided in two formats: as a the Flexible Image Transport
System\footnote{http://fits.gsfc.nasa.gov} (FITS) binary table, and as ASCII
plain text files.  The file formats are described briefly in this appendix,
and in more detail on the web page. 

\subsection{FITS files}
Each FITS file has the following format:
\begin{verbatim}
L                 Array[2500]        - multipole index (2..2501)
REDSHIFT          Array[50]          - redshift (z)
PC_EIGENVECTORS   Array[50, 5]       - PC as a function of redshift
EPSILON           2.0000000e-27      - energy deposition, see Section IIIB of paper
EIGENVALUES       Array[5]           - PC eigenvalues
PC_POWSPEC        Array[2500, 3, 5]  - PCs projected into Cl space, TT, EE, TE
PC_POWSPEC_PERP   Array[2500, 3, 5]  - same, but projected onto the space perpendicular to the 
                                       cosmological parameters
\end{verbatim}

We also provide the universal $e_\mathrm{WIMP}$ curve derived in \S \ref{subsec:dmpca}, following the same notation, and a list of efficiency coefficients for the WIMP models studied in \cite{Slatyer:2009yq}.

\subsection{ASCII files}
We also provide the mappings of the PCs into $\delta C_\ell$'s in ASCII files, one per principal
component.  File names are e.g.

\begin{verbatim}
epsilon_ann_Planck_PC01.dat
epsilon_ann_Planck_PC01_perp.dat
\end{verbatim}

etc., with one file for each PC, choice of binning, and experiment (CVL, \PLANCK and \WMAP7). Each of these ASCII files contains 4 columns: $\ell, PC_{TT}, PC_{EE},
PC_{TE}$.

For each experiment and choice of binning, one additional ASCII file contains the ordered eigenvalues for the principal components, and another holds the ordered eigenvectors / PCs as functions of redshift. These files are named e.g.

\begin{verbatim}
epsilon_ann_Planck_PC_eigenvalues.dat
epsilon_ann_Planck_PC_eigenvectors.dat
\end{verbatim}

\subsection{Units}
The PC vectors \texttt{PC\_POWSPEC} and \texttt{PC\_POWSPEC\_PERP}  are the changes in $ \ell (\ell + 1) C_\ell / 2 \pi$ (before and after projecting out the cosmological parameters, respectively) corresponding to an energy deposition history given by $\varepsilon_i = \varepsilon$ for all $i$ (where the $\varepsilon_i$ are dimensionful coefficients defined in Equation \ref{eq:decomposedpz}). Our convention,
described in \S \ref{sec:pcunits}, is that the $C_\ell$ have units of $\microK^2$, the principal components $e_i$ are dimensionless,
energy deposition from annihilations has
units cm$^3$/s/GeV, and energy deposited by decays has units of s$^{-1}$ (see Equation \ref{eq:pdef}). The energy deposition parameter $\varepsilon$ (labeled \texttt{EPSILON} in the files) is fixed at $2 \times 10^{-27}$ cm$^3$/s/GeV for the annihilation case, and $1 \times 10^{-26}\s^{-1}$ for
the decay case. 

\end{appendix}

\bibliography{epsilon}

\end{document}